\documentclass[10pt,journal,compsoc]{IEEEtran}
\ifCLASSOPTIONcompsoc
  \usepackage[nocompress]{cite}
\else
  \usepackage{cite}
\fi

\usepackage{amsmath,amsfonts,amssymb}
\usepackage{graphicx}
\usepackage{textcomp}
\usepackage{xcolor}
\def\BibTeX{{\rm B\kern-.05em{\sc i\kern-.025em b}\kern-.08em
		T\kern-.1667em\lower.7ex\hbox{E}\kern-.125emX}}

\usepackage{etoolbox}
\usepackage{multirow}
\usepackage{tabularx}

\newcolumntype{C}{>{\centering\arraybackslash}X}
\usepackage{booktabs}
\usepackage{lipsum}
\usepackage[aboveskip=0pt]{subcaption}
\usepackage{amsthm}
\usepackage{pgfplots}
\pgfplotsset{compat=1.16}
\newtheorem{lem}{Lemma}
\newtheorem{theorem}{Theorem}
\newtheorem{corollary}{Corollary}[theorem]
\usepackage{algorithmic,float}
\usepackage[ruled,linesnumbered,noend]{algorithm2e}
\let\oldnl\nl
\newcommand{\nonl}{\renewcommand{\nl}{\let\nl\oldnl}}
\usepackage{etoolbox}
\usepackage{caption} 
\makeatletter
\patchcmd\algocf@Vline{\vrule}{\vrule \kern-0.4pt}{}{}
\patchcmd\algocf@Vsline{\vrule}{\vrule \kern-0.4pt}{}{}
\makeatother

\SetKwInput{KwInput}{Input}               
\SetKwInput{KwOutput}{Output}         
\usepackage{tikz}
\usepackage{environ}
\newcommand\scalemath[2]{\scalebox{#1}{\mbox{\ensuremath{\displaystyle #2}}}}

\definecolor{ballblue}{rgb}{0.13, 0.67, 0.8}

\newlength\myindent
\makeatletter
\newcommand{\labeltext}[2]{
	\@bsphack
	\csname phantomsection\endcsname 
	\def\@currentlabel{#1}{\label{#2}}
	\@esphack
}
\makeatother
\newcommand*{\thead}[1]{\multicolumn{1}{|c|}{\bfseries #1}}

\usetikzlibrary{shapes.geometric,arrows,arrows.meta,calc,fit,matrix}
\definecolor{arrowcolor}{rgb}{.27,.45,.77}
\tikzstyle{arrow} = [arrowcolor,opacity=1,thin, {Triangle[angle=60:1.2mm]}-{Triangle[angle=60:1.2mm]}]
\tikzstyle{chart} = [rectangle, minimum width=3cm, minimum height=1cm, text centered,text width =3cm, draw=black, fill=white!30]
\tikzstyle{circlular} = [circle, minimum width=1cm, minimum height=1cm, text centered,text width =1cm, draw=black, fill=white!30]
\tikzstyle{circ} = [circle, minimum width=5mm, minimum height=5mm, text centered,text width =5mm, draw=black, fill=yellow!30]
\tikzstyle{roundrect3} = [rectangle,rounded corners, minimum width=3cm, minimum height=1.6cm, text centered,text width =1.4cm, draw=black, fill=red, text opacity=1]
\tikzstyle{roundrect4} = [rectangle,rounded corners, minimum width=3cm, minimum height=1.6cm, text centered,text width =1.4cm, draw=black, fill=blue, text opacity=1]
\tikzstyle{roundrect6} = [rectangle,rounded corners, minimum width=3cm, minimum height=1.6cm, text centered,text width =1.4cm, draw=black, fill=green, text opacity=1]
\tikzstyle{invisible} = [minimum width=3cm, minimum height=2cm, text centered,text width =1.4cm, draw=black, opacity = 1]
\tikzstyle{tinyarrow} = [thin,->,>=stealth]
\tikzstyle{strippedline} = [thick,dotted,>=stealth]

\BeforeBeginEnvironment{appendices}{\clearpage}

\makeatletter
\newcommand{\removelatexerror}{\let\@latex@error\@gobble}
\makeatother

\begin{document}

\title{Efficient Generalized Temporal Pattern Mining in Big Time Series Using Mutual Information}

\author{Van Long Ho,
	Nguyen Ho,
    Torben Bach Pedersen,~\IEEEmembership{CS Distinguished Contributor,~IEEE,}
        and Panagiotis Papapetrou
\IEEEcompsocitemizethanks{\IEEEcompsocthanksitem Van Long Ho, Nguyen Ho, and Torben Bach Pedersen are with the Department of Computer Science, Aalborg University, 9100 Aalborg, Denmark. \protect
 E-mail: \{vlh,ntth,tbp\}@cs.aau.dk.
\IEEEcompsocthanksitem Panagiotis Papapetrou is with the Department of Computer and Systems Sciences, Stockholm University, 7003 Stockholm, Sweden. \protect\\ E-mail: panagiotis@dsv.su.se.}
}

\IEEEtitleabstractindextext{
\begin{abstract}
Big time series are increasingly available from an ever wider range of IoT-enabled sensors deployed in various environments. Significant insights can be gained by mining temporal patterns from these time series. Temporal pattern mining (TPM) extends traditional pattern mining by adding event time intervals into extracted patterns, making them more expressive at the expense of increased time and space complexities. 
Besides frequent temporal patterns (FTPs), which occur frequently in the entire dataset, another useful type of temporal patterns are so-called {\em rare temporal patterns (RTPs)}, which appear rarely but with high confidence. Mining rare temporal patterns yields additional challenges. For FTP mining, the temporal information and complex relations between events already create an exponential search space. For RTP mining, the support measure is set very low, leading to a further combinatorial explosion and potentially producing too many uninteresting patterns. 
Thus, there is a need for a generalized approach which can mine both frequent and rare temporal patterns. 
This paper presents our {\em Generalized Temporal Pattern Mining from Time Series (GTPMfTS)} approach with the following specific contributions: (1) The end-to-end GTPMfTS process taking time series as input and producing frequent/rare temporal patterns as output. (2) The efficient {\em Generalized Temporal Pattern Mining (GTPM)} algorithm mines frequent and rare temporal patterns using efficient data structures for fast retrieval of events and patterns during the mining process, and employs effective pruning techniques for significantly faster mining. (3) An approximate version of GTPM that uses mutual information, a measure of data correlation, to prune unpromising time series from the search space. (4) An extensive experimental evaluation of GTPM for rare temporal pattern mining (RTPM) and frequent temporal pattern mining (FTPM), showing that RTPM and FTPM signficantly outperform the baselines on runtime and memory consumption, and can scale to big datasets. The approximate RTPM is up to one order of magnitude, and the approximate FTPM up to two orders of magnitude, faster than the baselines, while retaining high accuracy. 
\end{abstract}

}

\maketitle

\IEEEdisplaynontitleabstractindextext
\IEEEpeerreviewmaketitle

\section{Introduction}
\begin{figure}[!t]
	\resizebox{1\columnwidth}{!}{\begin{tikzpicture}

\newcommand{\DrawOn}[3]{
	\draw [black,thin] (axis cs: #1, #3+0.13) -- (axis cs: #2, #3+0.13);
	\node [black, scale=0.5] at (axis cs: #1/2+#2/2, #3+0.32) {\footnotesize On};
}

\newcommand{\DrawHigh}[3]{
	\draw [black,thin] (axis cs: #1, #3+0.13) -- (axis cs: #2, #3+0.13);
	\node [black, scale=0.5] at (axis cs: #1/2+#2/2, #3+0.32) {\footnotesize High};
}

\newcommand{\DrawOff}[3]{
	\draw [black,thin] (axis cs: #1, #3-0.12) -- (axis cs: #2, #3-0.12);
	\node [black, scale=0.5] at (axis cs: #1/2+#2/2, #3-0.3) {\footnotesize Off};
}
\newcommand{\DrawLow}[3]{
	\draw [black,thin] (axis cs: #1, #3-0.12) -- (axis cs: #2, #3-0.12);
	\node [black, scale=0.5] at (axis cs: #1/2+#2/2, #3-0.3) {\footnotesize Low};
}
\newcommand{\DrawMedium}[3]{
	\draw [black,thin] (axis cs: #1, #3-0.12) -- (axis cs: #2, #3-0.12);
	\node [black, scale=0.5] at (axis cs: #1/2+#2/2, #3-0.3) {\footnotesize Med};
}
\begin{axis}[axis line style={draw=none},
             tick style={draw=none},
             ymin = -3.0, ymax = 3.5,
             xmin = -130, xmax = 600,
             xticklabels={,,},
             yticklabels={,,},
             yscale=.5
             ]
\draw [black, thin, -latex] (axis cs: 0, -0.85) -- (axis cs: 600,-0.85) node [right] {};             

\addplot [mark=none, color=cyan, line width=0.6pt] table [x=x, y=co2_time, col sep=comma] {data/tikz/data_using.csv};
\addplot [mark=none, color=orange, line width=0.6pt] table [x=x, y=boiler_time, col sep=comma] {data/tikz/data_using.csv};

\node[black, scale=0.9, text width=2cm,align=center, font=\tiny\linespread{0.8}\selectfont] at (axis cs: -65, -0.15) {Water Boiler\\with Tank};
\node[black, scale=0.9, text width=2cm,align=center, font=\tiny\linespread{0.8}\selectfont] at (axis cs: -65, 1.25) {CO$_2$ \\Intensity};

\pgfplotsinvokeforeach{1, 32, 73, 85, 97, 121, 157, 185, 229, 253, 288, 315, 360, 373, 397, 444, 468, 504, 529, 552, 577}{
    \draw [black, thin] (axis cs: #1, -0.75) -- (axis cs: #1,-0.95);
}

\node [black, scale=0.45,rotate=270] at (axis cs: 1,-1.35) {\textbf \footnotesize 00:00};
\node [black, scale=0.45,rotate=270] at (axis cs: 32,-1.35) {\textbf \footnotesize 02:00};
\node [black, scale=0.45,rotate=270] at (axis cs: 73,-1.35) {\textbf \footnotesize 06:00};
\node [black, scale=0.45,rotate=270] at (axis cs: 85,-1.35) {\textbf \footnotesize 07:00};
\node [black, scale=0.45,rotate=270] at (axis cs: 97,-1.35) {\textbf \footnotesize 08:00};
\node [black, scale=0.45,rotate=270] at (axis cs: 121,-1.35) {\textbf \footnotesize 10:00};
\node [black, scale=0.45,rotate=270] at (axis cs: 157,-1.35) {\textbf \footnotesize 13:00};
\node [black, scale=0.45,rotate=270] at (axis cs: 185,-1.35) {\textbf \footnotesize 15:00};
\node [black, scale=0.45,rotate=270] at (axis cs: 229,-1.35) {\textbf \footnotesize 19:00};
\node [black, scale=0.45,rotate=270] at (axis cs: 253,-1.35) {\textbf \footnotesize 21:00};
\node [black, scale=0.45,rotate=270] at (axis cs: 288,-1.35) {\textbf \footnotesize 00:00};
\node [black, scale=0.45,rotate=270] at (axis cs: 315,-1.35) {\textbf \footnotesize 02:00};
\node [black, scale=0.45,rotate=270] at (axis cs: 360,-1.35) {\textbf \footnotesize 06:00};
\node [black, scale=0.45,rotate=270] at (axis cs: 373,-1.35) {\textbf \footnotesize 07:00};
\node [black, scale=0.45,rotate=270] at (axis cs: 397,-1.35) {\textbf \footnotesize 09:00};
\node [black, scale=0.45,rotate=270] at (axis cs: 444,-1.35) {\textbf \footnotesize 13:00};
\node [black, scale=0.45,rotate=270] at (axis cs: 468,-1.35) {\textbf \footnotesize 15:00};
\node [black, scale=0.45,rotate=270] at (axis cs: 504,-1.35) {\textbf \footnotesize 18:00};
\node [black, scale=0.45,rotate=270] at (axis cs: 529,-1.35) {\textbf \footnotesize 20:00};
\node [black, scale=0.45,rotate=270] at (axis cs: 552,-1.35) {\textbf \footnotesize 22:00};
\node [black, scale=0.45,rotate=270] at (axis cs: 577,-1.35) {\textbf \footnotesize 00:00};

\DrawLow{1}{27}{0.8}
\DrawMedium{33}{72}{1.55}
\DrawHigh{73}{121}{2.25}
\DrawMedium{122}{152}{1.55}
\DrawLow{157}{183}{1.25}
\DrawMedium{187}{228}{1.55}
\DrawHigh{229}{253}{2.8}
\DrawMedium{255}{286}{1.55}
\DrawLow{287}{319}{0.8}
\DrawMedium{326}{358}{1.55}
\DrawHigh{361}{397}{2.25}
\DrawMedium{399}{444}{1.55}
\DrawLow{446}{468}{1.25}
\DrawMedium{470}{504}{1.55}
\DrawHigh{505}{529}{2.8}
\DrawMedium{525}{552}{1.55}
\DrawLow{556}{577}{1.25}

\DrawOff{1}{84}{-0.2}
\DrawOn{83}{99}{0.15}
\DrawOff{99}{360}{-0.2}
\DrawOn{359}{375}{0.15}
\DrawOff{374}{577}{-0.2}

\draw[decorate,decoration={brace,mirror, raise=5pt}] (axis cs: 0,-1.36) --  node[scale=.9,below=5pt] {\scriptsize Day 1} (axis cs: 288,-1.36);
\draw[decorate,decoration={brace,mirror, raise=5pt}] (axis cs: 289,-1.36) --  node[scale=.9,below=5pt] {\scriptsize Day 2} (axis cs: 577,-1.36);

\end{axis}
\end{tikzpicture}}
	\vspace{-0.4in}
	\caption{CO2 intensity and water boiler electricity usage}
	\label{fig:TimeSeries}
\end{figure}
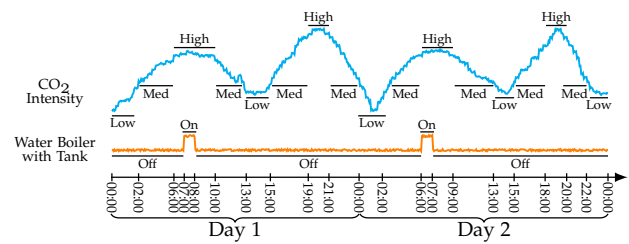

IoT-enabled sensors have enabled the collection of many big time series, e.g., from smart-meters, -plugs, and -appliances in households, weather stations, and GPS-enabled mobile devices. Extracting patterns from these time series can offer new domain insights for evidence-based decision making and optimization. 
As an example, consider Fig. \ref{fig:TimeSeries} that shows the electricity usage of a water boiler with a hot water tank collected by a $20$ euro  Wifi-enabled smart-plug, and accurate CO2 intensity (g/kWh) forecasts of local electricity, e.g., as supplied by the Danish Transmission System Operator \cite{co2}. 
From Fig. \ref{fig:TimeSeries}, we can identify several useful patterns. First, the water boiler switches \textit{On} once a day, for one hour between 6 and 7AM. This indicates that the resident takes only one hot shower per day which starts between 5.30 and 6.30AM. Second, all water boiler \textit{On} events are contained in CO2 \textit{High} events, i.e., the periods when CO2 intensity is high. Third, between two consecutive \textit{On} events of the boiler, there is a CO2 \textit{Low} event lasting for one or more hours which occurs at most 4 hours before the hot shower (so water heated during that event will still be hot at 6AM). 
Pattern mining can be used to extract the relations between CO2 intensity and water boiler events. However, traditional sequential patterns only capture the sequential occurrence of events, e.g., that one boiler \textit{On} event follows after another, but not that there is at least $23$ hours between them; or that there is a CO2 \textit{Low} event between the two boiler \textit{On} events, but not when or for how long it lasts. 
In contrast, \textit{temporal pattern mining} (TPM) adds temporal information into patterns, providing details on when certain relations between events happen, and for how long. 
For example, TPM expresses the above relations as: 
([7:00 - 8:00, Day X] BoilerOn $\rightarrow$ [6:00 - 7:00, Day X+1] BoilerOn) (meaning BoilerOn is followed by BoilerOn the next day), ([6:00 - 10:00, Day X] HighCO2 $\succcurlyeq$ [7:00 - 8:00, Day X] BoilerOn) (meaning HighCO2 contains BoilerOn), and ([7:00 - 8:00, Day X] BoilerOn $\rightarrow$ [0:00 - 2:00, Day X+1] LowCO2 $\rightarrow$ [6:00 - 7:00, Day X+1] BoilerOn) (meaning there is a LowCO2 event between two BoilerOn events).  
As the resident is very keen on reducing her CO2 footprint, we can rely on the above temporal patterns to automatically (using the smart-plug) delay turning on the boiler until the CO2 intensity is low again, saving CO2 without any loss of comfort for the resident.
In the smart city domain, temporal patterns extracted from vehicle GPS data \cite{torp2019traveltime} can reveal spatio-temporal correlations between traffic jams, advising drivers to take another route for their morning commute.

Finding frequent temporal patterns (FTPs) is useful; however, in many applications, some patterns appear rarely but are still very interesting and useful due to high confidence. We call such patterns {\em rare temporal patterns (RTPs)}. For example, considering smart city applications, a rare pattern could be: ([20:00, 22:00] Snow $\succcurlyeq$ [20:15, 21:15] HighWind $\rightarrow$ [21:20, 21:50] HighInjuryMotorist), which means that the coincidence of snow and strong winds leads to traffic accidents within an hour. This pattern occurs rarely but supports transportation coordinators in warning citizens about traffic accidents. In health care, identifying  symptoms and relations among them supports health experts in diagnosing diseases in the early phases.

\textit{Challenges of mining frequent temporal patterns.} Mining temporal patterns is much more expensive than mining sequential patterns. Not only does the temporal information add extra computation to the mining process, the complex relations between events also add an additional exponential factor O($3^{h^2}$) to the O($m^h$) search space complexity ($m$ is the number of events and $h$ is the length of temporal patterns), yielding an overall complexity of O($m^h3^{h^2}$) (see Lemma \ref{lem1} in Section \ref{sec:2freq}). 
Existing TPM methods \cite{tpminer,ieminer,hdfs} do not scale to big datasets, i.e., many time series and many sequences, and/or do not work directly on time series but only on pre-processed temporal events.

\textit{Challenges of mining rare temporal patterns.} The support measure represents the frequency of a temporal pattern across the entire dataset. However, to find rare temporal patterns, the support has to be set very low, which causes a combinatorial explosion, potentially producing too many patterns that are uninteresting to the user. Existing work proposes solutions to mine rare itemsets \cite{li2022asso, cui2022asso, ji2021asso, cai2021asso} and rare sequential patterns \cite{rahman2019sequential, iqbal2019sequential, samed2017sequential}. However, they do not consider the temporal aspect of items/events. Thus, addressing the explosion of rare temporal patterns with high confidence is still an open problem.

\textit{Generalized temporal pattern mining}. Since there are many joint challenges in mining frequent and rare temporal patterns, there is a need for a {\em generalized} approach that can mine both types of patterns efficiently. 

\textit{Contributions.} 
In this paper, we present our comprehensive {\em Generalized Temporal Pattern Mining from Time Series (GTPMfTS) approach} which solves the above challenges. The paper significantly extends a previous conference paper~\cite{ho2022efficientvldb}. 
Our key contributions are: 
(1) We present {\em end-to-end GTPMfTS process} that receives time series as input, and produces frequent/rare temporal patterns as output. Within this process, a splitting strategy is proposed to convert time series into event sequences while ensuring the preservation of temporal patterns. 
(2) We propose the {\em efficient Generalized Temporal Pattern Mining (GTPM) algorithm} to mine both frequent and rare temporal patterns. The novelties of GTPM are: a) the use of an efficient data structure, Hierarchical Hash Tables, to enable fast retrieval of events and patterns during the mining process; and b) pruning techniques based on the Apriori principle and the transitivity property of temporal relations to enable faster mining. 
(3) Based on the information theory concept of mutual information, which measures the correlation among time series, we propose a novel {\em approximate version of GTPM} that prunes unpromising time series to significantly reduce the search space and can scale on big datasets, i.e., many time series and many sequences. 
(4) We perform extensive experiments on synthetic and real-world datasets for both rare temporal pattern mining (RTPM) and frequent temporal pattern mining (FTPM), showing that our RTPM and FTPM significantly outperform the baselines on both runtime and memory usage. Compared to the baselines, the approximate RTPM has up to one order of magnitude speedup, and the approximate FTPM  up to two orders of magnitude speedup, while retaining high accuracy compared to the exact algorithms. 

Compared to the the conference version~\cite{ho2022efficientvldb}, this paper generalizes the TPM problem, to mine both frequent and (the novel proposal of) rare temporal patterns. For FTPM, this paper uses Hierarchical Hash Tables to retrieve events and patterns quickly, a significant improvement over the Hierarchical Pattern Graph in the conference version~\cite{ho2022efficientvldb}. Moreover, we now combine the lower bound of support and the lower bound of confidence from the conference version~\cite{ho2022efficientvldb} for the approximate FTPM to further accelerate the mining. For RTPM, we introduce the first exact and approximate algorithms to mine rare temporal patterns. In the present paper, we further provide a set of new experiments to compare our algorithms with the baselines.

\textit{Paper Outline.} The paper is structured as follows. Section 2 discusses the related work. Section 3 formulates the generalized temporal pattern mining problem. Section 4 describes the exact GTPM algorithm. Section 5 presents the approximate GTPM algorithm. Section 6 presents the experimental evaluation. Finally, Section 7 concludes and points to future work. 
\section{Related work}\label{sec:relatedwork}\vspace{-0.02in}
\textit{Temporal pattern mining:} Compared to sequential pattern mining, TPM is rather a new research topic. One of the first papers in this area is of Kam et al. that uses a hierarchical representation to manage temporal relations \cite{kam2000discovering}, and based on that mines temporal patterns. However, the approach in \cite{kam2000discovering} suffers from \textit{ambiguity} when presenting temporal relations. For example, using the representation in \cite{kam2000discovering}, it is possible to have two temporal patterns that involve the same set of temporal events, for example, (((a overlaps b) before c) overlaps d), and ((a overlaps b) before (c contains d)). Thus, the same set of events can be mapped to different temporal patterns that are semantically different. Our GTPM avoids this ambiguity by defining a temporal pattern as a set of pairwise temporal relations between two events.  
In \cite{wu2007mining}, Wu et al. develop TPrefix to mine temporal patterns from non-ambiguous temporal relations. However, TPrefix has several inherent limitations: it scans the database repeatedly, and the algorithm does not employ any pruning strategies to reduce the search space. 
In \cite{moskovitch2015fast}, Moskovitch et al. design a TPM algorithm using the transitivity property of temporal relations. They use this property to generate candidates by inferring new relations between events. In comparison, our GTPM uses the transitivity property for effective pruning. In \cite{batal2012mining}, Iyad et al. propose a TPM framework to detect events in time series. However, their focus is to find irregularities in the data. In \cite{wang2020mining}, Wang et al. propose a temporal pattern mining algorithm HUTPMiner to mine high-utility patterns. Different from our GTPM which uses \textit{support} and \textit{confidence} to measure the frequency of patterns, HUTPMiner uses \textit{utility} to measure the importance or profit of an event/ pattern, thereby addresses an orthogonal problem. In \cite{sharma2018stipa}, Amit et al. propose STIPA which uses a Hoeppner matrix representation to compress temporal patterns for memory savings. However, STIPA does not use any pruning/ optimization strategies and thus, despite the efficient use of memory, it cannot scale to large datasets, unlike our GTPM.
Other work \cite{batal2013temporal}, \cite{campbell2020temporal} proposes TPM algorithms to classify health record data.  
However, these methods are very domain-specific, thus cannot generalize to other domains. 

\begin{table*}
	\begin{minipage}{1\linewidth}
		\caption{A Symbolic Database $\mathcal{D}_{\text{SYB}}$}
		\vspace{-0.12in}
		\centering
		\resizebox{\textwidth}{1.2cm}{
			\setlength{\tabcolsep}{0.5mm} 
			\renewcommand{\arraystretch}{1.5}
			\begin{tabular}{|c|ccccccccc|ccccccccc|ccccccccc|ccccccccc|}
				\hline
				\textbf{Time} & \textbf{10:00} & \textbf{10:05} & \textbf{10:10} & \textbf{10:15} & \textbf{10:20} & \textbf{10:25} & \textbf{10:30} & \textbf{10:35} & \textbf{10:40} & \textbf{10:45} & \textbf{10:50} & \textbf{10:55} & \textbf{11:00} & \textbf{11:05} & \textbf{11:10} & \textbf{11:15} & \textbf{11:20} & \textbf{11:25} & \textbf{11:30} & \textbf{11:35} & \textbf{11:40} & \textbf{11:45} & \textbf{11:50} & \textbf{11:55} & \textbf{12:00} & \textbf{12:05} & \textbf{12:10} & \textbf{12:15} & \textbf{12:20} & \textbf{12:25} & \textbf{12:30} & \textbf{12:35} & \textbf{12:40} & \textbf{12:45} & \textbf{12:50} & \textbf{12:55}\\ 
				\hline 
				\textbf{S} & \normalsize On & \normalsize On & \normalsize On & \normalsize On & \normalsize Off & \normalsize Off & \normalsize Off & \normalsize On & \normalsize On & \normalsize Off & \normalsize Off & \normalsize Off & \normalsize Off & \normalsize Off & \normalsize Off & \normalsize On & \normalsize On & \normalsize On & \normalsize Off & \normalsize Off & \normalsize Off & \normalsize Off & \normalsize Off & \normalsize Off & \normalsize Off & \normalsize Off & \normalsize Off & \normalsize On & \normalsize On & \normalsize On & \normalsize On & \normalsize On & \normalsize On & \normalsize On & \normalsize On & \normalsize On\\ 			
				\hline 
				\textbf{T} & \normalsize Off & \normalsize Off & \normalsize Off & \normalsize Off & \normalsize Off & \normalsize Off & \normalsize Off & \normalsize On & \normalsize On & \normalsize Off & \normalsize Off & \normalsize On & \normalsize On & \normalsize Off & \normalsize Off & \normalsize On & \normalsize On & \normalsize On & \normalsize Off & \normalsize Off & \normalsize Off & \normalsize Off & \normalsize Off & \normalsize Off & \normalsize Off & \normalsize Off & \normalsize Off & \normalsize On & \normalsize On & \normalsize On & \normalsize On & \normalsize On & \normalsize On & \normalsize On & \normalsize On & \normalsize On\\ 
				\hline 
				\textbf{W} & \normalsize On & \normalsize On & \normalsize On & \normalsize On & \normalsize On & \normalsize On & \normalsize On & \normalsize On & \normalsize On & \normalsize Off & \normalsize Off & \normalsize Off & \normalsize Off & \normalsize On & \normalsize On & \normalsize On & \normalsize On & \normalsize On & \normalsize Off & \normalsize Off & \normalsize Off & \normalsize Off & \normalsize Off & \normalsize Off & \normalsize Off & \normalsize Off & \normalsize Off & \normalsize On & \normalsize On & \normalsize On & \normalsize On & \normalsize On & \normalsize On & \normalsize On & \normalsize On & \normalsize On\\ 
				\hline 
				\textbf{I} & \normalsize Off & \normalsize Off & \normalsize Off & \normalsize Off & \normalsize Off & \normalsize Off & \normalsize On & \normalsize On & \normalsize On & \normalsize Off & \normalsize Off & \normalsize Off & \normalsize On & \normalsize On & \normalsize Off & \normalsize Off & \normalsize On & \normalsize On & \normalsize Off & \normalsize Off & \normalsize Off & \normalsize Off & \normalsize Off & \normalsize Off & \normalsize Off & \normalsize Off & \normalsize Off & \normalsize On & \normalsize On & \normalsize Off & \normalsize Off & \normalsize Off & \normalsize Off & \normalsize Off & \normalsize On & \normalsize On\\ 
				\hline 
				
			\end{tabular}
		}
		\label{tbl:SymbolDatabase}
	\end{minipage}
	\vspace{-0.1in}
\end{table*}

The state-of-the-art TPM methods that currently achieve the best performance are our baselines: H-DFS \cite{hdfs}, TPMiner \cite{tpminer}, IEMiner \cite{ieminer}, and Z-Miner \cite{lee2020z}. H-DFS is a hybrid algorithm that uses breadth-first and depth-first search strategies to mine frequent arrangements of temporal intervals. H-DFS uses a data structure called ID-List to transform event sequences into vertical representations, and temporal patterns are generated by merging the ID-Lists of different events. This means that H-DFS does not scale well when the number of time series increases. In \cite{ieminer}, Patel et al. design a hierarchical lossless representation to model event relations, and propose IEMiner that uses Apriori-based optimizations to efficiently mine patterns from this new representation. In \cite{tpminer}, Chen et al. propose TPMiner that uses endpoint and endtime representations to simplify the complex relations among events. Similar to \cite{hdfs}, IEMiner and TPMiner do not scale to datasets with many time series. 
Z-Miner \cite{lee2020z}, proposed by Lee et al., is the most recent work addressing TPM. Z-Miner improves the mining efficiency over existing methods by employing two data structures: a hierarchical hash-based structure called Z-Table for time-efficient candidate generation and support count, and Z-Arrangement, a structure to efficiently store event intervals in temporal patterns for efficient memory consumption. Although using efficient data structures, Z-Miner neither employs the transitivity property of temporal relations nor mutual information for pruning. Thus, Z-Miner is less efficient than our exact and approximate GTPM in both runtimes and memory usage, and does not scale to large datasets with many sequences and many time series (see Section \ref{sec:experiment}).
Our GTPM algorithm improves on these methods by: 
(1) using efficient data structures and applying pruning techniques based on the Apriori principle and the transitivity property of temporal relations to enable fast mining, (2) the approximate GTPM can handle datasets with many time series and sequences, and (3), providing an end-to-end GTPMfTS process to mine temporal patterns directly from time series, a feature that is not supported by the baselines.

\textit{Rare pattern mining:} Finding rare patterns that occur infrequently in a given database has received some attention in recent years. Techniques to find rare patterns in time series, often called rare motifs, are proposed in \cite{wu2007mining, gao2017motif, begum2014raremotif}. However, since time series motifs are the repeated sub-sequences of the time series, rare motif discovery techniques cannot deal with temporal events, and thus, are insufficient for rare temporal pattern mining. A related approach concerns rare association rules \cite{li2022asso, cui2022asso, ji2021asso, cai2021asso, ali2021asso, bouasker2020asso, borah2020asso, fournier2020asso, biswas2019asso, piri2018asso, cappiello2015co, barkat2014open, ho2017towards, ho2013activity} that find rare associations between items in the database.  
However, all the mentioned work can only discover rare association rules built among itemsets, and cannot deal with temporal events and the complex temporal relations between them. Another research direction studies rare sequential patterns \cite{rahman2019sequential, iqbal2019sequential, samed2017sequential, rahman2016sequential, zhu2016sequential, ou2016sequential}. However, rare sequential patterns only consider sequential occurrence between events, and therefore, cannot model other complex relations such as overlapping or containing between temporal events. To the best of our knowledge, there is currently no existing work that studies rare temporal pattern mining which mines rare occurrences of temporal patterns in a time series database. 

\textit{Using correlations in TPM:} Different correlation measures such as expected support \cite{ahmed2016mining}, all-confidence \cite{lee2003comine}, and mutual information (MI) \cite{ke2008correlated, ho2019icde, ho2019amic,  blanchard2005using, cunjin2015mutual, honguyen2021efficient,  yao2003information, ho2020timedelay, ho2016corr, ho2015datavalue, energy2014, energy2016, ho2017improving} have been used to optimize the pattern mining process. However, these only support sequential patterns. To the best of our knowledge, our proposed approximate GTPM is the first that uses MI to optimize TPM.
\section{Preliminaries}\label{sec:preliminary}\vspace{-0.02in}
In this section, we introduce the notations and the main concepts that will be used throughout the paper.

\begin{table*}
	\begin{minipage}{.47\textwidth}
		\caption{Temporal Relations between Events}
		\vspace{-0.12in}
		\normalsize
		\resizebox{\textwidth}{2.5cm}{
			\begin{tabular}{|m{.45\columnwidth}| m{.94\columnwidth}|}
				\hline
				Follows: \hspace{0.2cm}$E_{i_{\triangleright e_i}} \rightarrow E_{j_{\triangleright e_j}}$ & \hspace{0.5cm} \begin{tikzpicture}
					
					\draw (0,0) -- node[below]{\textbf{e$_{i}$}} ++(0.7,0);
					\filldraw (0,0) circle (2pt)node[above]{\small t$_{s_i}$} ;
					\filldraw (0.7,0) circle (2pt)node[above ]{\small t$_{e_i}{\pm \epsilon}$};
					
					\filldraw (0.9,0) circle (2pt)node[below right]{\small t$_{s_j}$};
					\filldraw (2.3,0) circle (2pt)node[below]{\small t$_{e_j}$};
					\draw (0.9,0) -- node[above]{\textbf{e$_{j}$}} ++(1.4,0);
					
					\draw (3.5,0) -- node[below]{\textbf{e$_{i}$}} ++(0.7,0);
					\filldraw (3.5,0) circle (2pt)node[above]{\small t$_{s_i}$} ;
					\filldraw (4.2,0) circle (2pt)node[above]{\small t$_{e_i}{\pm \epsilon}$};
					
					\filldraw (4.7,0) circle (2pt)node[below right]{\small t$_{s_j}$};
					\filldraw (6.1,0) circle (2pt)node[below]{\small t$_{e_j}$};
					\draw (4.7,0) -- node[above]{\textbf{e$_{j}$}} ++(1.4,0);
					
					\node [align=center] at (3.5,-0.7) {\small t$_{e_i}{\pm \epsilon}$ $\le$ t$_{s_j}$};
					
				\end{tikzpicture}  \\ \hline
				
				Contains: \hspace{0.2cm}$E_{i_{\triangleright e_i}} \succcurlyeq E_{j_{\triangleright e_j}}$ & \hspace{0.5cm} 	\begin{tikzpicture}
					\draw (0,0) -- node[above]{\textbf{e$_i$}} ++(2.0,0);
					\filldraw (0,0) circle (2pt)node[above]{\small t$_{s_i}$} ;
					\filldraw (2,0) circle (2pt)node[above]{\small t$_{e_i}\pm \epsilon$};
					
					\draw (0,-0.5) -- node[above]{\textbf{e$_{j}$}} ++(2.0,0);
					\filldraw (0,-0.5) circle (2pt)node[below]{\small t$_{s_j}$} ;
					\filldraw (2,-0.5) circle (2pt)node[below]{\small t$_{e_j}$};
										
					\draw (3.5,0) -- node[above]{\textbf{e$_i$}} ++(2.1,0);
					\filldraw (3.5,0) circle (2pt)node[above]{\small t$_{s_i}$} ;
					\filldraw (5.6,0) circle (2pt)node[above]{\small t$_{e_i}\pm \epsilon$};	
					
					\draw (3.9,-0.5) -- node[above]{\textbf{e$_j$}} ++(1.4,0);
					\filldraw (3.9,-0.5) circle (2pt)node[below]{\small t$_{s_j}$} ;
					\filldraw (5.3,-0.5) circle (2pt)node[below]{\small t$_{e_j}$};	
					
					\draw (0,-1.5) -- node[above]{\textbf{e$_i$}} ++(2.0,0);
					\filldraw (0,-1.5) circle (2pt)node[above]{\small t$_{s_i}$} ;
					\filldraw (2.0,-1.5) circle (2pt)node[above]{\small t$_{e_i}\pm \epsilon$};
					
					\draw (-0,-2) -- node[above]{\textbf{e$_{j}$}} ++(1.5,0);
					\filldraw (0,-2) circle (2pt)node[below]{\small t$_{s_j}$} ;
					\filldraw (1.5,-2) circle (2pt)node[below]{\small t$_{e_j}$};
					
					\draw (3.6,-1.5) -- node[above]{\textbf{e$_i$}} ++(2.0,0);
					\filldraw (3.6,-1.5) circle (2pt)node[above]{t$_{s_i}$} ;
					\filldraw (5.6,-1.5) circle (2pt)node[above]{t$_{e_i}\pm \epsilon$};
					
					\draw (4.1,-2) -- node[above]{\textbf{e$_{j}$}} ++(1.5,0);
					\filldraw (4.1,-2) circle (2pt)node[below]{t$_{s_j}$} ;
					\filldraw (5.6,-2) circle (2pt)node[below]{t$_{e_j}$};
					
					\node [align=center] at (3.0,-2.7) {\small (t$_{s_{i}} \le$ t$_{s_j}$)  $\wedge$ (t$_{e_i}{\pm \epsilon}$ $\ge$ t$_{e_j}$)};
					
				\end{tikzpicture} \\ \hline
				
				Overlaps:\hspace{0.2cm} $E_{i_{\triangleright e_i}} \between E_{j_{\triangleright e_j}}$ & \hspace{0.5cm} \begin{tikzpicture}
					
					\draw (0,0) -- node[above]{\textbf{e$_i$}} ++(2,0);
					\filldraw (0,0) circle (2pt)node[above]{\small t$_{s_i}$} ;
					\filldraw (2,0) circle (2pt)node[above]{\small t$_{e_i}\pm \epsilon$};
					
					\draw (1,-0.75) -- node[above right]{\textbf{e$_{j}$}} ++(2.0,0);
					\filldraw (1,-0.75) circle (2pt)node[below]{\small t$_{s_j}$} ;
					\filldraw (3,-0.75) circle (2pt)node[below]{\small t$_{e_j}$};
					
					\draw[dashed] (1,0) -- (1,-0.75);
					\draw[dashed] (2,0) -- (2,-0.75);
					
					\draw[dashed,>=latex,thin,<->] (1,-0.325) -- node[above]{d$_{o}$} ++(1,0);
					
					\node [align=center] at (1.75,-1.45) {\small (t$_{s_i}<$ t$_{s_j}$) $\wedge$ (t$_{e_i}{\pm \epsilon}$ $<$ t$_{e_j}$) $\wedge$ 
						\small (t$_{e_i}$ $-$ t$_{s_j}$ $\ge$ d${_o}{\pm \epsilon}$)};
					
				\end{tikzpicture} \\
				\hline
		\end{tabular} }
		\label{tbl:relations}
	\end{minipage}
	\hspace{0.3in}
	\begin{minipage}{.48\textwidth}
	\caption{A Temporal Sequence Database $\mathcal{D}_{\text{SEQ}}$}
	\vspace{-0.12in}
	\label{tbl:SequenceDatabase}
	\resizebox{\textwidth}{2.5cm}{
		\begin{tabular}{ |c| m{7.5cm}| }
			\hline  \textbf{ID} & \textbf{\;\;\;\;\;\;\;\;\;\;\;\;\;\;\;\;\;\;\;\;\;\;\;\;\;\;\;\;\;\;\; Temporal sequences} \\
			\hline  
			1   & \scriptsize {(SOn,[10:00,10:15]), (TOff,[10:00,10:35]), (WOn,[10:00,10:40]), (IOff,[10:00,10:30]), (SOff,[10:15,10:35]), (IOn,[10:30,10:40]), (SOn, [10:35,10:40]), (TOn,[10:35,10:40]) } 
			\\
			\hline 
			2   &  \scriptsize {(SOff,[10:45,11:15]), (TOff,[10:45,10:55]), (WOff,[10:45,11:05]), (IOff,[10:45,11:00]),
			(TOn,[10:55,11:00]), (TOff,[11:00,11:15]), (IOn,[11:00,11:05]), (WOn,[11:05,11:25]), (IOff,[11:05,11:20]),
			(SOn,[11:15,11:25]), (TOn,[11:15,11:25]), (IOn,[11:20,11:25])}  
			\\ 
			\hline 
			3   &  \scriptsize {(SOff,[11:30,12:10]), (TOff,[11:30,12:10]), (WOff,[11:30,12:10]), (IOff,[11:30,12:10])}
			\\
			\hline 
			4   &  \scriptsize {(SOn,[12:15,12:55]), (TOn,[12:15,12:55]), (WOn,[12:15,12:55]), (IOn,[12:15,12:20]), IOff,[12:20,12:50]), (IOn,[12:50,12:55])}
			\\
			\hline
		\end{tabular} 
	}
	\end{minipage}	
	\vspace{-0.1in}
\end{table*}

\vspace{-0.1in}
\subsection{Temporal Event of Time Series}\vspace{-0.02in}
\textbf{Definition 3.1} (Time series) A \textit{time series} $X=  x_1, x_2, ..., x_n$ is a sequence of data values that measure the same phenomenon during an observation time period, and are chronologically ordered.  

\hspace{-0.2in}\textbf{Definition 3.2} (Symbolic time series) A \textit{symbolic time series} $X_S$ of a time series $X$ encodes the raw values of $X$ into a sequence of symbols. The finite set of permitted symbols used to encode $X$ is called the \textit{symbol alphabet} $\Sigma_X$ of $X$.

The symbolic time series $X_S$ is obtained using a mapping function $f$$:$ $X$$\rightarrow$$\Sigma_{X}$ that maps each value $x_i \in X$ to a symbol $\omega \in \Sigma_{X}$. For example, let $X$ = 1.61, 1.21, 0.41, 0.0 be a time series representing the energy usage of an electrical device. Using the symbol alphabet $\Sigma_X$ = \{On, Off\}, where On represents that the device is on and operating (e.g., $x_i \ge 0.5$), and Off that the device is off ($x_i < 0.5$), the symbolic representation of $X$ is: $X_S$ = On, On, Off, Off. The mapping function $f$ can be defined using existing time series representation techniques such as SAX \cite{lin2003symbolic}. 

\hspace{-0.2in}\textbf{Definition 3.3} (Symbolic database) 
Given a set of time series $\mathcal{X}=\{X_1,...,X_n\}$, the set of symbolic representations of the time series in $\mathcal{X}$ forms a \textit{symbolic database} $\mathcal{D_{\text{SYB}}}$.

An example of the symbolic database $\mathcal{D}_{\text{SYB}}$ is shown in Table \ref{tbl:SymbolDatabase}. There are $4$ time series representing the energy usage of $4$ electrical appliances: \{Stove, Toaster, Clothes Washer, Iron\}. For brevity, we name the appliances respectively as \{S, T, W, I\}. 
All appliances have the same alphabet $\Sigma$ = \{On, Off\}.

\hspace{-0.2in}\textbf{Definition 3.4} (Temporal event in a symbolic time series) A \textit{temporal event} $E$ in a symbolic time series $X_S$ is a tuple $E = (\omega, T)$ where $\omega \in \Sigma_X$ is a symbol, and $T=\{[t_{s_i}, t_{e_i}]\}$ is the set of time intervals during which $X_S$ is associated with the symbol $\omega$. 

Given a time series $X$, a temporal event is created by first converting $X$ into symbolic time series $X_S$, and then combining identical consecutive symbols in $X_S$ into one single time interval. For example, consider the symbolic representation of $S$ in Table \ref{tbl:SymbolDatabase}. By combining its consecutive On symbols, we form the temporal event \textit{``Stove is On''} as: (SOn, \{[10:00, 10:15], [10:35, 10:40], [11:15, 11:25], [12:15, 12:55]\}). 

\hspace{-0.2in}\textbf{Definition 3.5} (Instance of a temporal event) Let $E = (\omega, T)$ be a temporal event, and $[t_{s_i},t_{e_i}] \in T$ be a time interval. The tuple $e = (\omega, [t_{s_i},t_{e_i}])$ is called an \textit{instance} of the event $E$, representing a single occurrence of $E$ during $[t_{s_i},t_{e_i}]$. We use the notation $E_{\triangleright e}$ to say that event $E$ has an instance $e$. 

\subsection{Relations between Temporal Events}\vspace{-0.02in}
We adopt the popular Allen's relations model  \cite{allen} and define three
basic temporal relations between events. Furthermore, to avoid the exact time mapping problem in Allen's relations, we adopt the \textit{buffer} idea from \cite{hdfs}, adding a tolerance \textit{buffer} $\epsilon$ to the relation's endpoints. However, we change the way $\epsilon$ is used in \cite{hdfs} to ensure the relations are \textit{mutually exclusive} (proof is in the electronic appendix). 

Consider two temporal events $E_i$ and $E_j$, and their corresponding instances, $e_i=(\omega_i,[t_{s_i}, t_{e_i}])$ and $e_j=(\omega_j,[t_{s_j}, t_{e_j}])$. Let $\epsilon$ be a non-negative number ($\epsilon \ge 0$) representing the buffer size. The following relations can be defined between $E_i$ and $E_j$ through $e_i$ and $e_j$.

\hspace{-0.2in}\textbf{Definition 3.6} (Follows)  
$E_i$ and $E_j$ form a \textit{Follows} relation through $e_i$ and $e_j$, denoted as Follows($E_{i_{\triangleright e_i}}$,$E_{j_{\triangleright e_j}}$) or $E_{i_{\triangleright e_i}}$$\rightarrow$$E_{j_{\triangleright e_j}}$, iff \scalebox{0.9}{$t_{e_i}$$\pm$$\epsilon$$\le$$t_{s_j}$}.

\hspace{-0.2in}\textbf{Definition 3.7} (Contains) $E_i$ and $E_j$ form a \textit{Contains} relation through $e_i$ and $e_j$, denoted as Contains($E_{i_{\triangleright e_i}}$, $E_{j_{\triangleright e_j}}$) or $E_{i_{\triangleright e_i}}$$\succcurlyeq$$E_{j_{\triangleright e_j}}$, iff $(t_{s_i}$ $\le$ $t_{s_j})$ $\wedge$ $(t_{e_i} \pm \epsilon \ge t_{e_j})$.

\hspace{-0.2in}\textbf{Definition 3.8} (Overlaps) 
$E_i$ and $E_j$ form an \textit{Overlaps} relation through $e_i$ and $e_j$, denoted as Overlaps($E_{i_{\triangleright e_i}}$, $E_{j_{\triangleright e_j}}$) or $E_{i_{\triangleright e_i}}$ $\between$ $E_{j_{\triangleright e_j}}$, iff ${(t_{s_i} < t_{s_j})}$ $\wedge$ $(t_{e_i} \pm \epsilon < t_{e_j})$ $\wedge$ $(t_{e_i}-t_{s_j} \ge d_o \pm \epsilon)$, where $d_o$ is the minimal overlapping duration between two event instances, and $0 \le \epsilon \ll d_o$. 

The \textit{Follows} relation represents sequential occurrences of one event after another. For example, $E_{i_{\triangleright e_i}}$ is followed by $E_{j_{\triangleright e_j}}$ if the end time $t_{e_i}$ of $e_i$ occurs before the start time $t_{s_j}$ of $e_j$. Here, the buffer $\epsilon$ is used as a tolerance, i.e., the \textit{Follows} relation between $E_{i_{\triangleright e_i}}$ and $E_{j_{\triangleright e_j}}$ holds if $(t_{e_i} + \epsilon)$ or $(t_{e_i} - \epsilon)$ occurs before $t_{s_j}$. On the other hand, in a \textit{Contains} relation, one event occurs entirely within the timespan of another event. Finally, in an \textit{Overlaps} relation, the timespans of the two occurrences overlap each other. Table \ref{tbl:relations} illustrates the three temporal relations and their conditions.

\vspace{-0.18in}
\subsection{Temporal Pattern}\vspace{-0.02in}
\textbf{Definition 3.9} (Temporal sequence)
A list of $n$ event instances $S$$=$$<$$e_1,...,e_i,..., e_n$$>$ forms a \textit{temporal sequence} if the instances are chronologically ordered by their start times. 
Moreover, $S$ has size $n$, denoted as $|S| = n$. 

\hspace{-0.2in}\textbf{Definition 3.10} (Temporal sequence database)
A set of temporal sequences forms a \textit{temporal sequence database} $\mathcal{D}_{\text{SEQ}}$ where each row $i$ contains a temporal sequence $S_i$. 

Table \ref{tbl:SequenceDatabase} shows the temporal sequence database $\mathcal{D}_{\text{SEQ}}$, created from the symbolic database $\mathcal{D}_{\text{SYB}}$ in Table \ref{tbl:SymbolDatabase}.

\hspace{-0.2in}\textbf{Definition 3.11} (Temporal pattern) Let $\Re$$=$\{Follows, Contains, Overlaps\} be the set of temporal relations. A \textit{temporal pattern} $P$$=$$<$$(r_{12}, E_{1},$ $E_{2})$,...,$(r_{(n-1)(n)},E_{n-1},E_{n})$$>$ is a list of triples $(r_{\textit{ij}}$,$E_{i}$,$E_{j})$, each representing a relation $r_{\textit{ij}} \in \Re$ between two events $E_i$ and $E_j$.

Note that the relation $r_{\textit{ij}}$ in each triple is formed using the specific instances of $E_i$ and $E_j$. A temporal pattern that has $n$ events is called an $n$-event pattern. We use $E_i \in P$ to denote that the event $E_i$ occurs in $P$, and $P_1 \subseteq P$ to say that a pattern $P_1$ is a sub-pattern of $P$. 

\hspace{-0.2in}\textbf{Definition 3.12} (Temporal sequence supports a pattern) Let $S$$=$$<$$e_1$,...,$e_i$,...,$e_n$$>$ be a temporal sequence. We say that $S$ \textit{supports} a temporal pattern $P$, denoted as $P \in S$, iff $|S| \ge 2$ $\wedge$ $\forall (r_{\textit{ij}},E_i,E_j) \in P, $ $\exists (e_l, e_m) \in S$ such that $r_{\textit{ij}}$ holds between $E_{i_{\triangleright e_l}}$ and $E_{j_{\triangleright e_m}}$.

If $P$ is supported by $S$, $P$ can be written as $P$$=$$<$$(r_{12}$, $E_{1_{\triangleright e_1}}$, $E_{2_{\triangleright e_2}})$, ..., $(r_{(n-1)(n)}$,$E_{{n-1}_{\triangleright e_{n-1}}}$, $E_{{n}_{\triangleright e_{n}}})$$>$, where the relation between two events in each triple is expressed using the event instances.

In Fig. \ref{fig:TimeSeries}, consider the sequence $S=<$$e_1$=(HighCO2, [6:00, 10:00]), $e_2$$=$(BoilerOn, [7:00, 8:00]), $e_3$$=$(LowCO2, [13:00, 15:00])$>$ representing the order of CO2 intensity and boiler events. Here, $S$ supports a 3-event pattern $P$$=$$<$(Contains, HighCO2$_{\triangleright e_1}$, BoilerOn$_{\triangleright e_2}$), (Follows, HighCO2$_{\triangleright e_1}$, LowCO2$_{\triangleright e_3}$), (Follows, BoilerOn$_{\triangleright e_2}$, LowCO2$_{\triangleright e_3}$)$>$. 

\textit{Maximal duration constraint}: Let $P \in S$ be a temporal pattern supported by the sequence $S$. The duration between the start time of the instance $e_1$, and the end time of the instance $e_n$ in $S$ must not exceed the predefined maximal time duration $t_{\max}$: $t_{e_n} - t_{s_1} \leq t_{\max}$.

The maximal duration constraint guarantees that the relation between any two events is temporally valid. This enables the pruning of invalid patterns. For example, under this constraint, a \textit{Follows} relation between a \textit{``Washer On''} event and a \textit{``Dryer On''} event in Table \ref{tbl:SequenceDatabase} happening one year apart should be considered invalid. 

\subsection{Frequency and Likelihood Measures}\vspace{-0.02in}
Given a temporal sequence database $\mathcal{D}_{\text{SEQ}}$, we want to find patterns that occur at certain frequency in $\mathcal{D}_{\text{SEQ}}$. 
We use \textit{support} and \textit{confidence} \cite{omiecinski2003alternative} to measure the frequency and likelihood of a pattern. 

\hspace{-0.2in}\textbf{Definition 3.13} (Support of a temporal event) 
The \textit{support} of a temporal event $E$ in $\mathcal{D}_{\text{SEQ}}$ is the number of sequences $S \in \mathcal{D}_{\text{SEQ}}$ containing at least one instance $e$ of $E$. \vspace{-0.05in}
\begin{equation}
	\small
	\textit{supp}(E) = \lvert \{S \in \mathcal{D}_{\text{SEQ}}  \textit{ s.t. } \exists e \in S: E_{\triangleright e}  \} \rvert 
	\label{eq:support1}
	\vspace{-0.05in}
\end{equation}
\\
The \textit{relative support} of $E$ is the fraction between $\textit{supp}(E)$ and the size of $\mathcal{D}_{\text{SEQ}}$: \vspace{-0.07in}
\begin{equation}
	\small
	\textit{rel-supp}(E)= \textit{supp}(E) / |\mathcal{D}_{\text{SEQ}}| 
	\label{eq:support2}
\end{equation}

Similarly, the support of a group of events $(E_1,..., E_n)$, denoted as $\textit{supp}(E_1,..., E_n)$, is the number of sequences $S \in \mathcal{D}_{\text{SEQ}}$ which contain at least one instance $(e_1,...,  e_n)$ of the event group. 

\hspace{-0.2in}\textbf{Definition 3.14} (Support of a temporal pattern)
The \textit{support} of a pattern $P$ is the number of sequences $S \in \mathcal{D}_{\text{SEQ}}$ that support $P$. \vspace{-0.02in}
\begin{equation}
	\vspace{-0.05in}
	\small
	\textit{supp}(P) = \lvert \{S \in \mathcal{D}_{\text{SEQ}} \textit{ s.t. } P \in S\} \rvert 
	\label{eq:support3}
	\vspace{-0.02in}
\end{equation}
\\
The \textit{relative support} of $P$ in $\mathcal{D}_{\text{SEQ}}$ is the fraction \vspace{-0.05in}
\begin{equation}
	\vspace{-0.05in}
	\small
	\textit{rel-supp}(P)= \textit{supp(P)} / |\mathcal{D}_{\text{SEQ}}| 
	\label{eq:support4}
\end{equation}   

\hspace{-0.2in}\textbf{Definition 3.15} (Confidence of an event pair) 
The \textit{confidence} of an event pair $(E_i, E_j)$ in $\mathcal{D}_{\text{SEQ}}$ is the fraction between $\textit{supp}(E_i, E_j)$ and the support of its most frequent event: \vspace{-0.05in} 
\begin{equation}
	\vspace{-0.05in}
	\small
	\textit{conf}(E_i,E_j) = \frac{\textit{supp}(E_i, E_j)} {\max\{\textit{supp}(E_i), \textit{supp}(E_j) \}} 
	\label{eq:eventpairconf}
\end{equation}

\hspace{-0.2in}\textbf{Definition 3.16} (Confidence of a temporal pattern)
The \textit{confidence} of a temporal pattern $P$ in $\mathcal{D}_{\text{SEQ}}$ is the fraction between $\textit{supp}(P)$ and the support of its most frequent event: \vspace{-0.05in} 
\begin{equation}
	\vspace{-0.05in}
	\small
	\textit{conf}(P) = \frac{\textit{supp}(P)}{\max_{1 \leq k \leq |P|}\{\textit{supp}(E_k) \}} 
	\label{eq:confidence}
\end{equation}
where $E_k \in P$ is a temporal event. 
Since the denominator in Eq. \eqref{eq:confidence} is the maximum support of the events in $P$, the confidence computed in Eq. \eqref{eq:confidence} is the \textit{minimum confidence} of a pattern $P$ in $\mathcal{D}_{\text{SEQ}}$, which is also called the \textit{all-confidence} as in \cite{omiecinski2003alternative}. 
Note that unlike association rules, temporal patterns do not have antecedents and consequents. Instead, they represent pair-wise temporal relations between events based on their temporal occurrences. Thus, while the \textit{support} and \textit{relative support} of event(s)/ pattern(s) defined in Eqs. \eqref{eq:support1} $-$ \eqref{eq:support4} follow the same intuition as the traditional support concept, indicating how frequently an event/ pattern occurs in a given database, the \textit{confidence} computed in Eqs. \eqref{eq:eventpairconf} $-$ \eqref{eq:confidence} instead represents the minimum likelihood of an event pair/ pattern, knowing the likelihood of its most frequent event.

\textit{Frequent temporal patterns vs. Rare temporal patterns:} Consider a temporal pattern $P$ in a temporal sequence database $\mathcal{D}_{\text{SEQ}}$ with the support $\sigma = \textit{supp}(P)$ and the confidence $\delta = \textit{conf}(P)$. Pattern $P$ is considered to be \emph{frequent} in $\mathcal{D}_{\text{SEQ}}$ if both support $\sigma$ and confidence $\delta$ are high, representing the presence of pattern $P$ in a large fraction of sequences in the database. In contrast, pattern $P$ is considered to be \emph{rare} in $\mathcal{D}_{\text{SEQ}}$ if the support $\sigma$ is low and the confidence $\delta$ is high, indicating a type of pattern that occurs only in a small fraction of sequences but with high likelihood, given the occurrence evidence of the involved events.

\textbf{Problem Definition: Generalized Temporal Pattern Mining.} Given a set of univariate time series $\mathcal{X}=\{X_1,...,X_n\}$, let $\mathcal{D}_{\text{SEQ}}$ be the temporal sequence database obtained from ${\mathcal{X}}$, and $\sigma_{\min}$, $\sigma_{\max}$ and $\delta$ be the minimum support, maximum support and minimum confidence thresholds, respectively. The Generalized Temporal Pattern Mining from Time Series (GTPMfTS) problem aims to find all temporal patterns $P$ in $\mathcal{D}_{\text{SEQ}}$ such that $P$ satisfies the support and confidence constraints, i.e., $\sigma_{\min} \leq \textit{supp}(P) \leq \sigma_{\max}$ $\wedge$ $\textit{conf}(P) \geq \delta$.

Using the three constraints $\sigma_{\min}$, $\sigma_{\max}$ and $\delta$, GTPMfTS can mine frequent temporal patterns in $\mathcal{D}_{\text{SEQ}}$ by setting $\sigma_{\max} = \infty$, and assigning $\sigma_{\min}$ and $\delta$ to high threshold values. In contrast, to mine rare temporal patterns, GTPMfTS will assign low threshold values to $\sigma_{\min}$ and $\sigma_{\max}$, constraining on a low occurrence frequency, and a high value to $\delta$, constraining on a high likelihood of the patterns. 
\section{Generalized Temporal Pattern Mining} \label{sec:FTPMfTSMining}
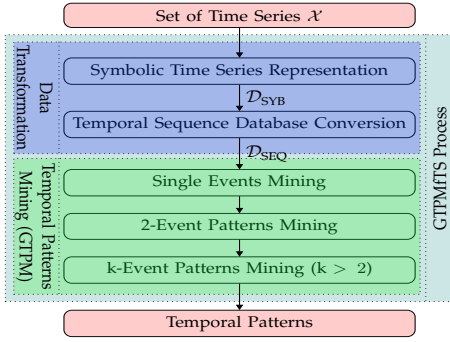
\begin{figure}[!t]
	\centering
	\captionsetup{justification=centering}
	\resizebox{.7\linewidth}{.5\linewidth}{
		\begin{tikzpicture}[node distance=-1.5cm, fill opacity=0.2]
	\tikzstyle{arrow} = [black,opacity=1,thin,-{Triangle[angle=60:1.2mm]}]
	\tikzstyle{line} = [black,opacity=1,thin,dotted]
	\tikzstyle{box} = [rectangle,rounded corners, minimum width=6cm, minimum height=0.5cm, text centered,text width = 6cm, draw=black, text opacity=1]
	\tikzstyle{bounding1} = [rectangle, dotted, minimum width=7.2cm, minimum height=2.3cm, draw=black, fill=blue, text opacity=1]
	\tikzstyle{bounding2} = [rectangle,dotted, minimum width=7.2cm, minimum height=2.8cm, draw=black, fill=green, text opacity=1]
	\tikzstyle{bounding3} = [rectangle,dotted, minimum width=8cm, minimum height=5.5cm, draw=black, fill=teal, text opacity=1]
	\tikzstyle{textlabel} = [text centered,text width = 3cm, text opacity=1, rotate=270, yshift = -12]
	\tikzstyle{textlabel2} = [text centered,text width = 3cm, text opacity=1, rotate=90, yshift = -7]
	
	\node (A) [box, fill=red] {\small Set of Time Series $\mathcal{X}$};
	
	\node (B) [box, below of=A, yshift=-2.65cm] {\small Symbolic Time Series Representation};
	\node (C) [box, below of=B, yshift=-2.6cm] {\small Temporal Sequence Database Conversion};
	\node (D) [box, below of=C, yshift=-2.7cm] {\small Single Events Mining};
	\node (E) [box, below of=D, yshift=-2.4cm] {\small 2-Event Patterns Mining};
	\node (F) [box, below of=E, yshift=-2.4cm] {\small k-Event Patterns Mining (k $>2$)};
	\node (G) [box, fill=red, below of=F, yshift=-2.6cm] {\small Temporal Patterns};
	
	\node (boundingC) [bounding3, fit=(B) (C) (D) (E) (F),  yshift=0.05cm, xshift=-0.15cm] {};
	\draw [line] ($(boundingC.north) + (3.45cm,0)$) -- node [black, textlabel2] {\small GTPMfTS Process}  ($(boundingC.south) + (3.45cm,0)$);
	
	\node (boundingA)[bounding1, fit=(B) (C), xshift=-.4cm] {};
	\draw [line] ($(boundingA.north) - (2.8cm,0)$) -- node [black, textlabel] {\small \phantom{abc}Data\phantom{abc} Transformation }  ($(boundingA.south) - (2.8cm,0)$);
	
	\node (boundingB)[bounding2, fit=(D) (E) (F), xshift=-.4cm] {};
	\draw [line] ($(boundingB.north) - (2.8cm,0)$) -- node [black, textlabel] {\small Temporal Patterns Mining (GTPM)}  ($(boundingB.south) - (2.8cm,0)$);

	\draw[arrow] (A) -- (B) ;
	\draw[arrow] (B) -- node[anchor=west] {$\mathcal{D}_{\text{SYB}}$} (C) ;
	\draw[arrow] (C) -- node[anchor=west] {$\mathcal{D}_{\text{SEQ}}$} (D) ;
	\draw[arrow] (D) -- (E) ;
	\draw[arrow] (E) -- (F) ;
	\draw[arrow] (F) -- (G) ;
\end{tikzpicture}
	}
	\caption{The GTPMfTS process} 
	\label{fig:framework}
\end{figure}

In this section, we present the Generalized Temporal Pattern Mining (GTPM) algorithm to mine both frequent and rare temporal patterns from time series. 
Fig. \ref{fig:framework} gives an overview of the GTPMfTS process which consists of two phases. The first phase, \textit{Data Transformation}, converts a set of time series $\mathcal{X}$ into a symbolic database $\mathcal{D}_{\text{SYB}}$, and then converts $\mathcal{D}_{\text{SYB}}$ into a temporal sequence database $\mathcal{D}_{\text{SEQ}}$. The second phase, \textit{Generalized Temporal Pattern Mining (GTPM)}, mines both frequent and rare temporal patterns, and consists of three steps: (1) \textit{Mining Single Events}, (2) \textit{Mining 2-Event Patterns}, and (3) \textit{Mining k-Event Patterns} (\text{k}$>$$2$). The final output is a set of all temporal patterns in $\mathcal{D}_{\text{SEQ}}$ that satisfy the minimum support, maximum support and minimum confidence constraints.

\vspace{-0.1in}
\subsection{Data Transformation}\vspace{-0.02in}
\subsubsection{Symbolic Time Series Representation}
Given a set of time series $\mathcal{X}$, the symbolic representation of each time series $X \in \mathcal{X}$ is obtained by using a mapping function as in Def. 3.2. 

\begin{figure}
	\centering
	\captionsetup{justification=centering}
	\begin{subfigure}[b]{0.6\linewidth}
		\resizebox{\linewidth}{.9cm}{
			\begin{tikzpicture}[fill opacity=.2]
\definecolor{arrowcolor}{rgb}{.27,.45,.77}
\tikzstyle{arrow} = [arrowcolor,opacity=1,thin, {Triangle[angle=60:1.2mm]}-{Triangle[angle=60:1.2mm]}]
\tikzstyle{roundrect} = [rectangle,rounded corners, minimum width=6cm, minimum height=0.8cm, text centered,text width =1.4cm, draw=white, fill=white, text opacity=1]
\tikzstyle{roundrect3} = [rectangle,rounded corners, minimum width=3cm, minimum height=0.8cm, text centered,text width =1.4cm, draw=black, fill=red, text opacity=1, fill opacity=.15]
\tikzstyle{roundrect4} = [rectangle,rounded corners, minimum width=3cm, minimum height=0.8cm, text centered,text width =1.4cm, draw=black, fill=blue, text opacity=1, fill opacity=.3]
\tikzstyle{line} = [black,opacity=1,thin, {Circle[scale=0.4]}-{Circle[scale=0.4]}]

\node (bbox) [roundrect] {};
\node (anode) [roundrect3, left of=bbox, xshift=-.5cm] {};
\node (cnode) [roundrect4,  left of=anode, xshift = 4cm] {};
\node [text opacity=1] at ($(anode.west)!0.5!(cnode.west)$) {$S_1$} ;
\node [text opacity=1] at ($(anode.east)!0.5!(cnode.east)$) {$S_2$} ;
\draw [arrow] ($(anode.west)+(0,.6)$) -- node [black, above, yshift=-2] {\small $t$}   ($(anode.east)+(0,.6)$);
\draw [arrow] ($(cnode.west)+(0,.6)$) -- node [black, above, yshift=-2] {\small $t$}   ($(cnode.east)+(0,.6)$);

\draw [line] ($(anode.east)+(-.01,.2)$) -- node [black, above,yshift=-2.5] {\tiny SOn}  ($(anode.east)+(-1,.2)$);
\draw [line] ($(anode.east)+(-.01,.0)$) -- node [black, above,yshift=-2.5] {\tiny TOn}  ($(anode.east)+(-1,.0)$);
\draw [line] ($(cnode.west)+(.01,-.1)$) -- node [black, above,yshift=-2.5] {\tiny WOn}  ($(cnode.west)+(.8,-.1)$);
\draw [line] ($(cnode.west)+(.1,-.3)$) -- node [black, above,yshift=-2.5] {\tiny IOn}  ($(cnode.west)+(.7,-.3)$);
\end{tikzpicture}
		}
		\vspace{-3mm}
		\caption{With no overlapping}
		\label{fig:basicSplit}
	\end{subfigure}
	
	\begin{subfigure}[b]{0.6\linewidth}
		\resizebox{\linewidth}{1.1cm}{%
			\begin{tikzpicture}[fill opacity=.2]
\definecolor{arrowcolor}{rgb}{.27,.45,.77}
\tikzstyle{arrow} = [arrowcolor,opacity=1,thin, {Triangle[angle=60:1.2mm]}-{Triangle[angle=60:1.2mm]}]
\tikzstyle{roundrect} = [rectangle,rounded corners, minimum width=6cm, minimum height=.8cm, text centered,text width =1.4cm, draw=white, fill=white, text opacity=1]
\tikzstyle{roundrect3} = [rectangle, rounded corners, minimum width=3cm, minimum height=.8cm, text centered,text width =1.4cm, draw=black, fill=red, text opacity=1, dashed, fill opacity=.15]
\tikzstyle{roundrect4} = [rectangle,rounded corners, minimum width=3cm, minimum height=.8cm, text centered,text width =1.4cm, draw=black, fill=blue, text opacity=1, fill opacity=.3]
\tikzstyle{line} = [black,opacity=1,thin, {Circle[scale=0.4]}-{Circle[scale=0.4]}]
	
\node (bbox) [roundrect] {};
\node (anode) [roundrect3, left of=bbox, xshift=-.5cm] {};
\draw [rounded corners] ($(anode.south east) + (-0.5,.007)$) --  ($(anode.south west) + (.007,.007)$) -- ($(anode.north west) + (.007,-.007)$)--  ($(anode.north east) - (-0.5,.007)$);

\node (cnode) [roundrect4,  left of=anode, xshift = 3cm] {};
\node [text opacity=1] at ($(anode.west)!0.6!(cnode.west)$) {$S_1$} ;
\node [text opacity=1] at ($(anode.east)!0.7!(cnode.east)$) {$S_2$} ;
\draw [arrow] ($(cnode.west)+(0,.48)$) -- node [black, above,yshift=-2] {\small $t_{\text{ov}}$}  ($(anode.east)+(0,.48)$);
\draw [arrow] ($(anode.west)+(0,.8)$) -- node [black, above, yshift=-2] {\small $t$}   ($(anode.east)+(0,.8)$);
\draw [arrow] ($(cnode.west)+(0,1)$) -- node [black, above, yshift=-2] {\small $t$}   ($(cnode.east)+(0,1)$);

\draw [line] ($(cnode.west)+(.01,.2)$) -- node [black, above,yshift=-2.5] {\tiny SOn}  ($(cnode.west)+(1,.2)$);
\draw [line] ($(cnode.west)+(.01,0)$) -- node [black, above,yshift=-2.5] {\tiny TOn}  ($(cnode.west)+(1,0)$);
\draw [line] ($(anode.east)+(.01,-.1)$) -- node [black, above,yshift=-2.5] {\tiny WOn}  ($(anode.east)+(.8,-.1)$);
\draw [line] ($(anode.east)+(.1,-.3)$) -- node [black, above,yshift=-2.5] {\tiny IOn}  ($(anode.east)+(.7,-.3)$);
\end{tikzpicture}
		}
		\vspace{-3mm}
		\caption{With overlapping}
		\label{fig:overlaptransaction2}
	\end{subfigure}
	\caption{Splitting strategy}
	\label{fig:splitting}
\end{figure}
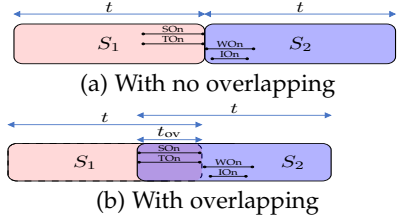

\vspace{-0.05in}\subsubsection{Temporal Sequence Database Conversion}
To convert $\mathcal{D}_{\text{SYB}}$ to $\mathcal{D}_{\text{SEQ}}$, a straightforward approach is to split the symbolic series in $\mathcal{D}_{\text{SYB}}$ into equal-length sequences, each belongs to a row in $\mathcal{D}_{\text{SEQ}}$. For example, if each symbolic series in Table \ref{tbl:SymbolDatabase} is split into $4$ sequences, then each sequence will last for $40$ minutes. The first sequence $S_1$ of $\mathcal{D}_{\text{SEQ}}$ therefore contains temporal events of S, T, W, and I from 10:00 to 10:40. The second sequence $S_2$ contains events from 10:45 to 11:25, and similarly for $S_3$ and $S_4$.

However, the splitting can lead to a potential loss of temporal patterns. The loss happens when a \textit{splitting point} accidentally divides a temporal pattern into different sub-patterns, and places these into separate sequences. We explain this situation in Fig. \ref{fig:basicSplit}. 
Consider $2$ sequences $S_1$ and $S_2$, each of length $t$. Here, the splitting point divides a pattern of $4$ events, \{SOn, TOn, WOn, IOn\}, into two sub-patterns, in which SOn and TOn are placed in $S_1$, and WOn and IOn in $S_2$. This results in the loss of this $4$-event pattern which can be identified only when all $4$ events are in the same sequence.

To prevent such a loss, we propose a \textit{splitting strategy} using overlapping sequences. Specifically, two consecutive sequences are overlapped by a duration $t_{\text{ov}}$: $0 \le t_{\text{ov}} \le t_{\max}$, where $t_{\max}$ is the \textit{maximal duration} of a temporal pattern. The value of $t_{\text{ov}}$ decides how large the overlap between $S_i$ and $S_{i+1}$ is: $t_{\text{ov}}=0$ results in no overlap, i.e., no redundancy, but with a potential loss of patterns, while $t_{\text{ov}}=t_{\max}$ creates large overlaps between sequences, i.e., high redundancy, but all patterns are preserved. 
As illustrated in Fig. \ref{fig:overlaptransaction2}, the overlapping between $S_1$ and $S_2$ keeps the $4$ events together in the same sequence $S_2$, and thus helps preserve the pattern. 

\subsection{Generalized Temporal Pattern Mining}\vspace{-0.02in}
\SetNlSty{}{}{:} 
\begin{algorithm}[!t]
	\algsetup{linenosize=\tiny}
	\SetInd{0.5em}{0.5em}
	\DontPrintSemicolon
	
	\caption{\mbox{Generalized Temporal Pattern Mining}} 
	\label{algorithmHTPGM}
	
	\KwInput{Temporal sequence database $\mathcal{D_{\text{SEQ}}}$, minimum support threshold $\sigma_{\min}$, maximum support threshold $\sigma_{\max}$, confidence threshold $\delta$}
	\KwOutput{The set of temporal patterns $P$ satisfying $\sigma_{\min}$, $\sigma_{\max}$, $\delta$}
	
	\nonl //Mining single events \;
	\ForEach{\textit{event} $E_i \in \mathcal{D_{\text{SEQ}}}$}{
		Compute $\textit{supp}(E_i)$;\;
		\If{$\textit{supp}(E_i) \geq \sigma_{\min}$}{
			Insert $E_i$ to \textit{1Freq};
		}
	}	
	
	\nonl //Mining 2-event patterns \;
	EventPairs $\leftarrow$ Cartesian(\textit{1Freq},\textit{1Freq});\;  
	FrequentPairs $\leftarrow$ $\emptyset$;\;
	\ForEach{$(E_i,E_j)$ in EventPairs}{
		Compute $\textit{supp}(E_i,E_j)$;\;
		\If{$\textit{supp}(E_i,E_j) \geq \sigma_{\min}$}{
			FrequentPairs $\leftarrow$ Apply\_Lemma4($E_i,E_j$);\;
		}
	}
	\ForEach{$(E_i,E_j)$ in FrequentPairs}{
		Retrieve event instances;\;  
		Check temporal relations against $\sigma_{\min}$, $\sigma_{\max}$, $\delta$;\;  
	}	
	
	\nonl //Mining k-event patterns \;
	Filtered1Freq $\leftarrow$ Transitivity\_Filtering(1Freq);\;
	kEvents $\leftarrow$ Cartesian(\textit{Filtered1Freq},\textit{(k-1)Freq});\;  
	FrequentkEvents $\leftarrow$ Apriori\_Filtering(kEvents);\;
	\ForEach{\textit{kEvents} in FrequentkEvents}{
		Retrieve relations;\;  
		Iteratively check relations against $\sigma_{\min}$, $\sigma_{\max}$, $\delta$;
	}	
\end{algorithm}

\begin{figure*}[!t]
	\setlength{\tabcolsep}{0pt}
	\begin{tabularx}{\linewidth}{ll}
		\begin{minipage}{.28\linewidth}
			\begin{minipage}{\linewidth}
				\captionsetup{justification=centering, font=small}
				\includegraphics[width=1\textwidth,height=3.1cm]{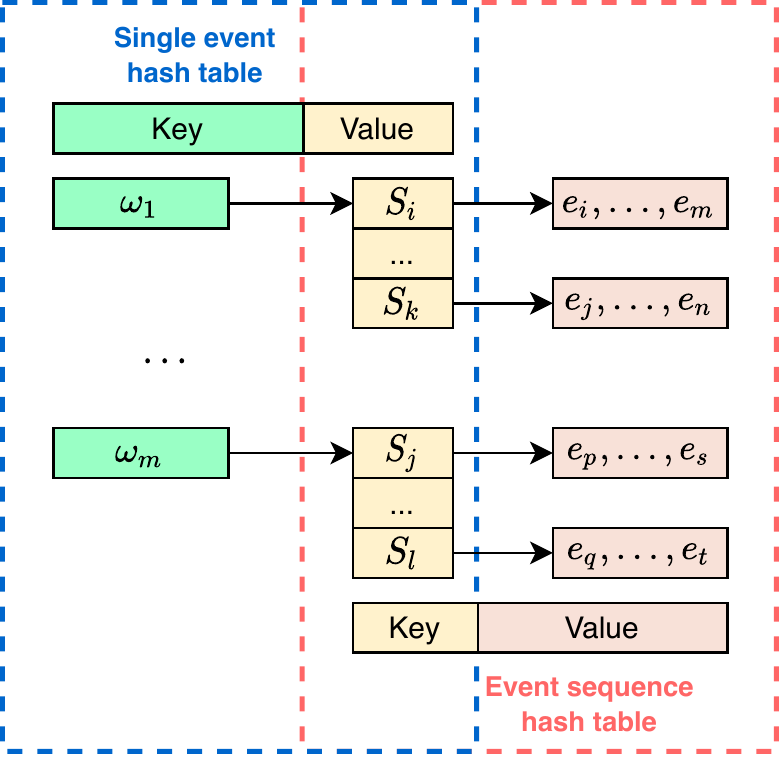}
				\vspace{-0.2in}
				\caption{The $HLH_1$ structure}
				\label{fig:hlh1}
			\end{minipage}
		\end{minipage}&
		\hspace{0.2in}
		\begin{minipage}{.68\linewidth}
			\begin{minipage}{\linewidth}
				\centering
				\captionsetup{justification=centering, font=small}
				\includegraphics[width=1\textwidth,height=3.1cm]{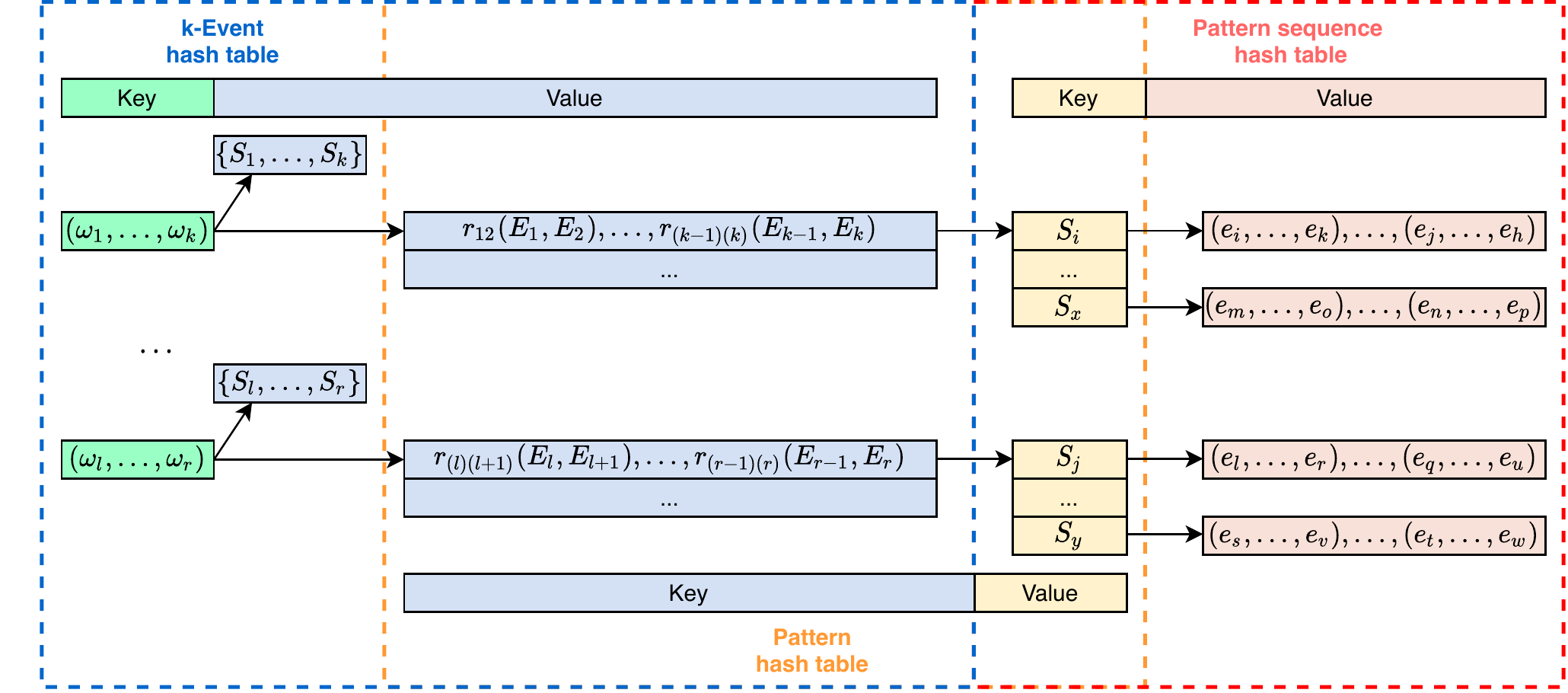}
				\vspace{-0.2in}
				\caption{The $HLH_k (k \geq 2)$ structure}
				\label{fig:hlhk}
			\end{minipage}
		\end{minipage}
	\end{tabularx}
	\vspace{-0.2in}
\end{figure*}

We now present the GTPM algorithm to mine temporal patterns, both frequent and rare, from $\mathcal{D}_{\text{SEQ}}$. We note that for frequent patterns, only two constraints $\sigma_{\min}$ and $\delta$ are used, whereas with rare patterns, all three constraints $\sigma_{\min}$, $\sigma_{\max}$, and $\delta$ are used. In the following when presenting the GTPM algorithm, the discussion applies to both frequent and rare patterns, with the implication that $\sigma_{\max}$ is set to $\infty$ when mining frequent patterns.

The main novelties of GTPM are: a) the use of efficient data structures, i.e., the \textit{Hierarchical Lookup Hash (HLH)} structure \cite{ho2022seasonal}, and b) the proposal of two groups of pruning techniques based on the Apriori principle and the temporal transitivity property of temporal events. Particularly, instead of using the Hierarchical Pattern Graph as in \cite{ho2022efficientvldb}, we use the Hierarchical Lookup Hash data structure to enable faster retrieval of events and patterns during the mining process. Algorithm \ref{algorithmHTPGM} provides the pseudo-code of our GTPM algorithm.

\vspace{-0.05in}
\subsection{Mining  Single Events}\label{sec:1freq}
\textbf{Hierarchical lookup hash structure $HLH_1$:}
We use the hierarchical lookup hash structure $HLH_1$, illustrated in Fig. \ref{fig:hlh1} to store single events. 
$HLH_1$ is a hierarchical data structure that consists of two hash tables: the \textit{single event hash table} $EH$, 
and the \textit{event sequence hash table} $SH$. 
Each hash table has a list of $<$key, value$>$ pairs. 
In $EH$, the key is the event symbol $\omega \in \Sigma_X$ representing the event $E_i$, and the value is the set of sequences $<S_i,...,S_k>$ (arranged in an increasing order) that contain $E_i$. In $SH$, the key is taken from the value component of $EH$, i.e., the set of sequences, while the value stores event instances of $E_i$ that occur in the corresponding sequence in $\mathcal{D}_{\text{SEQ}}$.
The $HLH_1$ structure enables faster retrieval of event sequences and instances when mining k-event patterns. 

\textbf{Mining Single Events:}  
The first step in GTPM is to find single events that satisfy the minimum support constraint $\sigma_{\min}$ (Alg. \ref{algorithmHTPGM}, lines 1-4). To do that, GTPM scans $\mathcal{D}_{\text{SEQ}}$ to compute the support of each event $E_i$, and checks whether \textit{supp}($E_i$)$\geq \sigma_{\min}$. Note that for single events, we do not consider the constraints on the confidence $\delta$, since confidence of single events is always $1$, and on maximum support $\sigma_{\max}$ because of the following lemma. 

\begin{lem}\label{lem0}
Let $P$ be a temporal pattern and $E_i$ be a single event such that $E_i \in P$. Then $\textit{supp}(P) \le \textit{supp}(E_i)$. 
\end{lem}
\textbf{Proof.} \textit{Detailed proofs of all lemmas, theorems, and complexities in this article can be found in the electronic appendix.}

From Lemma \ref{lem0}, a single event $E_i$ whose support \textit{supp}$(E_i) > \sigma_{\max}$ can form a pattern $P$ that has $\textit{supp}(P) \le \sigma_{\max}$. Thus, the constraint on $\sigma_{\max}$ is not considered for single events to avoid the loss of potential temporal patterns.

We provide a running example using data in Table \ref{tbl:SequenceDatabase}, with $\sigma_{\min}=0.7$, $\sigma_{\max} = 0.9$, 
 and $\delta=0.7$. The data structure $HLH_1$, shown in Fig. \ref{fig:patternTree}, stores $7$ single events satisfying $\sigma_{\min}$ constraint. The event WOff does not satisfy $\sigma_{\min}$ (only appears in sequences $2$ and $4$), and is thus omitted.  

\textbf{Complexity:} The complexity of finding single events is $O(m \cdot$$\mid$$\mathcal{D}_{\text{SEQ}}$$\mid)$, where $m$ is the number of distinct events.

\vspace{-0.1in}
\subsection{Mining  2-event Patterns}\label{sec:2freq} \vspace{-0.02in}
\textbf{Search space of GTPM}:
The next step in GTPM is to mine 2-event patterns. A straightforward approach would be to enumerate all possible event pairs, and check whether each pair can form patterns that satisfy the support and confidence constraints. However, this \textit{naive} approach is very expensive. Not only does it need to repeatedly scan $\mathcal{D}_{\text{SEQ}}$ to check each combination of events, the complex relations between events also add an extra exponential factor $3^{h^2}$ to the $m^h$ number of possible candidates, creating a very large search space that makes the approach infeasible.

\begin{lem}\label{lem1}\vspace{-0.05in}
Let $m$ be the number of distinct events in $\mathcal{D}_{\text{SEQ}}$, and $h$ be the longest length of a temporal pattern. The total number of temporal patterns is $O(m^h3^{h^2})$. 
\vspace{-0.05in}
\end{lem}
Lemma \ref{lem1} shows the driving factors of GTPM's exponential search space (proof in the electronic appendix): the number of events ($m$), the max pattern length ($h$), and the number of temporal relations ($3$). A dataset of just a few hundred events can create a very large search space with billions of candidate patterns. The optimizations and approximation proposed in the following sections will help mitigate this problem.

\begin{figure}
	\centering
	\captionsetup{justification=centering, font=small}
	\begin{minipage}{\linewidth}
		\includegraphics[width=\textwidth,height=2.5cm]{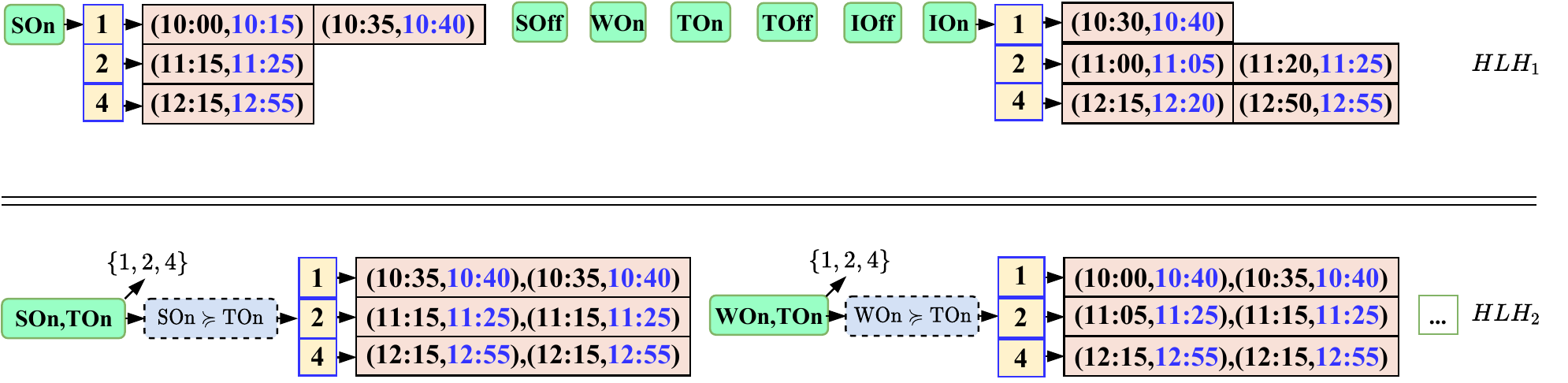}
		\vspace{-0.2in}
		\caption{A hierarchical lookup hash tables \\for the running example}
		\label{fig:patternTree}
	\end{minipage}
\end{figure}

\textbf{Hierarchical lookup hash structure $HLH_k$:} 
We maintain k-event groups and patterns found by GTPM using the $HLH_k$ $(k \geq 2)$ data structure, illustrated in Fig. \ref{fig:hlhk}. $HLH_k$ contains three hash tables, each has a list of $<$key, value$>$ pairs: the \textit{k-event hash table} $EH_k$, the \textit{pattern hash table} $PH_k$, and the \textit{pattern sequence hash table} $SH_k$. 
For each {\em $<$key, value$>$} pair of $EH_k$, {\em key} is the list of symbols $(\omega_1 ..., \omega_k)$ representing the k-event group $(E_1,...,E_k)$, and {\em value} is an \textit{object} structure which consists of two components: (1) the list of sequences $<S_i,...,S_k>$ (arranged in increasing order) where $(E_1,...,E_k)$ occurs, and (2), a list of k-event temporal patterns $P = \{(r_{12},E_1,E_2), ..., (r_{(k-1)(k)},E_{k-1},E_{k})\}$ created from the k-event group $(E_1,...,E_k)$.  
In $PH_k$, the {\em key} takes the {\em value} component of $EH_k$, i.e. the k-event pattern $P$, while the {\em value} is the list of sequences that support $P$.  
In $SH_k$, the {\em key} takes the {\em value} component of $PH_k$, i.e., the list of sequences that support $P$, while the {\em value} is the list of event instances from which the temporal relations in $P$ are formed. 
The $HLH_k$ hash structure helps speed up the mining of k-event groups through the use of sequences in $EH_k$, and enables faster search for temporal relations between $k$ events using the information in $PH_k$ and $SH_k$.

\vspace{-0.02in}\textbf{Two-steps filtering approach to mine 2-event patterns:} \label{sec:2filter}
Given the huge set of pattern candidates stated in Lemma \ref{lem1}, it is expensive to check their support and confidence. We propose a \textit{filtering approach} to reduce the unnecessary candidate checking. Specifically, the mining process is divided into two steps: (1) it first finds k-event groups that satisfy the minimum support and confidence constraints using $\sigma_{\min}$ and $\delta$, (2) it then generates temporal patterns only from those k-event groups. The correctness of this filtering approach is based on the Apriori-inspired lemmas below.
\begin{lem}\label{lem2}\vspace{-0.05in}
	Let $P$ be a 2-event pattern formed by an event pair $(E_i, E_j)$. Then, $\textit{supp}(P) \le \textit{supp}(E_i,E_j)$. 
	\vspace{-0.05in}
\end{lem} 

From Lemma \ref{lem2}, the support of a pattern is at most the support of its events. Thus, infrequent event pairs (those do not satisfy minimum support) cannot form frequent patterns and thereby, can be safely pruned.

\begin{lem}\label{lem3}	\vspace{-0.05in}
	Let $(E_i, E_j)$ be a pair of events forming a 2-event pattern $P$.  Then \textit{conf}($P$) $\le$ \textit{conf}($E_i,E_j$).
\vspace{-0.05in}
\end{lem}

From Lemma \ref{lem3}, the confidence of a pattern $P$ is always at most the confidence of its events. Thus, a low-confidence event pair cannot form any high-confidence patterns and therefore, can be safely pruned. 
We note that the Apriori principle has already been used in other work, e.g., \cite{tpminer,hdfs}, for mining optimization. However, they only apply this principle to the support (Lemma \ref{lem2}), while we further extend it to the confidence (Lemma \ref{lem3}). 
Applying Lemmas \ref{lem2} and \ref{lem3} to the event filtering step will remove infrequent or low-confidence event pairs, reducing the candidate patterns of GTPM. Furthermore, we do not consider the constraint on $\sigma_{\max}$ in this filtering step to avoid the loss of 2-event patterns, as event pairs that do not satisfy the $\sigma_{\max}$ constraint can still form 2-event patterns satisfying $\sigma_{\max}$ (Lemma \ref{lem2}).

\textbf{Step 2.1. Mining event pairs considering $\sigma_{\min}$ and $\delta$:}
This step finds event pairs in $\mathcal{D}_{\text{SEQ}}$ satisfying $\sigma_{\min}$ and $\delta$, using the set \textit{1Freq} found in $HLH_1$  (Alg. \ref{algorithmHTPGM}, lines 5-10). 
First, GTPM generates all possible event pairs by calculating the Cartesian product \textit{1Freq} $\times$ \textit{1Freq}. Next, for each pair $(E_i, E_j)$, the set $\mathcal{S}_{ij}$ (representing the set of sequences where both events occur) is computed by taking the \textit{intersection} between the set of sequences $\mathcal{S}_i$ of $E_i$ and the set of sequences $\mathcal{S}_j$ of $E_j$ in $HLH_1$. Finally, we compute the support $\textit{supp}(E_i, E_j)$ using $\mathcal{S}_{ij}$, and compare against $\sigma_{\min}$. 
If $\textit{supp}(E_i, E_j) \ge \sigma_{\min}$, $(E_i, E_j)$ has high enough support. Next, $(E_i, E_j)$ is further filtered using Lemma \ref{lem3}: $(E_i, E_j)$ is selected only if its confidence is at least $\delta$. After this step, only event pairs satisfying $\sigma_{\min}$ and $\delta$ are kept in $EH_2$ of $HLH_2$. 

\textbf{Step 2.2. Mining 2-event patterns:} This step mines 2-event patterns from the event pairs found in step 2.1 (Alg. \ref{algorithmHTPGM}, lines 11-13), considering three constraints $\sigma_{\min}$, $\sigma_{\max}$, and $\delta$. For each event pair $(E_i,E_j)$, we use the set of sequences $\mathcal{S}_{ij}$ to check the temporal relations between $E_i$ and $E_j$. Specifically, for each sequence $S \in \mathcal{S}_{ij}$, the pairs of event instances $(e_i,e_j)$ are extracted, and the relations between them are verified. The support and confidence of each relation $r(E_{i_{\triangleright e_i}},E_{j_{\triangleright e_j}})$ are computed and compared against $\sigma_{\min}$, and $\delta$ thresholds, after which only relations satisfying the two constraints are selected and stored in $PH_2$, while their event instances are stored in $SH_2$. Examples of the relations in $HLH_2$ can be seen in Fig. \ref{fig:patternTree}, e.g., event pair (SOn, TOn). We also emphasize that $HLH_2$ only stores patterns that satisfy the two constraints $\sigma_{\min}$, and $\delta$, thus, patterns in $PH_2$ are frequent temporal patterns. To mine rare temporal patterns from $HLH_2$, we take a further step by iterating through every 2-event pattern $P$ in $PH_2$, and checking the satisfaction of $P$ against the constraint $\sigma_{\max}$.

\textbf{Complexity:} Let $m$ be the number of single events in $HLH_1$, and $i$ be the average number of event instances of each event. The complexity of 2-event pattern mining is $O(m^2 i^2\mid$$\mathcal{D}_{\text{SEQ}}$$\mid$$^2)$. 

\vspace{-0.05in}
\subsection{Mining k-event Patterns} \label{sec:kfilter} \vspace{-0.02in}
Mining k-event patterns ($k \ge 3$) follows a similar process as 2-event patterns, with additional prunings based on the transitivity property of temporal relations.     

\textbf{Step 3.1. Mining k-event combinations considering $\sigma_{\min}$ and $\delta$:} This step finds k-event combinations that satisfy the minimum support and confidence constraints (Alg. \ref{algorithmHTPGM}, lines 14-16). 

Let \textit{(k-1)Freq} be the set of (k-1)-event combinations found in $HLH_{k-1}$, and \textit{1Freq} be the set of single events in $HLH_1$. To generate all k-event combinations, the typical process is to compute the Cartesian product: \textit{(k-1)Freq} $\times$ \textit{1Freq}. However, we observe that using \textit{1Freq} to generate k-event combinations at $HLH_k$ can create redundancy, since \textit{1Freq} might contain events that when combined with \textit{(k-1)Freq}, result in combinations that clearly cannot form any patterns satisfying the  minimum support constraint. To illustrate this observation, consider the event IOn in $HLH_1$ in Fig. \ref{fig:patternTree}. Here, IOn is a frequent event, and thus, can be combined with frequent event pairs in $HLH_2$ such as (SOn, TOn) to create a 3-event combination (SOn, TOn, IOn). However, (SOn, TOn, IOn) cannot form any 3-event patterns whose support is greater than $\sigma_{\min}$, since IOn is not present in any frequent 2-event patterns in $HLH_2$. To reduce the redundancy, the combination (SOn, TOn, IOn) should not be created in the first place. We rely on the \textit{transitivity property} of temporal relations to identify such event combinations.  

\begin{lem} \label{lem:transitivity}\vspace{-0.05in}
Let $S=<e_1$,..., $e_{n-1}>$ be a temporal sequence that supports an (n-1)-event pattern $P=<(r_{12}$, $E_{1_{\triangleright e_1}}$, $E_{2_{\triangleright e_2}})$,..., $(r_{(n-2)(n-1)}$, $E_{{n-2}_{\triangleright e_{n-2}}}$, $E_{{n-1}_{\triangleright e_{n-1}}})>$. Let $e_n$ be a new event instance added to $S$ to create the temporal sequence $S^{'}$$=$$<e_1, ..., e_{n}>$. 

The set of temporal relations $\Re$ is transitive on $S^{'}$: $\forall e_i \in S^{'}$, $i < n$, $\exists r \in \Re$ s.t. $r(E_{i_{\triangleright e_i}}$,$E_{n_{\triangleright e_n}})$ holds.
		
\vspace{-0.05in}
\end{lem}

Lemma \ref{lem:transitivity} says that given a temporal sequence $S$, a new event instance added to $S$ will always form at least one temporal relation with existing instances in $S$. This is due to the temporal transitivity property, which can be used to prove the following lemma.

\begin{lem}\label{lem:filter}\vspace{-0.05in}
	Let $N_{k-1}=(E_1,...,E_{k-1})$ be a (\textit{k-1})-event combination and $E_k$ be a single event, both satisfying the $\sigma_{\min}$ constraint. The combination $N_k= N_{k-1} \cup E_k$ can form k-event temporal patterns whose support is at least $\sigma_{\min}$ if $\forall E_i \in N_{k-1}$, $\exists r \in \Re$ s.t. $r(E_i,E_k)$ is a frequent temporal relation. 
	\vspace{-0.05in}
\end{lem}
From Lemma \ref{lem:filter}, only single events in $HLH_1$ that appear in $HLH_{k-1}$ should be used to create k-event combinations.
Using this result, a filtering on \textit{1Freq} is performed before calculating the Cartesian product. Specifically, from the events in $HLH_{k-1}$, we extract distinct single events $D_{k-1}$, and \textit{intersect} $D_{k-1}$ with \textit{1Freq} to remove redundant single events: \textit{Filtered1Freq} = $D_{k-1}$ $\cap$ \textit{1Freq}. 
Next, the Cartesian product \textit{(k-1)Freq} $\times$ \textit{Filtered1Freq} is calculated to generate k-event combinations. Finally, we apply Lemmas \ref{lem2} and \ref{lem3} to select k-event combinations \textit{kFreq} which upheld the $\sigma_{\min}$ and $\delta$ constraints. Similar to step 2.1, we do not consider $\sigma_{\max}$ when generating the k-event combination. 

\textbf{Step 3.2. Mining k-event patterns:} This step mines k-event patterns that satisfy the three constraints of $\sigma_{\min}$, $\sigma_{\max}$, and $\delta$ (Alg. \ref{algorithmHTPGM}, lines 17-19). 
Unlike 2-event patterns, verifying the relations in a k-event combination ($k \ge 3$) is much more expensive, as it requires to compute the frequency of $\frac{1}{2}k(k-1)$ triples of temporal relations. To reduce the cost of relation checking, we propose an iterative verification method that relies on the \textit{transitivity property} and the Apriori principle. 

\begin{lem}\label{lem5}\vspace{-0.05in}
	Let $P$ and $P^{'}$ be two temporal patterns. If $P^{'} \subseteq P$, then \textit{conf}($P^{'}$) $ \geq$ \textit{conf}($P$).
	\vspace{-0.05in}
\end{lem} 

\begin{lem} \label{lem6}\vspace{-0.05in}
Let $P$ and $P^{'}$ be two temporal patterns. If $P^{'} \subseteq P$ and $\frac{\textit{supp}(P^{'})}{\max_{1 \leq k \leq \mid P \mid}\{\textit{supp}(E_k)\}}_{E_k \in P} \leq \delta$, then \textit{conf($P$)} $\leq \delta$.
\vspace{-0.05in}
\end{lem}

Lemma \ref{lem5} says that, the confidence of a pattern $P$ is always at most the confidence of its sub-patterns. Consequently, from Lemma \ref{lem6}, a temporal pattern $P$ cannot be high-confidence if any of its sub-patterns are low-confidence.

Let $N_{k-1}=(E_1,...,E_{k-1})$ be a (k-1)-event combination in $HLH_{k-1}$, $N_1=(E_k)$ be an event in $HLH_1$, and $N_k=N_{k-1} \cup N_1 = (E_1,...,E_k)$ be a k-event combination in $HLH_k$. To find k-event patterns for $N_k$, we first retrieve the set $P_{k-1}$ containing (k-1)-event patterns of $N_{k-1}$ by accessing the $EH_{k-1}$ table. 
Each $p_{k-1} \in P_{k-1}$ is a list of $\frac{1}{2}(k-1)(k-2)$ triples: $\{(r_{12}$, $E_{1_{\triangleright e_1}}$, $E_{2_{\triangleright e_2}})$,...,$(r_{(k-2)(k-1)}$, $E_{{k-2}_{\triangleright e_{k-2}}}$, $E_{{k-1}_{\triangleright e_{k-1}}})\}$. We iteratively verify the possibility of $p_{k-1}$ forming a k-event pattern with $E_k$ that can satisfy the $\sigma_{\min}$ constraint as follows. 
We first check whether the triple $(r_{(k-1)k}$, $E_{{k-1}_{\triangleright e_{k-1}}}$, $E_{k_{\triangleright e_{k}}})$ satisfies the constraints of $\sigma_{\min}$, $\sigma_{\max}$, and $\delta$ by accessing the $HLH_2$ table. If the triple does not satisfy the minimum and maximum support constraints (using Lemmas \ref{lem:transitivity} and \ref{lem:filter}), or the confidence constraint (using Lemmas \ref{lem:transitivity}, \ref{lem5}, and \ref{lem6}), the verifying process stops immediately for $p_{k-1}$. Otherwise, it continues on the triple $(r_{(k-2)k}$, $E_{{k-2}_{\triangleright e_{k-2}}}$, $E_{k_{\triangleright e_{k}}})$, until it reaches $(r_{1k}$, $E_{1_{\triangleright e_1}}$, $E_{k_{\triangleright e_{k}}})$. 

We note that the transitivity property of temporal relations has been exploited in \cite{moskovitch2015fast} to generate new relations. Instead, we use this property to prune unpromising candidates (Lemmas \ref{lem:transitivity}, \ref{lem:filter}, \ref{lem5}, \ref{lem6}).

\textbf{Complexity:} 
Let $r$ be the average number of (k-1)-event patterns in $HLH_{k-1}$. The complexity of k-event pattern mining is $O($$\mid$$\textit{1Freq}$$\mid$ $\cdot$ $\mid$$\textit{(k-1)Freq}$$\mid$ $\cdot$ $r$ $\cdot$ $k^2$$\cdot$$\mid$$\mathcal{D}_{\text{SEQ}}$$\mid$$)$.

\textbf{GTPM overall complexity:} Throughout this section, we have seen that GTPM complexity depends on the size of the search space $(O(m^h3^{h^2}))$ and the complexity of the mining process itself, i.e., $O(m \cdot$$\mid$$\mathcal{D}_{\text{SEQ}}$$\mid)$ $+$ $O(m^2 i^2\mid$$\mathcal{D}_{\text{SEQ}}$$\mid$$^2)$ $+$ $O($$\mid$$\textit{1Freq}$$\mid$ $\cdot$ $\mid$$\textit{(k-1)Freq}$$\mid$ $\cdot$ $r$ $\cdot$ $k^2$$\cdot$$\mid$$\mathcal{D}_{\text{SEQ}}$$\mid$$)$. 
While the parameters $m$, $h$, $i$, $r$ and $k$ depend on the number of time series, others such as $\mid$$\textit{1Freq}$$\mid$, $\mid$$\textit{(k-1)Freq}$$\mid$ and $\mid$$\mathcal{D}_{\text{SEQ}}$$\mid$ also depend on the number of temporal sequences. Thus, given a dataset, GTPM complexity is driven by two main factors: the number of time series and the number of temporal sequences.

\section{Approximate GTPM}\label{sec:MI}\vspace{-0.02in}
\subsection{Mutual Information of Symbolic Time Series}\vspace{-0.02in}
Let $X_S$ and $Y_S$ be the symbolic series representing the time series $X$ and $Y$, respectively, and $\Sigma_X$, $\Sigma_Y$ be their alphabets.

\hspace{-0.2in}\textbf{Definition 5.1} (Entropy) The \textit{entropy} of $X_S$, denoted as $H(X_S)$, is defined as\vspace{-0.1in}
\begin{equation}
	\small
	H(X_S)= - \sum_{x \in \Sigma_X} p(x) \cdot \log p(x) 
	\vspace{-0.05in}
\end{equation}	
Intuitively, the entropy measures the amount of information or the inherent uncertainty in the possible outcomes of a random variable. The higher the $H(X_S)$, the more uncertain the outcome of $X_S$.

The conditional entropy $H(X_S\vert Y_S)$ quantifies the amount of information needed to describe the outcome of $X_S$, given the value of $Y_S$, and is defined as\vspace{-0.07in}
\begin{equation}
	\small
	H(X_S\vert Y_S) =  - \sum_{x \in \Sigma_X} \sum_{y \in \Sigma_Y} p(x,y) \cdot \log \frac{p(x,y)}{p(y)} 
\end{equation}

\hspace{-0.2in}\textbf{Definition 5.2} (Mutual information) The \textit{mutual information} (MI) of two symbolic series $X_S$ and $Y_S$, denoted as $I(X_S;Y_S)$, is defined as\vspace{-0.02in}
\begin{equation}	
	\small
	I(X_S;Y_S)=\sum_{x \in \Sigma_X} \sum_{y \in \Sigma_Y} p(x,y) \cdot \log \frac{p(x,y)}{p(x) \cdot p(y)} 
	\label{eq:MI}
\end{equation}
The MI represents the reduction of uncertainty of one variable (e.g., $X_S$), given the knowledge of another variable (e.g., $Y_S$). The larger $I(X_S;Y_S)$, the more information is shared between $X_S$ and $Y_S$, and thus, the less uncertainty about one variable given the other.  

Since $0 \le I(X_S;Y_S) \leq \min(H(X_S), H(Y_S))$ \cite{thomas}, MI has no upper bound. To scale the MI into the range $[0-1]$, we use normalized mutual information as defined below. 

\hspace{-0.2in}\textbf{Definition 5.3} (Normalized mutual information) The \textit{normalized mutual information} (NMI) of two symbolic time series $X_S$ and $Y_S$, denoted as $\widetilde{I}(X_S;Y_S)$, is defined as\vspace{-0.05in}
\begin{equation}	
	\small
	\widetilde{I}(X_S;Y_S)= \frac{I(X_S;Y_S)}{H(X_S)} =1-\frac{H(X_S|Y_S)}{H(X_S)}
	\label{eq:NMI}
	\vspace{-0.05in}
\end{equation}
$\widetilde{I}(X_S;Y_S)$ represents the reduction (in percentage) of the uncertainty of $X_S$ due to knowing $Y_S$. Based on Eq. \eqref{eq:NMI}, a pair of variables $(X_S,Y_S)$ holds a mutual dependency if $\widetilde{I}(X_S;Y_S) > 0$. Eq. \eqref{eq:NMI} also shows that NMI is not symmetric, i.e., $\widetilde{I}(X_S;Y_S)  \neq \widetilde{I}(Y_S;X_S)$.

\subsection{Lower Bound of the Support of an Event Pair}\vspace{-0.02in} \label{section:lowerbound}
Consider two symbolic series $X_S$ and $Y_S$. Let $X_1$ be an event in $X_S$, $Y_1$ be an event in $Y_S$, and $\mathcal{D}_{\text{SYB}}$ and $\mathcal{D}_{\text{SEQ}}$ be the symbolic and the sequence databases created from $X_S$ and $Y_S$, respectively. 
We first study the relationship between the support of $(X_1,Y_1)$ in $\mathcal{D}_{\text{SYB}}$ and $\mathcal{D}_{\text{SEQ}}$. 	\vspace{-0.05in}

\begin{lem}\label{lem:supportconnection1}
	Let $\textit{supp}(X_1,Y_1)_{\mathcal{D}_{\text{SYB}}}$ and $\textit{supp}(X_1,Y_1)_{\mathcal{D}_{\text{SEQ}}}$ be the support of $(X_1,Y_1)$ in $\mathcal{D}_{\text{SYB}}$ and $\mathcal{D}_{\text{SEQ}}$, respectively. Then  $\textit{supp}(X_1,Y_1)_{\mathcal{D}_{\text{SYB}}} \leq \textit{supp}(X_1,Y_1)_{\mathcal{D}_{\text{SEQ}}}$ holds.
	\vspace{-0.05in}
\end{lem}
\begin{proof} (Sketch - Detailed proof in the electronic appendix).
		Let $n$ be the length of each symbolic time series in $\mathcal{D}_{\text{SYB}}$, and $m$ be the length of each temporal sequence. 
		The number of temporal sequences obtained in $\mathcal{D}_{\text{SEQ}}$ is: $\lceil \frac{n}{m} \rceil$.
		
		The support of $(X_1,Y_1)$ in $\mathcal{D}_{\text{SYB}}$ is computed as: 
		\begin{equation}
			supp(X_1,Y_1)_{\mathcal{D}_{\text{SYB}}} = \frac{\sum_{i=1}^{\lceil \frac{n}{m} \rceil} \sum_{j=1}^{m} s_{ij}} {n}
			\label{eq:supportSYBSEQ1}
		\end{equation}
		where
		\begin{equation}
			\scalemath{0.94}{s_{ij} = \begin{cases}
					\small
					1, & \parbox[l]{1\columnwidth}{\text{if $(X_1,Y_1)$ occurs in row j of the sequence $s_i$ in  $\mathcal{D}_{\text{SYB}}$}}\\
					0, &\text{otherwise}
				\end{cases} \nonumber
				\label{eq:sij}
			}
		\end{equation}
		Moreover, we have: 
		\begin{equation}
			\textit{supp}(X_1,Y_1)_{\mathcal{D}_{\text{SEQ}}} = \frac{\sum_{i=1}^{\lceil \frac{n}{m} \rceil} g_i} {n/m} = \frac{m \cdot \sum_{i=1}^{\lceil \frac{n}{m} \rceil} g_i} {n} 
			\label{eq:supportSYBSEQ2}
		\end{equation}
		where 
		\[g_i = \begin{cases}
			1,& \text{if $(X_1,Y_1)$ occurs in the sequence $g_i$ in $\mathcal{D}_{\text{SEQ}}$}\\
			0,              & \text{otherwise}
		\end{cases}\]
		We also get:
		\begin{align} 
			\textit{supp}(X_1,Y_1)_{\mathcal{D}_{\text{SEQ}}} &= \frac{m \cdot \sum_{i=1}^{\lceil \frac{n}{m} \rceil} g_i} {n} = \frac{\sum_{i=1}^{\lceil \frac{n}{m} \rceil} m \cdot g_i} {n} \nonumber \\ 
			& = \frac{ \sum_{i=1}^{\lceil \frac{n}{m} \rceil} \left( \sum_{j=1}^{m} s_{ij} + \vartheta_i \right) } {n}
			\label{eq:supportDSEQ1}
		\end{align}
		where $s_{ij}$ is defined as in Eq. \eqref{eq:supportSYBSEQ1}, and
		\begin{equation}
			\vartheta_{i} =  \begin{cases}
				m - \sum_{j=1}^{m} s_{ij}, & \parbox[l]{1\columnwidth}{ \text{if $\sum_{j=1}^{m} s_{ij} \neq 0 $}}\\
				0,  & \text{otherwise}
			\end{cases} \nonumber
		\end{equation}
	\\
		From Eq. \eqref{eq:supportDSEQ1}, we have:
		\begin{align}
			\textit{supp}(X_1,Y_1)_{\mathcal{D}_{\text{SEQ}}} &=  \frac{\sum_{i=1}^{\lceil \frac{n}{m} \rceil} \sum_{j=1}^{m} s_{ij}} {n} +  \frac{\sum_{i=1}^{\lceil \frac{n}{m} \rceil} \vartheta_{i}} {n} \nonumber \\
			& =	supp(X_1,Y_1)_{\mathcal{D}_{\text{SYB}}} + \vartheta 
			\label{eq:supportDSEQ2}
		\end{align}
		where
		$\vartheta =  \frac{\sum_{i=1}^{\lceil \frac{n}{m} \rceil} \vartheta_{i}} {n}$ is the difference between the probabilities of $(X_1, Y_1)$ in $\mathcal{D}_{\text{SEQ}}$ and $\mathcal{D}_{\text{SYB}}$.
		\\
		From Eq. \eqref{eq:supportDSEQ2}, we have: 
		\begin{equation}
			\textit{supp}(X_1,Y_1)_{\mathcal{D}_{\text{SYB}}} \leq \textit{supp}(X_1,Y_1)_{\mathcal{D}_{\text{SEQ}}}
		\end{equation}
		
	\end{proof}

From Lemma \ref{lem:supportconnection1}, a frequent event pair in $\mathcal{D}_{\text{SYB}}$ is also frequent in $\mathcal{D}_{\text{SEQ}}$. We now investigate the relation between $\widetilde{I}(X_S;Y_S)$ in $\mathcal{D}_{\text{SYB}}$ and the support of $(X_1,Y_1)$ in $\mathcal{D}_{\text{SEQ}}$.

\begin{theorem} (Lower bound of the support) \textit{Let $\mu_{\min}$ be the minimum mutual information threshold. If the NMI\hspace{0.04in} $\widetilde{I}(X_S$;$Y_S) \ge \mu_{\min}$, then the lower bound of the support of $(X_1,Y_1)$ in $\mathcal{D}_{\text{SEQ}}$ is: \vspace{-0.1in}
			\begin{align} \vspace{-0.1in}
				\small
				\textit{supp}(X_1,Y_1)_{\mathcal{D}_{\text{SEQ}}} \ge \lambda_2 \cdot e^{W\left( \frac{\log{\lambda_{1}^{1-\mu_{\min}}} \cdot ln2}{\lambda_2} \right)}
				\label{eq:lowerboundsupport}
			\end{align}
			where $\lambda_1$ is the minimum support of $X_i \in X_S$, $\lambda_2$ is the support of $Y_1 \in Y_S$, and $W$ is the Lambert function \cite{corless1996lambertw}.} 
		\label{theorem:boundSupport}   
	\end{theorem}	
	\begin{proof} (Sketch - Detailed proof in the electronic appendix). From Eq. \eqref{eq:NMI}, we have:\vspace{-0.05in}
		\begin{align}
			\small
			\widetilde{I}(X_S;Y_S)= 1 - \frac{H(X_S \vert Y_S)}{H(X_S)} \ge \mu_{\min} 
		\end{align}
		\begin{align}
			\small
			\Rightarrow \frac{H(X_S \vert Y_S)}{H(X_S)} & = \frac{p(X_1, Y_1) \cdot \log p(X_1 \vert Y_1) }   {\sum_{i} p(X_i) \cdot \log p(X_i)} \nonumber \\ &\scalemath{0.95}{ + \frac{\sum_{i \neq 1 \wedge j \neq 1} p(X_i, Y_j) \cdot \log \frac{p(X_i , Y_j)} {p(Y_j)}}{\sum_{i} p(X_i) \cdot \log p(X_i)}
				\le  1-\mu_{\min} }
			\label{eq:sketchsupp1}
		\end{align}
		Let $\lambda_1 = p(X_k)$ such that $p(X_k) = \min \lbrace p(X_i)\rbrace, \forall i$, 
		and $\lambda_2=p(Y_1)$. 
		We obtain:\vspace{-0.05in}
		\begin{align}
			\small
			\frac{H(X_S \vert Y_S)}{H(X_S)} &\geq  \frac{p(X_1, Y_1) \cdot \log \frac{p(X_1 , Y_1)} {\lambda_2} }{\log \lambda_1} 
			\label{eq:sketchsupp2}
		\end{align}	
		\\
		From Eqs. \eqref{eq:sketchsupp1}, \eqref{eq:sketchsupp2}, the support lower bound of $(X_1,Y_1)$ in $\mathcal{D}_{\text{SYB}}$ is derived as:\vspace{-0.05in}
		\begin{align}
			\small
			\textit{supp}(X_1,Y_1)_{\mathcal{D}_{\text{SYB}}} \geq  \lambda_2 \cdot e^{W\left(\frac{\log \lambda_{1}^{1-\mu_{\min}} \cdot \ln 2}{\lambda_2}  \right)} 
		\end{align}	
		\vspace{-0.05in}Since: \vspace{-0.1in}
		\begin{align}
			\small
			\textit{supp}(X_1,Y_1)_{\mathcal{D}_{\text{SEQ}}} \ge \textit{supp}(X_1,Y_1)_{\mathcal{D}_{\text{SYB}}}  
			\vspace{-0.05in}
		\end{align}
		\vspace{-0.05in}It follows that:\vspace{-0.1in}
		\begin{align}
			\small
			\textit{supp}(X_1,Y_1)_{\mathcal{D}_{\text{SEQ}}} \ge \lambda_2 \cdot e^{W\left(\frac{\log \lambda_{1}^{1-\mu_{\min}} \cdot \ln 2}{\lambda_2}  \right)} 
		\end{align}
	\end{proof}

From Theorem \ref{theorem:boundSupport}, we can derive the minimum MI threshold $\mu_{\min}$ such that the support of $(X_1,Y_1)$ is at least $\sigma_{\min}$.
	\begin{corollary} 
		\label{corollary:lowerboundsupport} 
		The support of an event pair $(X_1,Y_1) \in (X_S,Y_S)$ in $\mathcal{D}_{\text{SEQ}}$ is at least $\sigma_{\min}$ if \hspace{0.01in} $\widetilde{I}(X_S;Y_S)$ is at least $\mu_{\min}$, where: \vspace{-0.07in} 
		\begin{align}
			\vspace{-0.1in}
			\small
			\mu_{\min} \geq \begin{cases}
				\small
				1 - \frac{\lambda_2}{e \cdot \ln 2 \cdot \log \frac{1}{\lambda_1} }, & \text{if $\quad 0 \leq \frac{\sigma_{\min}}{\lambda_2} \leq \frac{1}{e}$}\\
				1 - \frac{\sigma_{\min} \cdot \log \frac{\sigma_{\min}}{\lambda_2} }{\ln 2 \cdot \log \lambda_1},              &\text{otherwise}
			\end{cases}
			\label{eq:muSupportSetting}
		\end{align}
		\vspace{-0.15in}
	\end{corollary}

\textbf{Interpretation of the support lower bound:} 
Given two symbolic series $X_S$ and $Y_S$, and a minimum mutual information threshold $\mu_{\min}$. Theorem \ref{theorem:boundSupport} says that, if $X_S$ and $Y_S$ are mutually dependent with the minimum MI value $\mu_{\min}$, then the support of an event pair in ($X_S$, $Y_S$) is at least the lower bound in Eq. \eqref{eq:lowerboundsupport}. Combining Theorem \ref{theorem:boundSupport} and Lemma \ref{lem2}, we can conclude that if an event pair of ($X_S$,$Y_S$) has a support less than the lower bound in Eq. \eqref{eq:lowerboundsupport}, then any pattern $P$ formed by that event pair also has support less than that lower bound. This allows us to construct an approximate version of GTPM (discussed in Section \ref{sec:approximateTPMfTS}).

\subsection{Lower bound of the Confidence of an Event Pair}
Consider two events $X_1$, $Y_1$ of two symbolic series $X_S$ and $Y_S$. We derive the confidence lower bound of $(X_1,Y_1)$ in the sequence database $\mathcal{D}_{\text{SEQ}}$ as follows. 

\begin{theorem} (Lower bound of the confidence) Let $\sigma_{\min}$ and $\mu_{\min}$ be the minimum support and minimum mutual information thresholds, respectively. Assume that $ \textit{supp}(X_1,Y_1)_{\mathcal{D}_{\text{SEQ}}} \ge \sigma_{\min}$. If the NMI\hspace{0.04in} $\widetilde{I}(X_S$;$Y_S) \ge \mu_{\min}$, then the lower bound of the confidence of $(X_1,Y_1)$ in $\mathcal{D}_{\text{SEQ}}$ is: \vspace{-0.05in} 
	\begin{align} \vspace{-0.15in}
		\small
		\textit{conf}(X_1,Y_1)_{\mathcal{D}_{\text{SEQ}}} \ge \sigma_{\min}  \cdot  \lambda_1^{\frac{1-\mu_{\min}}{\sigma_{\min}} } \cdot \left(\frac{n_x - 1}{1 - \sigma_{\min}}\right)^{\frac{\lambda_3}{\sigma_{\min}}}   
		\vspace{-0.05in}
		\label{eq:lowerboundconfidence}
	\end{align}
	where $n_x$ is the number of symbols in $\Sigma_X$, $\lambda_1$ is the minimum support of $X_i \in X_S$, and $\lambda_3$ is the support of $(X_i, Y_j) \in (X_S, Y_S)$ such that $p(X_i|Y_j)$ is minimal, $\forall (i \neq 1$ $\wedge$ $j \neq 1)$. 
	\label{theorem:boundconfidence}   
	\vspace{-0.05in}
\end{theorem}
\vspace{-0.15in}
\begin{proof} (Sketch - Detailed proof in the electronic appendix).  
	Let $\lambda_1 = p(X_k)$ such that $p(X_k) = \min \lbrace p(X_i)\rbrace, \forall i$,  
	and $\lambda_3=p(X_m,Y_n)$ such that $p(X_m\vert Y_n) = \min \lbrace p(X_i\vert Y_j) \rbrace, \forall (i \neq 1 \wedge j \neq 1)$.
	Then, by applying the min-max inequality theorem for the sum of ratio \cite{beckenbach1961introduction} to the numerator of Eq. \eqref{eq:sketchsupp1}, we obtain:\vspace{-0.05in}
	\begin{align}
		\small
		\frac{H(X_S \vert Y_S)}{H(X_S)} &\geq  \frac{p(X_1, Y_1) \cdot \log p(X_1 \vert Y_1) + \lambda_3 \cdot \log \frac{1-p(X_1,Y_1)}{n_x-p(Y_1)}}{\log \lambda_1}
		\nonumber \\
		&\geq \frac{\sigma_{\min} \cdot \log \frac{p(X_1,Y_1)}{p(Y_1)} + \lambda_3 \cdot \log \frac{1-\sigma_{\min}}{n_x - 1}} {\log \lambda_1} 
		\label{eq:sketch33}
	\end{align}	
	\\
	Next, assume that $\textit{supp}(Y_{1})_{\mathcal{D}_{\text{SYB}}} \ge \textit{supp}(X_{1})_{\mathcal{D}_{\text{SYB}}}$. From Eqs. \eqref{eq:sketchsupp1}, \eqref{eq:sketch33}, the confidence lower bound of $(X_1,Y_1)$ in $\mathcal{D}_{\text{SYB}}$ is derived as:\vspace{-0.05in}
	\begin{equation}
		\scalemath{0.9}{\textit{conf}(X_1,Y_1)_{\mathcal{D}_{\text{SYB}}} = 
			\frac{\textit{supp}(X_1,Y_1)_{\mathcal{D}_{\text{SYB}}}}{\textit{supp}(Y_{1})_{\mathcal{D}_{\text{SYB}}}} \geq  \lambda_1^{\frac{1-\mu_{\min}}{\sigma_{\min}} } \cdot \left(\frac{n_x - 1}{1-\sigma_{\min}}\right)^\frac{\lambda_3}{\sigma_{\min}}}
	\end{equation}	
	\\
	Since: 
	\vspace{-0.24in}
	\begin{align}
		\vspace{-0.2in}
		\small
		\textit{conf}(X_1,Y_1)_{\mathcal{D}_{\text{SEQ}}} \ge \sigma_{\min} \cdot \textit{conf}(X_1,Y_1)_{\mathcal{D}_{\text{SYB}}}  
		\vspace{-0.05in}
	\end{align}
	\\
	It follows that:\vspace{-0.15in}
	\begin{align}
		\small
		\textit{conf}(X_1,Y_1)_{\mathcal{D}_{\text{SEQ}}} \ge \sigma_{\min} \cdot \lambda_1^{\frac{1-\mu_{\min}}{\sigma_{\min}} } \cdot \left(\frac{n_x - 1}{1-\sigma_{\min}}\right)^\frac{\lambda_3}{\sigma_{\min}} 
	\end{align}
\end{proof}

From Theorem \ref{theorem:boundconfidence}, we can derive the minimum MI threshold $\mu_{\min}$ such that the confidence of $(X_1,Y_1)$ is at least $\delta$.

\begin{corollary} \label{corollary:lowerboundconfidence} 
	The confidence of an event pair $(X_1,Y_1) \in (X_S,Y_S)$ in $\mathcal{D}_{\text{SEQ}}$ is at least $\delta$ if \hspace{0.01in} $\widetilde{I}(X_S;Y_S)$ is at least $\mu_{\min}$, where: \vspace{-0.07in}
	\begin{align}
		\vspace{-0.1in}
		\small
		\mu_{\min} &\geq 1-\sigma_{\min} \cdot  \log_{\lambda_1}{ \left(\frac{\delta}{\sigma_{\min}} \cdot \left(\frac{1-\sigma_{\min}}{n_x - 1}\right)^{\frac{\lambda_3}{\sigma_{\min}}}\right)}
		\label{eq:muConfidenceSetting}
	\end{align}
	\vspace{-0.15in}
\end{corollary}

\textbf{Interpretation of the confidence lower bound:} 
	Given two symbolic series $X_S$ and $Y_S$, and a minimum mutual information threshold $\mu_{\min}$. Theorem \ref{theorem:boundconfidence} says that, if $X_S$ and $Y_S$ are mutually dependent with the minimum MI value $\mu_{\min}$, then the confidence of an event pair in ($X_S$, $Y_S$) is at least the lower bound in Eq. \eqref{eq:lowerboundconfidence}. Combining Theorem \ref{theorem:boundconfidence} and Lemma \ref{lem3}, if an event pair of ($X_S$,$Y_S$) has a confidence less than the lower bound in Eq. \eqref{eq:lowerboundconfidence}, then any pattern $P$ formed by that event pair also has a confidence less than that lower bound. This allows us to construct an approximate version of GTPM (discussed in Section \ref{sec:approximateTPMfTS}).

\subsection{Upper Bound of the Support of an Event Pair}
We derive the support upper bound of the event pair $(X_1,Y_1)$ of $X_S$ and $Y_S$ in $\mathcal{D}_{\text{SEQ}}$ as follows. 

\begin{theorem} (Upper bound of the support) Let $\sigma_{\min}$ be the minimum support threshold, and $\mu_{\max}$ be the maximum mutual information threshold, respectively. Assume that $\textit{supp}(X_1,Y_1)_{\mathcal{D}_{\text{SEQ}}} \ge \sigma_{\min}$. If the NMI\hspace{0.04in} $\widetilde{I}(X_S$;$Y_S) \le \mu_{\max}$, then the upper bound of the support of $(X_1,Y_1)$ in $\mathcal{D}_{\text{SEQ}}$ is: 
	\begin{align} \vspace{-0.15in}
		\small
		\textit{supp}(X_1,Y_1)_{\mathcal{D}_{\text{SEQ}}} \le \lambda_2  \cdot e^{W \left( \frac{\log \frac{\lambda_5^{1-\mu_{\max}}}{\lambda_4^{1-\sigma_{\min}}} \cdot \ln2} {\lambda_2}  \right)} + \vartheta
		\vspace{-0.05in}
		\label{eq:upperboundSupport}
	\end{align}
	where:  $\lambda_2$ is the support of $Y_1 \in Y_S$, $\lambda_4$ is the fraction between the support of $(X_i, Y_j) \in (X_S, Y_S)$ and the support of $Y_j \in Y_S$ such that $p(X_i|Y_j)$ is minimal, $\forall i \neq 1$ $\wedge$ $j \neq 1$, $\lambda_5$ is the maximum support of $X_i \in X_S$, and $\vartheta$ is the difference between the probabilities of $(X_1, Y_1)$ in $\mathcal{D}_{\text{SEQ}}$ and $\mathcal{D}_{\text{SYB}}$.
	\label{theorem:upperboundSupport}   
	\vspace{-0.05in}
\end{theorem}

\begin{proof} (Sketch - Detailed proof in the electronic appendix).  
	Let $\lambda_2 = p(Y_1)$, $\lambda_4 = \min \lbrace p(X_i|Y_j) \rbrace$ $\forall (i\neq 1 \wedge j \neq 1)$, and $\lambda_5 = \max \lbrace p(X_i) \rbrace$ $\forall i$. We obtain:
	\begin{align}
		\scalemath{0.9} {\frac{H(X_S \vert Y_S)}{H(X_S)} \leq \frac{p(X_1,Y_1) \cdot \log \frac{p(X_1,Y_1)}{\lambda_2} + (1-\sigma_{\min}) \cdot \log \lambda_4}{\log \lambda_5} }
		\label{eq:upper1}
	\end{align} 
	From Eqs. \eqref{eq:NMI}, we have:
	\begin{align}
		\scalemath{0.9}{\widetilde{I}(X_S;Y_S)= 1 - \frac{H(X_S \vert Y_S)}{H(X_S)} \leq \mu_{\max}  \Rightarrow \frac{H(X_S \vert Y_S)}{H(X_S)} \geq 1-\mu_{\max} }
		\label{eq:upper2}
	\end{align}
	From Eqs. \eqref{eq:upper1} and \eqref{eq:upper2}, we have:
	\begin{align}
		&\scalemath{0.92}{\frac{p(X_1,Y_1) \cdot \log \frac{p(X_1,Y_1)}{\lambda_2} + (1-\sigma_{\min}) \cdot \log \lambda_4}{\log \lambda_5} \geq 1- \mu_{\max} } \\
		&\Leftrightarrow p(X_1,Y_1) \leq \lambda_2 \cdot e^{W \left( \frac{\log \frac{\lambda_5^{1-\mu_{\max}}}{\lambda_4^{1-\sigma_{\min}}} \cdot \ln2} {\lambda_2}  \right)}
		\label{eq:upper3}
	\end{align}
	From Eq. \eqref{eq:supportDSEQ2}, we have:
	\begin{align}
		\scalemath{0.95}{p(X_1, Y_1) = supp(X_1,Y_1)_{\mathcal{D}_{\text{SYB}}} = supp(X_1,Y_1)_{\mathcal{D}_{\text{SEQ}}} - \vartheta }
		\label{eq:upper4}
	\end{align}
	From Eqs. \eqref{eq:upper3} and \eqref{eq:upper4}, we have:
	\begin{align}
		supp(X_1,Y_1)_{\mathcal{D}_{\text{SEQ}}} \leq \lambda_2 \cdot e^{W \left( \frac{\log \frac{\lambda_5^{1-\mu_{\max}}}{\lambda_4^{1-\sigma_{\min}}} \cdot \ln2} {\lambda_2}  \right)} + \vartheta
	\end{align}
\end{proof}

From Theorem \ref{theorem:upperboundSupport}, we can derive the maximum MI threshold $\mu_{\max}$ such that the support of $(X_1,Y_1)$ is at most $\sigma_{\max}$.

\begin{corollary} \label{corollary:upperboundsupport} 
	The support of an event pair $(X_1,Y_1) \in (X_S,Y_S)$ in $\mathcal{D}_{\text{SEQ}}$ is at most $\sigma_{\max}$ if \hspace{0.01in} $\widetilde{I}(X_S;Y_S)$ is at most $\mu_{\max}$, where: \vspace{-0.07in} 
	\begin{align}
		\vspace{-0.1in}
		\small
		\mu_{\max} \leq 1- \frac{\frac{\sigma_{\max} - \vartheta}{\lambda_2} \cdot \log\frac{\sigma_{\max}-\vartheta}{\lambda_2} + \log \lambda_4^{1-\sigma_{\min}}}{\log \lambda_5}
		\label{eq:muUpperSupportSetting}
	\end{align}
	\vspace{-0.15in}
\end{corollary}

\textbf{Interpretation of the support upper bound:} 
Given a maximum MI threshold $\mu_{\max}$, let $X_S$ and $Y_S$ be two symbolic series. Theorem \ref{theorem:upperboundSupport} says that, if the NMI of $X_S$ and $Y_S$ is at most $\mu_{\max}$, then the support of an event pair in ($X_S$, $Y_S$) is at most the upper bound in Eq. \eqref{eq:upperboundSupport}. Combining Theorem \ref{theorem:upperboundSupport} and Lemma \ref{lem2}, we can conclude that if an event pair in ($X_S$,$Y_S$) has a support less than the upper bound, then any pattern $P$ formed by that event pair also has support less than that upper bound. 

\textbf{Setting the values of $\mu_{\min}$ and $\mu_{\max}$:}
GTPM uses three user-defined parameters, the minimum support $\sigma_{\min}$, the maximum support $\sigma_{\max}$, and the minimum confidence $\delta$ to mine both frequent and rare temporal patterns (with $\sigma_{\max}$ is set to $\infty$ in case of frequent patterns). To mine frequent patterns that satisfy both $\sigma_{\min}$ and $\delta$ constraints, we select $\mu_{\min}$ such that both Eqs. \eqref{eq:muSupportSetting} and \eqref{eq:muConfidenceSetting} hold, i.e., the maximum value of $\mu_{\min}$ provided by the two equations. On the other hand, to mine rare patterns that also have to satisfy $\sigma_{\max}$ constraint, $\mu_{\max}$ is chosen using Eq. \eqref{eq:muUpperSupportSetting}. 

\removelatexerror
\begin{minipage}{\columnwidth}
	\hspace{-0.27in}
	\SetNlSty{}{}{:} 
	\begin{algorithm}[H]
		\algsetup{linenosize=\tiny}
		\SetInd{0.5em}{0.5em}
		\small
		\DontPrintSemicolon
		\caption{Approximate GTPM using MI}
		\label{algorithmMI-RTPM}
		\KwInput{A set of time series $\mathcal{X}$, a minimum support threshold $\sigma_{\min}$, a maximum support threshold $\sigma_{\max}$, a minimum confidence threshold $\delta$}
		\KwOutput{The set of temporal patterns $P$}
		
		Convert $\mathcal{X}$ to $\mathcal{D}_{\text{SYB}}$ and $\mathcal{D}_{\text{SEQ}}$;\;
		Scan $\mathcal{D}_{\text{SYB}}$ to compute the probability of each event, event pair, and plus $\vartheta$ value;\;
		\ForEach{\textit{pair of symbolic time series $(X_S,Y_S)$} $\in \mathcal{D}_{\text{SYB}}$}{
			Compute $\widetilde{I}(X_S;Y_S)$ and $\widetilde{I}(Y_S;X_S)$;\;
			Compute $\mu_{\min}$ using Eqs. \eqref{eq:muSupportSetting} and \eqref{eq:muConfidenceSetting};\;
			Compute $\mu_{\max}$ using Eqs. \eqref{eq:muUpperSupportSetting};\;
			\If {$\min \{\widetilde{I}(X_S; Y_S), \widetilde{I}(Y_S; X_S)\} \geq \mu_{\min}$}{
				\If {$\min \{\widetilde{I}(X_S; Y_S), \widetilde{I}(Y_S; X_S)\} \leq \mu_{\max}$}{
					Insert $X_S$ and $Y_S$ into $\mathsf{X}_C$;\;
				}
			} 	
		}
		\ForEach{$X_{S}$ $\in \mathsf{X}_C$}{
			Mine single events from $X_{S}$ as in Section \ref{sec:1freq};\;
		}
		\ForEach{$(X_{S},Y_{S}) \in \mathsf{X}_C$}{
			Mine 2-event patterns from  $(X_{S},Y_{S})$ as in Section \ref{sec:2filter};
		}
		\If{$k \ge 3$}{
			Mine k-event patterns similar to the exact GTPM in Section \ref{sec:kfilter};
		}
	\end{algorithm}
\end{minipage}

\subsection{Using the Bounds for Approximate GTPM} \label{sec:approximateTPMfTS}
\textit{Approximate GTPM:} Approximate GTPM is based on the exact GTPM and performs the mining only on \textit{the set of mutually dependent symbolic series $\mathsf{X}_C \in \mathcal{X}$ with minimum and maximum MI thresholds $\mu_{\min}$ and $\mu_{\max}$}. Algorithm \ref{algorithmMI-RTPM} describes the approximate GTPM. First, $\mathcal{D_{\text{SYB}}}$ is scanned once to compute the probability of each single event, pair of events, and plus $\vartheta$ value (line 2). Next, NMI, $\mu_{\min}$, and $\mu_{\max}$ are computed for each symbolic series pair (lines 4-6). The pairs of symbolic series whose $\min \{\widetilde{I}(X_S; Y_S), \widetilde{I}(Y_S; X_S)\}$ is at least $\mu_{\min}$, and $\min \{\widetilde{I}(X_S; Y_S), \widetilde{I}(Y_S; X_S)\}$ is at most $\mu_{\max}$ are inserted into $\mathsf{X}_C$ (lines 7-9). Then, we traverse each series in $\mathsf{X}_C$ to mine the single events (lines 10-11). Next, each event pair in corresponding series in $\mathsf{X}_C$ is employed to mine the 2-event patterns (lines 12-13). For k-event pattern ($k \ge 3$), the mining process is similar to GTPM (lines 14-15).

\textbf{Complexity analysis of Approximate GTPM:} To compute NMI, $\mu_{\min}$, and $\mu_{\max}$, we only have to scan $\mathcal{D}_{\text{SYB}}$ once to calculate the probability for each single event, pair of events, and plus $\vartheta$ value. Thus, the cost of NMI, $\mu_{\min}$, and $\mu_{\max}$ computations is $\mid$$\mathcal{D}_{\text{SYB}}$$\mid$. On the other hand, the complexity of the exact GTPM at $HLH_1$ and $HLH_2$ are $O(m^2 i^2\mid$$\mathcal{D}_{\text{SEQ}}$$\mid$$^2) + O(m \cdot$$\mid$$\mathcal{D}_{\text{SEQ}}$$\mid)$ (Sections \ref{sec:1freq} and \ref{sec:2freq}). Thus, the approximate GTPM is significantly faster than the exact GTPM.

\section{Experimental Evaluation}\label{sec:experiment}\vspace{-0.02in}
We evaluate GTPM in two different settings: to mine rare temporal patterns, named as RTPM, and to mine frequent temporal patterns, named as FTPM. Note that for RTPM, all three constraints $\sigma_{\min}$, $\sigma_{\max}$ and $\delta$ are used, whereas for FTPM, only $\sigma_{\min}$ and $\delta$ are used. In each setting, the performance of both exact and approximate versions are assessed. We use real-world datasets from four application domains: smart energy, smart city, sign language, and health. 
Due to space limitations, we only present here the most important results, and discuss other findings in the electronic appendix.

\subsection{Experimental Setup}\vspace{-0.02in}
\hspace{0.17in}\textbf{Datasets:}
We use three \textit{smart energy} (SE) datasets, NIST \cite{nist}, UKDALE \cite{ukdale}, and DataPort \cite{pecan} that measure the energy consumption of electrical appliances in residential households. 
For the \textit{smart city} (SC), we use weather and vehicle collision data obtained from NYC Open Data Portal \cite{smartcity}. 
For \textit{sign language}, we use the American Sign Language (ASL) datasets \cite{neidle2018new} containing annotated video sequences of different ASL signs and gestures. For \textit{health}, we combine the \textit{influenza} (INF) dataset \cite{diseasedata} and weather data \cite{openweather} from Kawasaki, Japan. 
Table \ref{tbl:datasetCharacteristic} summarizes their characteristics. 

\begin{table}[!t]
	\centering
	\begin{minipage}{\columnwidth}
		\caption{Characteristics of the Datasets}
		\vspace{-0.1in}
		\resizebox{\columnwidth}{1cm}{
			\begin{tabular}{ |c|c|c|c|c|c|c|}
				\hline &&&&&&\\[0.01em]  \thead{} & {\bfseries NIST} & {\bfseries UKDALE} & {\bfseries DataPort} & {\bfseries SC} & {\bfseries ASL} & {\bfseries INF} \\ &&&&&&\\[0.01em]
				\hline 
				\centering \# sequences & 1460 & 1520 & 1460 & 1216 & 1908 &608\\
				\hline 
				\centering \# variables & 49 & 24 & 21 & 26 & 25 &25\\
				\hline 
				\centering \# distinct events & 98 & 48 & 42 & 130 & 173 &124\\
				\hline 
				\centering \# instances/seq. & 55 & 190 & 49 & 162 & 20 &48\\
				\hline
			\end{tabular}
		}
		\label{tbl:datasetCharacteristic}
	\end{minipage}
\end{table}

\textbf{Baseline methods:} Our exact RTPM version is referred to as E-RTPM, and the approximate one as A-RTPM. Since our work is the first that studies rare temporal pattern mining, there is not an exact baseline to compare against RTPM. However, we adapt the state-of-the-art method for frequent temporal pattern mining Z-Miner \cite{lee2020z} to find rare temporal patterns. The Adapted Rare Z-Miner is referred to as ARZ-Miner. Similarly, we denote the exact FTPM version as E-FTPM, and the approximate one as A-FTPM. We use $4$ baselines (detailed in Section \ref{sec:relatedwork}) to compare with our FTPM: Z-Miner \cite{lee2020z}, TPMiner \cite{tpminer}, IEMiner \cite{ieminer}, and H-DFS \cite{hdfs}. Since the exact versions (E-RTPM and E-FTPM) and the baselines provide the same exact solutions, we use the baselines only for quantitative evaluation. 

\textbf{Infrastructure:} We use a VM with 32 AMD EPYC cores (2GHz), 512 GB RAM, and 1 TB storage. 

\textbf{Parameters:} Table \ref{tbl:params} lists the parameters and their values used in our experiments.

\begin{table}[!t]
	\begin{minipage}{\columnwidth}
		\caption{Parameters and values}
		\vspace{-0.1in}
		\small
			\resizebox{\columnwidth}{3cm}{
				\begin{tabular}{ |m{2.0cm}|m{6.0cm}| }
					\hline {\bfseries  Params} & {\bfseries  Values} \\ 
					\hline 
					\multirow{2}{2.0 cm}{Minimum support $\sigma_{\min}$} & User-defined: \\ &   \phantom{i} $\sigma_{\min}$ $=$ 0.2\%, 0.4\%, 0.6\% 1\%, 3\% ... \\
					\hline 
					\multirow{2}{2.0 cm}{Maximum support $\sigma_{\max}$} & User-defined: \\ & \phantom{i} $\sigma_{\max}$ $=$ 2\%, 6\%, 10\%, 15\%, 20\%, ... \\
					\hline 
					\multirow{2}{2.0 cm}{Minimum confidence $\delta$} & User-defined: \\ & \phantom{i} $\delta$ $=$ 40\%, 50\%, 60\%, 70\%, 80\%, ... \\
					\hline
						\multirow{5}{2.0 cm}{Overlapping duration $t_{\text{ov}}$} & User-defined: \\
					& \phantom{i} $t_{\text{ov}}$ (hours) $=$ 0, 1, 2, 3 (NIST,\;\; UKDALE, \phantom{i} 
					DataPort, and SC) \\
					& \phantom{i} $t_{\text{ov}}$ (frames) $=$ 0, 150, 300, 450 (ASL) \\
					& \phantom{i} $t_{\text{ov}}$ (days) $=$ 0, 7, 10, 14 (INF) \\
					\hline
						\multirow{6}{2.0 cm}{Tolerance \\buffer $\epsilon$} & User-defined:\\
						& \phantom{i} $\epsilon$ (mins) $=$ 0, 1, 2, 3 (NIST, UKDALE, and \phantom{i} DataPort) \\
						& \phantom{i} $\epsilon$ (mins) $=$ 0, 5, 10, 15 (SC) \\
						& \phantom{i} $\epsilon$ (frames) $=$ 0, 30, 45, 60 (ASL) \\
						& \phantom{i} $\epsilon$ (days) $=$ 0, 1, 2, 3 (INF) \\					
					\hline
				\end{tabular}
		}
		\label{tbl:params}
	\end{minipage}
\end{table}

\subsection{Qualitative Evaluation}
\textit{Rare temporal patterns:} Table \ref{tbl:rarePatterns} shows several interesting rare temporal patterns extracted by RTPM. Patterns P1-P5 are from SC and P6-P8 are from INF. Analyzing these patterns can reveal some rare but interesting relations between temporal events. For example, P1-P5 show there exists an association between extreme weather conditions and high accident numbers, such as high pedestrian injury during a heavy snowing day, which is very important to act on even though it occurs rarely. 

\textit{Frequent temporal patterns:} Table \ref{tbl:interestingPatterns} lists some interesting frequent temporal patterns extracted by FTPM. 
Patterns P9-P15 are from SEs and P16-P18 are from ASL. Analyzing these patterns will reveal useful information about the domains. For example, P9-P15 show how the residents interact with electrical appliances in their houses. Specifically, P9 shows that a resident turns on the light upstairs in the early morning, and goes to the bathroom. Then, within a minute later, the microwave in the kitchen is turned on. This pattern occurs with minimum support of 20\%, reflecting a living habit of the residents. Moreover, P9 also implies that there might be more than one person living in the house, in which one resident is in the bathroom while the other is downstairs preparing breakfast. 

\begin{table*}[!t]	
	\caption{Summary of Interesting Rare Patterns}
	\vspace{-0.1in}
	\centering
	\resizebox{0.98\textwidth}{!}{
		\begin{tabular}{ |p{13.5cm}|c|c|c| }
			\hline  {\bfseries \;\;\;\;\;\;\;\;\;\;\;\;\;\;\;\;\;\;\;\;\;\;\;\;\;\;\;\;\;\;\;\;\;\;\;\;\;\;\;\;\;\;\;\;\;\;\;\;\;\;\;\;\;\;\;\;\;\;\;\;\;\;\;\;\;\;\;\;\;\;\;\;\;\;\;\;\;\;\;\;\;\;\;\;\;\;\;\;\;\;\;\;\;\;\;\; Patterns} & {\bfseries $\sigma_{\min}$ (\%)} & {\bfseries $\delta$ (\%)} & {\bfseries $\sigma_{\max}$ (\%)}\\	\hline  
			(P1) Heavy Rain $\succcurlyeq$ Unclear Visibility $\succcurlyeq$ Overcast Cloudiness $\rightarrow$ High Motorist Injury   &  5 & 30 & 9
			\\ \hline 
			(P2) Heavy Rain $\between$ Strong Wind $\rightarrow$ High Motorist Injury   &  2& 40 & 6 \\
			\hline
			(P3) Very Strong Wind $\rightarrow$ High Motorist Injury & 5 & 40 & 9 \\
			\hline
			(P4) Strong Wind $\between$ High Pedestrian Injury & 4 & 30 & 8\\
			\hline
			(P5) Extremely Unclear Visibility $\succcurlyeq$ High Snow $\succcurlyeq$ High Pedestrian Injury   &  3 & 45 & 7 
			\\ \specialrule{1.5pt}{1pt}{1pt}
			(P6) Frost Temperature $\between$ High Snow $\succcurlyeq$ High Influenza & 1 & 42 & 6
			\\ \hline 
			(P7) Low Temperature $\succcurlyeq$High Influenza & 1 & 42 & 6
			\\ \hline 
			(P8) Heavy Rain $\succcurlyeq$ High Influenza  & 3 & 35 & 8\\
			\hline
		\end{tabular} 
	}
	\label{tbl:rarePatterns}
\end{table*}
\begin{table*}[!t]	
	\caption{Summary of Interesting Frequent Patterns}
	\vspace{-0.1in}
	\centering
	\resizebox{0.98\linewidth}{!}{
		\begin{tabular}{ |p{16.5cm}|c|c| }
			\hline  {\bfseries \;\;\;\;\;\;\;\;\;\;\;\;\;\;\;\;\;\;\;\;\;\;\;\;\;\;\;\;\;\;\;\;\;\;\;\;\;\;\;\;\;\;\;\;\;\;\;\;\;\;\;\;\;\;\;\;\;\;\;\;\;\;\;\;\;\;\;\;\;\;\;\;\;\;\;\;\;\;\;\;\;\;\;\;\;\;\;\;\;\;\;\;\;\;\;\;\;\;\;\; Patterns} & {\bfseries $\sigma_{\min}$ (\%)} & {\bfseries $\delta$ (\%)}\\
			\hline  
			(P9) ([05:58, 08:24] First Floor Lights) $\succcurlyeq$ ([05:58, 06:59] Upstairs Bathroom Lights) $\succcurlyeq$ ([05:59, 06:06] Microwave) &  20 & 30  \\
			\hline 
			(P10) ([18:00, 18:30] Lights Dining Room) $\rightarrow$ ([18:31, 20:16] Children Room Plugs) $\between$ ([19:00, 22:31] Lights Living Room) & 20  &  20 \\
			\hline
			(P11) ([15:59, 16:05] Hallway Lights) $\rightarrow$ ([17:58, 18:29] Kitchen Lights $\succcurlyeq$ ([18:00, 18:18] Plug In Kitchen) $\succcurlyeq$ ([18:08, 18:15] Microwave) &  20 & 25 \\
			\hline
			(P12) ([06:02, 06:19] Kitchen Lights) $\rightarrow$ ([06:05, 06:12] Microwave) $\between$ ([06:09, 06:11] Kettle) & 20 & 35 \\
			\hline 
			(P13) ([16:45, 17:30] Washer) $\rightarrow$ ([17:40,18:55] Dryer) $\rightarrow$ ([19:05, 20:10] Dining Room Lights)  $\succcurlyeq$ ([19:10, 19:30] Cooktop)
			&  10 & 30
			\\ \hline 
			(P14) ([06:10, 07:00] Kitchen Lights) $\succcurlyeq$ ([06:10, 06:15] Kettle) $\rightarrow$ ([06:30, 06:40] Toaster) $\rightarrow$ ([06:45, 06:48] Microwave)   &  25 & 40
			\\ \hline 
			(P15) ([18:00, 18:25] Kitchen Lights) $\succcurlyeq$ ([18:00, 18:05] Kettle) $\rightarrow$ ([18:05, 18:10] Microwave) $\rightarrow$ ([19:35, 20:50] Washer)   &  20 & 40
			\\ \specialrule{1.5pt}{1pt}{1pt}
			(P16) [2.12 seconds] Negation  $\succcurlyeq$ [0.27 seconds] Lowered Eye-brows  & 10 & 10\\
			\hline 
			(P17) [2.04 seconds] Negation $\succcurlyeq$ [0.52 seconds] Rapid Shake-head & 10 & 10\\
			\hline
			(P18) [1.53 seconds] Wh-question $\succcurlyeq$ [0.36 seconds] Lowered Eye-brows   $\rightarrow$ [0.05 seconds] Blinking Eye-aperture  & 10 & 15\\
			\hline
		\end{tabular} 
	}
	\label{tbl:interestingPatterns}
\end{table*}

\subsection{Quantitative Evaluation of RTPM}
\begin{table*}[!t]
	\centering
	\begin{minipage}{.7\linewidth}
		\captionsetup{justification=centering, font=small}
		\caption{Pruned Time Series and Events from A-RTPM}
		\vspace{-0.1in}
		\resizebox{\columnwidth}{1.3cm}{
			\begin{tabular}{|c|m{1.1cm}|m{1.1cm}|m{1.1cm}|c|c|c|c|c|c|}
				\hline
				\multirow{4}{*}{\bfseries \# Attr.} & \multicolumn{9}{c|}{\bfseries $\sigma_{\min}$ (\%) - $\delta$ (\%) - $\sigma_{\max}$ (\%)} 
				\\ \cline{2-10}   
				& \multicolumn{3}{c|}{\bfseries NIST} & \multicolumn{6}{c|}{\bfseries SC}
				\\  \cline{2-10}  
				& \multicolumn{3}{c|}{\bfseries Pruned Time Series / Events (\%)} & \multicolumn{3}{c|}{\bfseries Pruned Time Series (\%)}  & \multicolumn{3}{c|}{\bfseries Pruned Events (\%)}    
				\\  \cline{2-10} 
				& {\bfseries 6-80-20} & {\bfseries \;\;3-70-15} & {\bfseries 1-60-10}  & {\bfseries 6-80-20} & {\bfseries 3-70-15} & {\bfseries 1-60-10}  & {\bfseries 6-80-20} & {\bfseries 3-70-15} & {\bfseries 1-60-10} \\
				\hline
				200 & \;\;59.50 & \;\;39.50 & \;\;22.50 & 48.50 & 30.50 & 15.50 & 39.10 & 25.10 & 11.90 \\  \hline
				
				400 & \;\;58.50 & \;\;38.25 & \;\;21.25 & 45.75 & 29.75 & 14.75 & 37.55 & 24.30 & 11.45 \\  \hline
				
				600 & \;\;56.50 & \;\;36.17 & \;\;19.83 & 43.17 & 27.17 & 14.33 & 36.43 & 23.03 & 10.57 \\  \hline
				
				800 & \;\;51.63  & \;\;35.88 & \;\;19.63 & 42.38 & 23.88 & 14.25 & 33.55 & 21.28 & 10.30 \\  \hline
				
				1000 & \;\;49.70 & \;\;34.10 & \;\;19.40 & 41.30 & 22.70 & 13.80 & 32.94 & 20.14 & 9.96 \\  \hline
			\end{tabular}
		}
		\label{tbl:rarePrunedAttributesEventsPercent}
	\end{minipage}
	\begin{minipage}{0.29\linewidth}
		\caption{RTPM Accuracy (\%)}
		\vspace{-0.1in}
		\resizebox{\linewidth}{1.3cm}{
			\small
			\begin{tabular}{|c|c|c|c|c|c|c|}
				\hline 
				\multirow{3}{*}{$\sigma_{\max}$ (\%)} & \multicolumn{6}{c|}{\bfseries $\sigma_{\min}$ (\%) - $\delta$ (\%)} \\  
				\cline{2-7}  
				& \multicolumn{3}{c|}{\bfseries NIST} & \multicolumn{3}{c|}{\bfseries SC}
				\\  \cline{2-7}
				& {\bfseries 1-60} & {\bfseries 3-70} & {\bfseries 6-80}  & {\bfseries 1-60} & {\bfseries 3-70} & {\bfseries 6-80}  \\
				\hline
				10 & 93  & 96   & 100  & 91  & 93  & 100  \\  \hline				
				15 & 86  & 92   & 95  & 86   & 91  & 100  \\  \hline					
				20 & 84  & 92   & 92  & 83 & 87  & 90  \\  \hline								
			\end{tabular}
		}			
		\label{tbl:rareAccuracyReal}
	\end{minipage}
\end{table*}
\begin{figure*}[!t]
	\hspace{2.59in}
	\ref{legendcomparisonrare}
	\clearpage
	\vspace{-0.15in}
	\begin{minipage}[t]{1\columnwidth} 
		\centering
		\begin{subfigure}{0.32\columnwidth}
			\centering
			\resizebox{\linewidth}{!}{
				\begin{tikzpicture}[scale=0.2]
					\begin{axis}[
						compat=newest,
						xlabel={$\sigma_{min}$ (\%)},
						ylabel={Runtime (sec)}, 
						label style={font=\Huge},
						ticklabel style = {font=\Huge},
						xmin=1, xmax=5,
						xtick={1,2,3,4,5},
						xticklabels = {1,3,6,9,12},
						ymin=10, ymax=10000,
						legend columns=-1,
						legend entries = {A-RTPM, E-RTPM, ARZ-Miner},
						legend style={nodes={scale=0.5,  transform shape}, font=\Large},
						legend to name={legendcomparisonrare},
						ymode=log,
						log basis y={10},
						ymajorgrids=true,
						grid style=dashed,
						line width=1.75pt
						]						
						\addplot[
						color=blue,
						mark=asterisk,
						mark size=4pt,
						]
						coordinates {
							(1,150)(2,125)(3,60)(4,46)(5,34)
						};
						\addplot[
						color=teal,
						mark=diamond,
						mark size=4pt,
						]
						coordinates {
							(1,340)(2,187)(3,120)(4,93)(5,77)
						};
						\addplot[
						color=red,
						mark=pentagon*,
						mark size=4pt,
						]
						coordinates {
							(1,1502)(2,926)(3,734)(4,616)(5,452)
						};
					\end{axis}
				\end{tikzpicture}
			}
			\captionsetup{justification=centering, font=scriptsize}
			\caption{\scriptsize Varying $\sigma_{min}$}
		\end{subfigure}
		\begin{subfigure}{0.32\columnwidth}
			\centering
			\resizebox{\linewidth}{!}{
				\begin{tikzpicture}[scale=0.2]
					\begin{axis}[
						compat=newest,
						xlabel={$\delta$ (\%)},
						ylabel={Runtime (sec)}, 
						label style={font=\Huge},
						ticklabel style = {font=\Huge},
						xmin=60, xmax=100,
						ymin=0.1, ymax=10000,
						xtick={60,70,80,90,100},
						legend columns=-1,
						legend entries = {A-RTPM, E-RTPM, ARZ-Miner},
						legend style={nodes={scale=0.5,  transform shape}, font=\Large},
						legend to name={legendcomparisonrare},
						ymode=log,
						log basis y={10},
						ymajorgrids=true,
						grid style=dashed,
						line width=1.75pt
						]
						\addplot[
						color=blue,
						mark=asterisk,
						mark size=4pt,
						]
						coordinates {
							(60,150)(70,135)(80,81)(90,20)(100,0.9)
						};
						\addplot[
						color=teal,
						mark=diamond,
						mark size=4pt,
						]
						coordinates {
							(60,340)(70,267)(80,164)(90,50)(100,1.5)
						};
						\addplot[
						color=red,
						mark=pentagon*,
						mark size=4pt,
						]
						coordinates {
							(60,1502)(70,1225)(80,909)(90,461)(100,10)
						};
					\end{axis}
				\end{tikzpicture}
			}
		\captionsetup{justification=centering, font=scriptsize}
		\caption{\scriptsize Varying $\delta$}
		\end{subfigure}
		\begin{subfigure}{0.32\columnwidth}
			\centering
			\resizebox{\linewidth}{!}{
				\begin{tikzpicture}[scale=0.2]
					\begin{axis}[
						compat=newest,
						xlabel={$\sigma_{max}$ (\%)},
						ylabel={Runtime (sec)}, 
						label style={font=\Huge},
						ticklabel style = {font=\Huge},
						xmin=1, xmax=5,
						ymin=10, ymax=10000,
						xtick={1,2,3,4,5},
						xticklabels = {15,20,25,30,35},
						legend columns=-1,
						legend entries = {A-RTPM, E-RTPM, ARZ-Miner},
						legend style={nodes={scale=0.5,  transform shape}, font=\Large},
						legend to name={legendcomparisonrare},
						ymode=log,
						log basis y={10},
						ymajorgrids=true,
						grid style=dashed,
						line width=1.75pt
						]
						\addplot[
						color=blue,
						mark=asterisk,
						mark size=4pt,
						]
						coordinates {
							(1,106)(2,125)(3,150)(4,174)(5,187)
						};
						\addplot[
						color=teal,
						mark=diamond,
						mark size=4pt,
						]
						coordinates {
							(1,194)(2,218)(3,340)(4,382)(5,401)
						};
						\addplot[
						color=red,
						mark=pentagon*,
						mark size=4pt,
						]
						coordinates {
							(1,926)(2,1024)(3,1902)(4,2802)(5,3967)
						};
					\end{axis}
				\end{tikzpicture}
			}
			\captionsetup{justification=centering, font=scriptsize}
			\caption{\scriptsize Varying $\sigma_{max}$}
		\end{subfigure}
		\vspace{-0.1in}
		\captionsetup{justification=centering, font=small}
		\caption{RTPM-Runtime Comparison on {\footnotesize NIST} (real-world)}
		\label{fig:rareruntimebaselineNIST}
	\end{minipage}
	\hspace{0.2in}  
	\begin{minipage}[t]{1\columnwidth} 
		\centering
		\begin{subfigure}{0.32\columnwidth}
			\centering
			\resizebox{\linewidth}{!}{
				\begin{tikzpicture}[scale=0.2]
					\begin{axis}[
						compat=newest,
						xlabel={$\sigma_{min}$ (\%)},
						ylabel={Runtime (sec)}, 
						label style={font=\Huge},
						ticklabel style = {font=\Huge},
						xmin=1, xmax=5,
						xtick={1,2,3,4,5},
						xticklabels = {1,3,6,9,12},
						ymin=0.1, ymax=100,
						legend columns=-1,
						legend entries = {A-RTPM, E-RTPM, ARZ-Miner},
						legend style={nodes={scale=0.5,  transform shape}, font=\Large},
						legend to name={legendcomparisonrare},
						ymode=log,
						log basis y={10},
						ymajorgrids=true,
						grid style=dashed,
						line width=1.75pt
						]
						\addplot[
						color=blue,
						mark=asterisk,
						mark size=4pt,
						]
						coordinates {
							(1,3)(2,1.8)(3,1.5)(4,1.2)(5,0.5)
						};
						\addplot[
						color=teal,
						mark=diamond,
						mark size=4pt,
						]
						coordinates {
							(1,7)(2,5)(3,4.2)(4,2.5)(5,1.2)
						};
						\addplot[
						color=red,
						mark=pentagon*,
						mark size=4pt,
						]
						coordinates {
							(1,28)(2,23)(3,20)(4,15)(5,7)
						};
					\end{axis}
				\end{tikzpicture}
			}
			\captionsetup{justification=centering, font=scriptsize}
			\caption{\scriptsize Varying $\sigma_{min}$}
		\end{subfigure}
		\begin{subfigure}{0.32\columnwidth}
			\centering
			\resizebox{\linewidth}{!}{
				\begin{tikzpicture}[scale=0.2]
					\begin{axis}[
						compat=newest,
						xlabel={$\delta$ (\%)},
						ylabel={Runtime (sec)}, 
						label style={font=\Huge},
						ticklabel style = {font=\Huge},
						xmin=60, xmax=100,
						ymin=0.1, ymax=100,
						xtick={60,70,80,90,100},
						legend columns=-1,
						legend entries = {A-RTPM, E-RTPM, ARZ-Miner},
						legend style={nodes={scale=0.5,  transform shape}, font=\Large},
						legend to name={legendcomparisonrare},
						ymode=log,
						log basis y={10},
						ymajorgrids=true,
						grid style=dashed,
						line width=1.75pt
						]
						\addplot[
						color=blue,
						mark=asterisk,
						mark size=4pt,
						]
						coordinates {
							(60,3)(70,2.4)(80,2)(90,1.5)(100,1)
						};
						\addplot[
						color=teal,
						mark=diamond,
						mark size=4pt,
						]
						coordinates {
							(60,7)(70,6)(80,5.5)(90,4)(100,2)
						};
						\addplot[
						color=red,
						mark=pentagon*,
						mark size=4pt,
						]
						coordinates {
							(60,28)(70,25)(80,24)(90,18)(100,12)
						};
					\end{axis}
				\end{tikzpicture}
			}
			\captionsetup{justification=centering, font=scriptsize}
			\caption{\scriptsize Varying $\delta$}
		\end{subfigure}
		\begin{subfigure}{0.32\columnwidth}
			\centering
			\resizebox{\linewidth}{!}{
				\begin{tikzpicture}[scale=0.2]
					\begin{axis}[
						compat=newest,
						xlabel={$\sigma_{max}$ (\%)},
						ylabel={Runtime (sec)}, 
						label style={font=\Huge},
						ticklabel style = {font=\Huge},
						xmin=1, xmax=5,
						ymin=0.1, ymax=100,
						xtick={1,2,3,4,5},
						xticklabels = {15,20,25,30,35},
						legend columns=-1,
						legend entries = {A-RTPM, E-RTPM, ARZ-Miner},
						legend style={nodes={scale=0.5,  transform shape}, font=\Large},
						legend to name={legendcomparisonrare},
						ymode=log,
						log basis y={10},
						ymajorgrids=true,
						grid style=dashed,
						line width=1.75pt
						]
						\addplot[
						color=blue,
						mark=asterisk,
						mark size=4pt,
						]
						coordinates {
							(1,2)(2,2.5)(3,3)(4,7)(5,8.5)
						};
						\addplot[
						color=teal,
						mark=diamond,
						mark size=4pt,
						]
						coordinates {
							(1,5)(2,6.5)(3,7)(4,15)(5,17)
						};
						\addplot[
						color=red,
						mark=pentagon*,
						mark size=4pt,
						]
						coordinates {
							(1,21)(2,25)(3,28)(4,40)(5,46)
						};
					\end{axis}
				\end{tikzpicture}
			}
			\captionsetup{justification=centering, font=scriptsize}
			\caption{\scriptsize Varying $\sigma_{max}$}
		\end{subfigure}
		\vspace{-0.1in}
		\captionsetup{justification=centering, font=small}
		\caption{RTPM-Runtime Comparison on SC (real-world)}
		\label{fig:rareruntimebaselineSC}
	\end{minipage}
	\vspace{-0.02in}  
\end{figure*}  

\begin{figure*}[!t]
	\begin{minipage}[t]{1\columnwidth} 
		\centering
		\begin{subfigure}{0.32\columnwidth}
			\centering
			\resizebox{\linewidth}{!}{
				\begin{tikzpicture}[scale=0.2]
					\begin{axis}[
						compat=newest,
						xlabel={$\sigma_{min}$ (\%)},
						ylabel={Memory Usage (MB)}, 
						label style={font=\Huge},
						ticklabel style = {font=\Huge},
						xmin=1, xmax=5,
						xtick={1,2,3,4,5},
						xticklabels = {1,3,6,9,12},
						ymin=100, ymax=50000,
						legend columns=-1,
						legend entries = {A-RTPM, E-RTPM, ARZ-Miner},
						legend style={nodes={scale=0.5,  transform shape}, font=\Large},
						legend to name={legendcomparisonrare},
						ymode=log,
						log basis y={10},
						ymajorgrids=true,
						grid style=dashed,
						line width=1.75pt
						]
						\addplot[
						color=blue,
						mark=asterisk,
						mark size=4pt,
						]
						coordinates {
							(1,426)(2,356)(3,279)(4,224)(5,204)
						};
						\addplot[
						color=teal,
						mark=diamond,
						mark size=4pt,
						]
						coordinates {
							(1,747)(2,581)(3,507)(4,364)(5,307)
						};
						\addplot[
						color=red,
						mark=pentagon*,
						mark size=4pt,
						]
						coordinates {
							(1,27206)(2,12620)(3,7684)(4,2708)(5,1706)
						};
					\end{axis}
				\end{tikzpicture}
			}
			\captionsetup{justification=centering, font=scriptsize}
			\caption{\scriptsize Varying $\sigma_{min}$}
		\end{subfigure}
		\begin{subfigure}{0.32\columnwidth}
			\centering
			\resizebox{\linewidth}{!}{
				\begin{tikzpicture}[scale=0.2]
					\begin{axis}[
						compat=newest,
						xlabel={$\delta$ (\%)},
						ylabel={Memory Usage (MB)}, 
						label style={font=\Huge},
						ticklabel style = {font=\Huge},
						xmin=60, xmax=100,
						ymin=10, ymax=50000,
						xtick={60,70,80,90,100},
						legend columns=-1,
						legend entries = {A-RTPM, E-RTPM, ARZ-Miner},
						legend style={nodes={scale=0.5,  transform shape}, font=\Large},
						legend to name={legendcomparisonrare},
						ymode=log,
						log basis y={10},
						ymajorgrids=true,
						grid style=dashed,
						line width=1.75pt
						]
						\addplot[
						color=blue,
						mark=asterisk,
						mark size=4pt,
						]
						coordinates {
							(60,426)(70,402)(80,352)(90,130)(100,95)
						};
						\addplot[
						color=teal,
						mark=diamond,
						mark size=4pt,
						]
						coordinates {
							(60,747)(70,614)(80,558)(90,376)(100,184)
						};
						\addplot[
						color=red,
						mark=pentagon*,
						mark size=4pt,
						]
						coordinates {
							(60,27206)(70,16247)(80,9006)(90,3216)(100,1026)
						};
					\end{axis}
				\end{tikzpicture}
			}
			\captionsetup{justification=centering, font=scriptsize}
			\caption{\scriptsize Varying $\delta$}
		\end{subfigure}
		\begin{subfigure}{0.32\columnwidth}
			\centering
			\resizebox{\linewidth}{!}{
				\begin{tikzpicture}[scale=0.2]
					\begin{axis}[
						compat=newest,
						xlabel={$\sigma_{max}$ (\%)},
						ylabel={Memory Usage (MB)}, 
						label style={font=\Huge},
						ticklabel style = {font=\Huge},
						xmin=1, xmax=5,
						ymin=100, ymax=50000,
						xtick={1,2,3,4,5},
						xticklabels = {15,20,25,30,35},
						legend columns=-1,
						legend entries = {A-RTPM, E-RTPM, ARZ-Miner},
						legend style={nodes={scale=0.5,  transform shape}, font=\Large},
						legend to name={legendcomparisonrare},
						ymode=log,
						log basis y={10},
						ymajorgrids=true,
						grid style=dashed,
						line width=1.75pt
						]
						\addplot[
						color=blue,
						mark=asterisk,
						mark size=4pt,
						]
						coordinates {
							(1,219)(2,338)(3,426)(4,506)(5,589)
						};
						\addplot[
						color=teal,
						mark=diamond,
						mark size=4pt,
						]
						coordinates {
							(1,452)(2,506)(3,747)(4,852)(5,957)
						};
						\addplot[
						color=red,
						mark=pentagon*,
						mark size=4pt,
						]
						coordinates {
							(1,4247)(2,6004)(3,27206)(4,39851)(5,48028)
						};
					\end{axis}
				\end{tikzpicture}
			}
			\captionsetup{justification=centering, font=scriptsize}
			\caption{\scriptsize Varying $\sigma_{max}$}
		\end{subfigure}
		\vspace{-0.1in}
		\captionsetup{justification=centering, font=small}
		\caption{RTPM-Memory Usage Comparison on {\footnotesize NIST (real-world)}}
		\label{fig:rarememorybaselineNIST}
	\end{minipage}
	\hspace{0.2in}  
	\begin{minipage}[t]{1\columnwidth} 
		\centering
		\begin{subfigure}{0.32\columnwidth}
			\centering
			\resizebox{\linewidth}{!}{
				\begin{tikzpicture}[scale=0.2]
					\begin{axis}[
						compat=newest,
						xlabel={$\sigma_{min}$ (\%)},
						ylabel={Memory Usage (MB)}, 
						label style={font=\Huge},
						ticklabel style = {font=\Huge},
						xmin=1, xmax=5,
						xtick={1,2,3,4,5},
						xticklabels = {1,3,6,9,12},
						ymin=10, ymax=1000,
						legend columns=-1,
						legend entries = {A-RTPM, E-RTPM, ARZ-Miner},
						legend style={nodes={scale=0.5,  transform shape}, font=\Large},
						legend to name={legendcomparisonrare},
						ymode=log,
						log basis y={10},
						ymajorgrids=true,
						grid style=dashed,
						line width=1.75pt
						]
						\addplot[
						color=blue,
						mark=asterisk,
						mark size=4pt,
						]
						coordinates {
							(1,70)(2,50)(3,34)(4,27)(5,20)
						};
						\addplot[
						color=teal,
						mark=diamond,
						mark size=4pt,
						]
						coordinates {
							(1,110)(2,90)(3,75)(4,45)(5,35)
						};
						\addplot[
						color=red,
						mark=pentagon*,
						mark size=4pt,
						]
						coordinates {
							(1,428)(2,390)(3,302)(4,256)(5,177)
						};
					\end{axis}
				\end{tikzpicture}
			}
			\captionsetup{justification=centering, font=scriptsize}
			\caption{\scriptsize Varying $\sigma_{min}$}
		\end{subfigure}
		\begin{subfigure}{0.32\columnwidth}
			\centering
			\resizebox{\linewidth}{!}{
				\begin{tikzpicture}[scale=0.2]
					\begin{axis}[
						compat=newest,
						xlabel={$\delta$ (\%)},
						ylabel={Memory Usage (MB)}, 
						label style={font=\Huge},
						ticklabel style = {font=\Huge},
						xmin=60, xmax=100,
						ymin=10, ymax=1000,
						xtick={60,70,80,90,100},
						legend columns=-1,
						legend entries = {A-RTPM, E-RTPM, ARZ-Miner},
						legend style={nodes={scale=0.5,  transform shape}, font=\Large},
						legend to name={legendcomparisonrare},
						ymode=log,
						log basis y={10},
						ymajorgrids=true,
						grid style=dashed,
						line width=1.75pt
						]
						\addplot[
						color=blue,
						mark=asterisk,
						mark size=4pt,
						]
						coordinates {
							(60,70)(70,55)(80,42)(90,33)(100,28)
						};
						\addplot[
						color=teal,
						mark=diamond,
						mark size=4pt,
						]
						coordinates {
							(60,110)(70,92)(80,82)(90,64)(100,55)
						};
						\addplot[
						color=red,
						mark=pentagon*,
						mark size=4pt,
						]
						coordinates {
							(60,428)(70,406)(80,382)(90,361)(100,304)
						};
					\end{axis}
				\end{tikzpicture}
			}
			\captionsetup{justification=centering, font=scriptsize}
			\caption{\scriptsize Varying $\delta$}
		\end{subfigure}
		\begin{subfigure}{0.32\columnwidth}
			\centering
			\resizebox{\linewidth}{!}{
				\begin{tikzpicture}[scale=0.2]
					\begin{axis}[
						compat=newest,
						xlabel={$\sigma_{max}$ (\%)},
						ylabel={Memory Usage (MB)}, 
						label style={font=\Huge},
						ticklabel style = {font=\Huge},
						xmin=1, xmax=5,
						ymin=10, ymax=1000,
						xtick={1,2,3,4,5},
						xticklabels = {15,20,25,30,35},
						legend columns=-1,
						legend entries = {A-RTPM, E-RTPM, ARZ-Miner},
						legend style={nodes={scale=0.5,  transform shape}, font=\Large},
						legend to name={legendcomparisonrare},
						ymode=log,
						log basis y={10},
						ymajorgrids=true,
						grid style=dashed,
						line width=1.75pt
						]
						\addplot[
						color=blue,
						mark=asterisk,
						mark size=4pt,
						]
						coordinates {
							(1,35)(2,50)(3,70)(4,80)(5,96)
						};
						\addplot[
						color=teal,
						mark=diamond,
						mark size=4pt,
						]
						coordinates {
							(1,70)(2,85)(3,110)(4,130)(5,150)
						};
						\addplot[
						color=red,
						mark=pentagon*,
						mark size=4pt,
						]
						coordinates {
							(1,302)(2,358)(3,428)(4,542)(5,598)
						};
					\end{axis}
				\end{tikzpicture}
			}
			\captionsetup{justification=centering, font=scriptsize}
			\caption{\scriptsize Varying $\sigma_{max}$}
		\end{subfigure}
		\vspace{-0.1in}
		\captionsetup{justification=centering, font=small}
		\caption{RTPM-Memory Usage Comparison on SC (real-world)}
		\label{fig:rarememorybaselineSC}
	\end{minipage}
	\vspace{-0.02in}  
\end{figure*}  
\input{graph/rare_scalabilitySequenceAttribute}
\begin{figure*}[!t]
	\vspace{-0.1in}
	\begin{minipage}[t]{1\columnwidth} 
		\centering
		\begin{subfigure}{0.32\columnwidth}
			\centering
			\resizebox{\linewidth}{!}{
				\begin{tikzpicture}[scale=0.6]
					\begin{axis}[
						compat=newest,
						xlabel={$\sigma_{min}$ (\%)},
						ylabel={Runtime (sec)}, 
						label style={font=\Huge},
						ticklabel style = {font=\huge},
						xticklabel style = {xshift=2mm}, 
						ymin=1, ymax=10000,
						xmin=1, xmax=5,
						xtick={1,2,3,4,5},
						xticklabels = {1,3,6,9,12},
						legend columns=-1,
						legend entries = {(NoPrune)-E-RTPM, (Apriori)-E-RTPM, (Trans)-E-RTPM, (All)-E-RTPM},
						legend style={nodes={scale=0.55,   transform shape}, font=\small},
						legend to name={legendpruning},
						ymode=log,
						log basis y={10},
						ymajorgrids=true,
						grid style=dashed,
						line width=1.75pt
						]
						\addplot[
						color=blue,
						mark=pentagon,
						mark size=4pt,
						] 	
						coordinates {
							(1,1724)(2,1405)(3,1000)(4,700)(5,525)
						};
						
						\addplot[
						color=black,
						mark=triangle*,
						mark size=4pt,
						]	
						coordinates {
							(1,299)(2,250)(3,195)(4,58)(5,30)
						};
						
						\addplot[
						color=teal,
						mark=*,
						mark size=4pt,
						] 
						coordinates {
							(1,195)(2,180)(3,115)(4,39)(5,20)
						};
						
						\addplot[
						color=red,
						mark=diamond,
						mark size=4pt,
						] 
						coordinates {
							(1,160)(2,117)(3,75)(4,20)(5,8)
						};
					\end{axis}
				\end{tikzpicture}
			}
			\captionsetup{justification=centering, font=scriptsize}
			\caption{Varying $\sigma_{min}$}
		\end{subfigure}
		\begin{subfigure}{0.32\columnwidth}
			\centering
			\resizebox{\linewidth}{!}{
				\begin{tikzpicture}[scale=0.6]
					\begin{axis}[
						compat=newest,
						xlabel={$\delta$ (\%)},
						ylabel={Runtime (sec)}, 
						label style={font=\Huge},
						ticklabel style = {font=\huge},
						xticklabel style = {xshift=2mm}, 
						xmin=60, xmax=100,
						ymin=1, ymax=10000,
						xtick={60,70,80,90,100},
						legend columns=-1,
						legend entries = {(NoPrune)-E-RTPM, (Apriori)-E-RTPM, (Trans)-E-RTPM, (All)-E-RTPM},
						legend style={nodes={scale=0.55,  transform shape}, font=\small},
						legend to name={legendpruning},
						ymode=log,
						log basis y={10},
						ymajorgrids=true,
						grid style=dashed,
						line width=1.75pt
						]
						\addplot[
						color=blue,
						mark=pentagon,
						mark size=4pt,
						] 	
						coordinates {
							(60,2054)(70,1800)(80,1500)(90,1005)(100,115)
						};
						
						\addplot[
						color=black,
						mark=triangle*,
						mark size=4pt,
						]	
						coordinates {
							(60,372)(70,300)(80,250)(90,70)(100,7)
						};
						
						\addplot[
						color=teal,
						mark=*,
						mark size=4pt,
						] 
						coordinates {
							(60,250)(70,200)(80,130)(90,40)(100,5)
						};
						
						\addplot[
						color=red,
						mark=diamond,
						mark size=4pt,
						] 
						coordinates {
							(60,190)(70,157)(80,100)(90,30)(100,4)
						};
					\end{axis}
				\end{tikzpicture}
			}
			\captionsetup{justification=centering, font=scriptsize}
			\caption{Varying $\delta$}
		\end{subfigure}
		\begin{subfigure}{0.32\columnwidth}
			\centering
			\resizebox{\linewidth}{!}{
				\begin{tikzpicture}[scale=0.6]
					\begin{axis}[
						compat=newest,
						xlabel={$\sigma_{max}$ (\%)},
						ylabel={Runtime (sec)}, 
						label style={font=\Huge},
						ticklabel style = {font=\huge},
						xticklabel style = {xshift=2mm}, 
						xmin=1, xmax=5,
						ymin=100, ymax=10000,
						xtick={1,2,3,4,5},
						xticklabels = {15,20,25,30,35},
						legend columns=-1,
						legend entries = {(NoPrune)-E-RTPM, (Apriori)-E-RTPM, (Trans)-E-RTPM, (All)-E-RTPM},
						legend style={nodes={scale=0.55,  transform shape}, font=\small},
						legend to name={legendpruning},
						ymode=log,
						log basis y={10},
						ymajorgrids=true,
						grid style=dashed,
						line width=1.75pt
						]
						\addplot[
						color=blue,
						mark=pentagon,
						mark size=4pt,
						] 
						coordinates {
							(1,2969)(2,3416)(3,4958)(4,5561)(5,7169)
						};
						
						\addplot[
						color=black,
						mark=triangle*,
						mark size=4pt,
						] 
						coordinates {
							(1,400)(2,526)(3,653)(4,751)(5,958)
						};
						
						\addplot[
						color=teal,
						mark=*,
						mark size=4pt,
						]	
						coordinates {
							(1,250)(2,301)(3,423)(4,520)(5,622)
						};
						
						\addplot[
						color=red,
						mark=diamond,
						mark size=4pt,
						] 
						coordinates {
							(1,120)(2,188)(3,200)(4,300)(5,401)
						};
					\end{axis}
				\end{tikzpicture}
			}
			\captionsetup{justification=centering, font=scriptsize}
			\caption{Varying $\sigma_{max}$}
		\end{subfigure}
		\vspace{-0.1in}
		\ref{legendpruning}
		\caption{Runtimes of E-RTPM on NIST (real-world)} 
		\label{fig:rarepruningExactRTPM1}
	\end{minipage}%
	\hspace{0.2in}  
	\begin{minipage}[t]{1\columnwidth} 
		\centering
		\begin{subfigure}{0.32\columnwidth}
			\centering
			\resizebox{\linewidth}{!}{
				\begin{tikzpicture}[scale=0.6]
					\begin{axis}[
						compat=newest,
						xlabel={$\sigma_{min}$ (\%)},
						ylabel={Runtime (sec)}, 
						label style={font=\Huge},
						ticklabel style = {font=\huge},
						xticklabel style = {xshift=2mm}, 
						ymin=1, ymax=100,
						xmin=1, xmax=5,
						xtick={1,2,3,4,5},
						xticklabels = {1,3,6,9,12},
						legend columns=-1,
						legend entries = {(NoPrune)-E-RTPM, (Apriori)-E-RTPM, (Trans)-E-RTPM, (All)-E-RTPM},
						legend style={nodes={scale=0.55,  transform shape}, font=\small},
						legend to name={legendpruning},
						ymode=log,
						log basis y={10},
						ymajorgrids=true,
						grid style=dashed,
						line width=1.75pt
						]
						\addplot[
						color=blue,
						mark=pentagon,
						mark size=4pt,
						] 	
						coordinates {
							(1,85)(2,72)(3,60)(4,50)(5,21)
						};
						
						\addplot[
						color=black,
						mark=triangle*,
						mark size=4pt,
						]	
						coordinates {
							(1,27)(2,22)(3,15)(4,8)(5,4)
						};
						
						\addplot[
						color=teal,
						mark=*,
						mark size=4pt,
						] 
						coordinates {
							(1,16)(2,12)(3,10)(4,5)(5,2.5)
						};
						
						\addplot[
						color=red,
						mark=diamond,
						mark size=4pt,
						] 
						coordinates {
							(1,7)(2,5)(3,4.2)(4,2.5)(5,1.2)
						};
					\end{axis}
				\end{tikzpicture}
			}
			\captionsetup{justification=centering, font=scriptsize}
			\caption{Varying $\sigma_{min}$}
		\end{subfigure}
		\begin{subfigure}{0.32\columnwidth}
			\centering
			\resizebox{\linewidth}{!}{
				\begin{tikzpicture}[scale=0.6]
					\begin{axis}[
						compat=newest,
						xlabel={$\delta$ (\%)},
						ylabel={Runtime (sec)}, 
						label style={font=\Huge},
						ticklabel style = {font=\huge},
						xticklabel style = {xshift=2mm},
						xmin=60, xmax=100,
						ymin=1, ymax=100,
						xtick={60,70,80,90,100},
						legend columns=-1,
						legend entries = {(NoPrune)-E-RTPM, (Apriori)-E-RTPM, (Trans)-E-RTPM, (All)-E-RTPM},
						legend style={nodes={scale=0.55,  transform shape}, font=\small},
						legend to name={legendpruning},
						ymode=log,
						log basis y={10},
						ymajorgrids=true,
						grid style=dashed,
						line width=1.75pt
						]
						\addplot[
						color=blue,
						mark=pentagon,
						mark size=4pt,
						] 	
						coordinates {
							(60,95)(70,80)(80,75)(90,65)(100,36)
						};
						
						\addplot[
						color=black,
						mark=triangle*,
						mark size=4pt,
						]	
						coordinates {
							(60,20)(70,18)(80,14)(90,7)(100,5)
						};
						
						\addplot[
						color=teal,
						mark=*,
						mark size=4pt,
						] 
						coordinates {
							(60,17)(70,14)(80,11)(90,6)(100,3)
						};
						
						\addplot[
						color=red,
						mark=diamond,
						mark size=4pt,
						] 
						coordinates {
							(60,7)(70,6)(80,5)(90,4)(100,2)
						};
					\end{axis}
				\end{tikzpicture}
			}
			\captionsetup{justification=centering, font=scriptsize}
			\caption{Varying $\delta$}
		\end{subfigure}
		\begin{subfigure}{0.32\columnwidth}
			\centering
			\resizebox{\linewidth}{!}{
				\begin{tikzpicture}[scale=0.6]
					\begin{axis}[
						compat=newest,
						xlabel={$\sigma_{max}$ (\%)},
						ylabel={Runtime (sec)}, 
						label style={font=\Huge},
						ticklabel style = {font=\huge},
						xticklabel style = {xshift=2mm}, 
						xmin=1, xmax=5,
						ymin=1, ymax=1000,
						xtick={1,2,3,4,5},
						xticklabels = {15,20,25,30,35},
						legend columns=-1,
						legend entries = {(NoPrune)-E-RTPM, (Apriori)-E-RTPM, (Trans)-E-RTPM, (All)-E-RTPM},
						legend style={nodes={scale=0.55,  transform shape}, font=\small},
						legend to name={legendpruning},
						ymode=log,
						log basis y={10},
						ymajorgrids=true,
						grid style=dashed,
						line width=1.75pt
						]
						\addplot[
						color=blue,
						mark=pentagon,
						mark size=4pt,
						] 
						coordinates {
							(1,200)(2,225)(3,327)(4,384)(5,527)
						};
						
						\addplot[
						color=black,
						mark=triangle*,
						mark size=4pt,
						] 
						coordinates {
							(1,27)(2,32)(3,47)(4,55)(5,80)
						};
						
						\addplot[
						color=teal,
						mark=*,
						mark size=4pt,
						]	
						coordinates {
							(1,14)(2,18)(3,20)(4,30)(5,35)
						};
						
						\addplot[
						color=red,
						mark=diamond,
						mark size=4pt,
						] 
						coordinates {
							(1,5)(2,6.5)(3,7)(4,15)(5,17)
						};
					\end{axis}
				\end{tikzpicture}
			}
			\captionsetup{justification=centering, font=scriptsize}
			\caption{Varying $\sigma_{max}$}
		\end{subfigure}
		\vspace{-0.1in}
		\ref{legendpruning}
		\caption{Runtimes of E-RTPM on SC (real-world)} 
		\label{fig:rarepruningExactRTPM2}
	\end{minipage}      
	\vspace{-0.15in}  
\end{figure*}

\subsubsection{RTPM: Baseline comparison on real world datasets}\label{sec:rare_baselines}
We compare E-RTPM and A-RTPM with the adapted baseline ARZ-Miner in terms of runtime and memory usage. 
Figs. \ref{fig:rareruntimebaselineNIST}, \ref{fig:rareruntimebaselineSC}, \ref{fig:rarememorybaselineNIST}, and \ref{fig:rarememorybaselineSC} show the comparison results on NIST and SC. Note that Figs. \ref{fig:rareruntimebaselineNIST}-\ref{fig:rare_scaleAttribute_SC} use the same legend and log-scale y axes. 

As shown in Figs. \ref{fig:rareruntimebaselineNIST} and \ref{fig:rareruntimebaselineSC}, A-RTPM achieves the best runtime among all methods, and E-RTPM has better runtime than the baseline. The range and average speedups of A-RTPM compared to other methods are: $[1.9$-$7.2]$ and $3.4$ (E-RTPM), $[5.4$-$48.9]$ and $16.5$ (ARZ-Miner). The speedup of E-RTPM compared to the baseline is $[2.9$-$24.7]$ and $7.4$ on average.
Note that the time to compute MI, $\mu_{\min}$, and $\mu_{\max}$ for NIST and SC in Figs. \ref{fig:rareruntimebaselineNIST} and \ref{fig:rareruntimebaselineSC} are $35.4$ and $28.7$ seconds, respectively, i.e., negligible compared to the total runtime.

In terms of memory consumption, as shown in Figs. \ref{fig:rarememorybaselineNIST} and \ref{fig:rarememorybaselineSC}, A-RTPM uses the least memory, while E-RTPM uses less memory than the baseline. A-RTPM consumes $[1.6$-$3.9]$ (on average $2.1$) times less memory than E-RTPM, and $[7.2$-$120.6]$ (on average $24.1$) times less  than ARZ-Miner. E-RTPM uses $[4.6$-$61.8]$ (on average $14.7$) times less memory than ARZ-Miner.
\subsubsection{RTPM: Scalability evaluation on synthetic datasets}\label{sec:rarescalability}
As discussed in Section \ref{sec:FTPMfTSMining}, the complexity of GTPM in general (and RTPM in particular) is driven by two main factors: (1) the number of temporal sequences, and (2) the number of time series. The  evaluation on real-world datasets has shown that E-RTPM and A-RTPM outperform the baseline significantly in both runtimes and memory usage. 
However, to further assess the scalability of RTPM, we scale these two factors using synthetic datasets. Specifically, starting from the real-world datasets, we generate $10$ times more sequences, and create up to $1000$ synthetic time series. We then evaluate the scalability of RTPM in two scenarios: varying the number of sequences, and varying the number of time series. 
 
Figs. \ref{fig:rare_scaleSequence_Energy} and \ref{fig:rare_scaleSequence_SC} show the runtimes of A-RTPM, E-RTPM and the baseline when the number of sequences changes. We can see that A-RTPM and E-RTPM outperform and scale better than the baseline in this configuration.
The range and average speedups of A-RTPM w.r.t. other methods are: [$2.3$-$5.7$] and $3.2$ (E-RTPM), [$5.1$-$19.8$] and $12.5$ (ARZ-Miner). Similarly, the range and average speedups of E-RTPM compared to ARZ-Miner are [$2.7$-$7.6$] and $5.3$. 

Figs. \ref{fig:rare_scaleAttribute_Energy} and \ref{fig:rare_scaleAttribute_SC} compare the runtimes of A-RTPM with other methods when changing the number of time series. It is seen that, A-RTPM achieves highest speedup in this configuration. 
The range and average speedups of A-RTPM are [$3.5$-$7.4$] and $4.6$ (E-RTPM), [$7.2$-$24.8$] and $15.2$ (ARZ-Miner), and of E-RTPM is [$3.6$-$9.5$] and $6.4$ (ARZ-Miner). 

On average, E-RTPM consumes $17.2$ times less memory than the baseline, while A-RTPM uses $20.6$ times less memory than E-RTPM and the baseline in the scalability study.
Furthermore, Fig. \ref{fig:rare_scaleAttribute_Energy_sub1} shows that A-RTPM and E-RTPM can scale well on big datasets while the baseline cannot. Specifically, the baseline fails for large configurations as it runs out of memory, e.g., when \# Time Series $\geq 1000$ on the synthetic NIST. 
We add an additional bar chart for A-RTPM, including the time to compute MI, $\mu_{\min}$, and $\mu_{\max}$ (top red) and the mining time (bottom blue) for comparison, showing that this time is negligible.

Finally, the percentage of time series and events pruned by A-RTPM in the scalability test are provided in Table \ref{tbl:rarePrunedAttributesEventsPercent}. Note that for the NIST dataset, every time series has two events, On and Off. Thus, the percentage of pruned time series and the percentage of pruned events are the same in NIST.
We can see that the higher $\sigma_{\min}$, $\delta$, and $\sigma_{\max}$, the more time series (events) are pruned. This is because higher $\sigma_{\min}$ and $\delta$ result in higher $\mu_{\min}$, and higher $\sigma_{\max}$ results in lower $\mu_{\max}$, and thus, more pruned time series.
\subsubsection{E-RTPM:  Evaluation of different pruning techniques}\label{rare_sec_exact}
We evaluate the following combinations of E-RTPM pruning techniques: (1) NoPrune: E-RTPM with no pruning, (2) Apriori: E-RTPM with Apriori-based pruning (Lemmas \ref{lem2}, \ref{lem3}), (3) Trans: E-RTPM with transitivity-based pruning (Lemmas \ref{lem:transitivity}, \ref{lem:filter}, \ref{lem5}, \ref{lem6}), and (4) All: E-RTPM applied both pruning techniques. 

We use $3$ different scenarios that vary: the minimum support, the minimum confidence, and the maximum support. Figs. \ref{fig:rarepruningExactRTPM1}, \ref{fig:rarepruningExactRTPM2} 
show the results. We see that (All)-E-RTPM has the best performance of all versions, with a speedup over (NoPrune)-E-RTPM ranging from $15$ up to $74$, depending on the configurations. Thus, the proposed prunings are very effective in improving E-RTPM performance. Furthermore, (Trans)-E-RTPM delivers a larger speedup than (Apriori)-E-RTPM, with the average speedup between $12$ and $28$ for (Trans)-E-RTPM, and between $7$ and $19$  for (Apriori)-E-RTPM, but applying both yields the best speedup.
\begin{figure*}[!t]
	\hspace{1in}
	\ref{legendcomparison}
	\clearpage
	\vspace{-0.1in}
	\begin{minipage}[t]{1\columnwidth} 
		\centering
		\begin{subfigure}{0.45\columnwidth}
			\centering
			\resizebox{\linewidth}{!}{
				\begin{tikzpicture}[scale=0.2]
					\begin{axis}[
						compat=newest,
						xlabel={$\sigma_{\min}$ (\%)},
						ylabel={Runtime (sec)}, 
						label style={font=\Huge},
						ticklabel style = {font=\huge},
						xticklabel style = {xshift=2mm}, 
						xmin=20, xmax=100,
						ymin=0.1, ymax=100000,
						xtick={20,40,60,80,100},
						legend columns=-1,
						legend entries = {A-FTPM, E-FTPM, A-HTPGM, E-HTPGM, Z-Miner, TPMiner, IEMiner, H-DFS},
						legend style={nodes={scale=0.5,  transform shape}, font=\Large},
						legend to name={legendcomparison},
						ymode=log,
						log basis y={10},
						ymajorgrids=true,
						grid style=dashed,
						line width=1.75pt
						]
						\addplot[
						color=blue,
						mark=square,
						mark size=4pt,
						]
						coordinates {
							(20,352)(40,126)(60,50)(80,16)(100,0.2)
						};
						\addplot[
						color=teal,
						mark=asterisk,
						mark size=4pt,
						]
						coordinates {
							(20,1661)(40,609)(60,121)(80,47)(100,0.5)
						};
						\addplot[
						color=red,
						dashed,
						mark=asterisk,
						mark size=4pt,
						]
						coordinates {
							(20,1174)(40,426)(60,100)(80,30)(100,0.3)
						};
						\addplot[
						color=brown,
						dashed,
						mark=oplus*,
						mark size=3pt,
						]
						coordinates {
							(20,3968)(40,1265)(60,556)(80,79)(100,0.8)
						};
						\addplot[
						color=cyan,
						mark=pentagon,
						mark size=4pt,
						]
						coordinates {
							(20,19063)(40,3078)(60,1660)(80,160)(100,3)
						};
						\addplot[
						color=red,
						mark=triangle,
						mark size=4pt,
						]
						coordinates {
							(20,31445)(40,5817)(60,2957)(80,433)(100,9)
						};
						\addplot[
						color=black,
						mark=*,
						mark size=4pt,
						]
						coordinates {
							(20,63440)(40,8479)(60,4571)(80,822)(100,18)
						};
						\addplot[
						color=violet,
						mark=diamond,
						mark size=5pt,
						]
						coordinates {
							(20,80864)(40,16567)(60,7628)(80,1657)(100,31)
						};		
					\end{axis}
				\end{tikzpicture}
			}
			\captionsetup{justification=centering, font=scriptsize}
			\caption{\scriptsize Varying $\sigma_{\min}$}
		\end{subfigure}
		\begin{subfigure}{0.45\columnwidth}
			\centering
			\resizebox{\linewidth}{!}{
				\begin{tikzpicture}[scale=0.2]
					\begin{axis}[
						compat=newest,
						xlabel={$\delta$ (\%)},
						ylabel={Runtime (sec)}, 
						label style={font=\Huge},
						ticklabel style = {font=\huge},
						xticklabel style = {xshift=2mm}, 
						xmin=20, xmax=100,
						ymin=0.1, ymax=100000,
						xtick={20,40,60,80,100},
						legend columns=-1,
						legend entries = {A-FTPM, E-FTPM, A-HTPGM, E-HTPGM, Z-Miner, TPMiner, IEMiner, H-DFS},
						legend style={nodes={scale=0.5,  transform shape}, font=\Large},
						legend to name={legendcomparison},
						ymode=log,
						log basis y={10},
						ymajorgrids=true,
						grid style=dashed,
						line width=1.75pt
						]
						\addplot[
						color=blue,
						mark=square,
						mark size=4pt,
						]
						coordinates {
							(20,952)(40,384)(60,101)(80,20)(100,0.4)
						};
						\addplot[
						color=teal,
						mark=asterisk,
						mark size=4pt,
						]
						coordinates {
							(20,2061)(40,751)(60,306)(80,51)(100,1)
						};
						\addplot[
						color=red,
						dashed,
						mark=asterisk,
						mark size=4pt,
						]
						coordinates {
							(20,1174)(40,516)(60,200)(80,41)(100,0.7)
						};
						\addplot[
						color=brown,
						dashed,
						mark=oplus*,
						mark size=3pt,
						]
						coordinates {
							(20,3968)(40,1465)(60,836)(80,109)(100,1.5)
						};
						\addplot[
						color=cyan,
						mark=pentagon,
						mark size=4pt,
						]
						coordinates {
							(20,19063)(40,4038)(60,2330)(80,291)(100,5)
						};
						\addplot[
						color=red,
						mark=triangle,
						mark size=4pt,
						]
						coordinates {
							(20,31445)(40,8967)(60,4036)(80,609)(100,12)
						};
						\addplot[
						color=black,
						mark=*,
						mark size=4pt,
						]
						coordinates {
							(20,63440)(40,16284)(60,9216)(80,954)(100,20)
						};
						\addplot[
						color=violet,
						mark=diamond,
						mark size=5pt,
						]
						coordinates {
							(20,80864)(40,29967)(60,15967)(80,1738)(100,35)
						};		
					\end{axis}
				\end{tikzpicture}
			}
		\captionsetup{justification=centering, font=scriptsize}
		\caption{\scriptsize Varying $\delta$}
		\end{subfigure}
		\vspace{-0.1in}
		\captionsetup{justification=centering, font=small}
		\caption{FTPM-Runtime Comparison on NIST (real-world)}
		\label{fig:runtimebaselineRE}
	\end{minipage}%
	\hspace{0.2in}  
	\begin{minipage}[t]{1\columnwidth} 
		\centering
		\begin{subfigure}{0.45\columnwidth}
			\centering
			\resizebox{\linewidth}{!}{
				\begin{tikzpicture}[scale=0.2]
					\begin{axis}[
						compat=newest,
						xlabel={$\sigma_{\min}$ (\%)},
						ylabel={Runtime (sec)}, 
						label style={font=\Huge},
						ticklabel style = {font=\huge},
						xticklabel style = {xshift=2mm}, 
						xmin=20, xmax=100,
						ymin=0.1, ymax=10000,
						xtick={20,40,60,80,100},
						legend columns=-1,
						legend entries = {A-FTPM, E-FTPM, A-HTPGM, E-HTPGM, Z-Miner, TPMiner, IEMiner, H-DFS},
						legend style={nodes={scale=0.5,  transform shape}, font=\Large},
						legend to name={legendcomparison},
						ymode=log,
						log basis y={10},
						ymajorgrids=true,
						grid style=dashed,
						line width=1.75pt
						]
						\addplot[
						color=blue,
						mark=square,
						mark size=4pt,
						]
						coordinates {
							(20,30)(40,14)(60,7)(80,0.4)(100,0.2)
						};
						\addplot[
						color=teal,
						mark=asterisk,
						mark size=4pt,
						]
						coordinates {
							(20,75)(40,46)(60,14)(80,1)(100,0.5)
						};
						\addplot[
						color=red,
						dashed,
						mark=asterisk,
						mark size=4pt,
						]
						coordinates {
							(20,60)(40,26)(60,10)(80,0.7)(100,0.3)
						};
						\addplot[
						color=brown,
						dashed,
						mark=oplus*,
						mark size=3pt,
						]
						coordinates {
							(20,87)(40,58)(60,16)(80,1.4)(100,0.9)
						};
						\addplot[
						color=cyan,
						mark=pentagon,
						mark size=4pt,
						]
						coordinates {
							(20,194)(40,116)(60,30)(80,4)(100,3)
						};
						\addplot[
						color=red,
						mark=triangle,
						mark size=4pt,
						]
						coordinates {
							(20,418)(40,255)(60,90)(80,7)(100,6)
						};
						\addplot[
						color=black,
						mark=*,
						mark size=4pt,
						]
						coordinates {
							(20,1419)(40,657)(60,128)(80,10)(100,8)
						};
						\addplot[
						color=violet,
						mark=diamond,
						mark size=5pt,
						]
						coordinates {
							(20,2516)(40,952)(60,240)(80,15)(100,12)
						};		
					\end{axis}
				\end{tikzpicture}
			}
			\captionsetup{justification=centering, font=scriptsize}
			\caption{\scriptsize Varying $\sigma_{\min}$}
		\end{subfigure}
		\begin{subfigure}{0.45\columnwidth}
			\centering
			\resizebox{\linewidth}{!}{
				\begin{tikzpicture}[scale=0.2]
					\begin{axis}[
						compat=newest,
						xlabel={$\delta$ (\%)},
						ylabel={Runtime (sec)}, 
						label style={font=\Huge},
						ticklabel style = {font=\huge},
						xticklabel style = {xshift=2mm}, 
						xmin=20, xmax=100,
						ymin=0.1, ymax=10000,
						xtick={20,40,60,80,100},
						legend columns=-1,
						legend entries = {A-FTPM, E-FTPM, A-HTPGM, E-HTPGM, Z-Miner, TPMiner, IEMiner, H-DFS},
						legend style={nodes={scale=0.5,  transform shape}, font=\Large},
						legend to name={legendcomparison},
						ymode=log,
						log basis y={10},
						ymajorgrids=true,
						grid style=dashed,
						line width=1.75pt
						]
						\addplot[
						color=blue,
						mark=square,
						mark size=4pt,
						]
						coordinates {
							(20,30)(40,22)(60,10)(80,1.4)(100,0.5)
						};
						\addplot[
						color=teal,
						mark=asterisk,
						mark size=4pt,
						]
						coordinates {
							(20,75)(40,57)(60,18)(80,3)(100,1.1)
						};
						\addplot[
						color=red,
						dashed,
						mark=asterisk,
						mark size=4pt,
						]
						coordinates {
							(20,60)(40,48)(60,15)(80,2.2)(100,1)
						};
						\addplot[
						color=brown,
						dashed,
						mark=oplus*,
						mark size=3pt,
						]
						coordinates {
							(20,87)(40,81)(60,30)(80,5.1)(100,2.5)
						};
						\addplot[
						color=cyan,
						mark=pentagon,
						mark size=4pt,
						]
						coordinates {
							(20,194)(40,124)(60,63)(80,8)(100,6)
						};
						\addplot[
						color=red,
						mark=triangle,
						mark size=4pt,
						]
						coordinates {
							(20,418)(40,367)(60,104)(80,12)(100,10)
						};
						\addplot[
						color=black,
						mark=*,
						mark size=4pt,
						]
						coordinates {
							(20,1419)(40,759)(60,217)(80,24)(100,15)
						};
						\addplot[
						color=violet,
						mark=diamond,
						mark size=5pt,
						]
						coordinates {
							(20,2516)(40,1126)(60,308)(80,36)(100,20)
						};		
					\end{axis}
				\end{tikzpicture}
			}
			\captionsetup{justification=centering, font=scriptsize}
			\caption{\scriptsize Varying $\delta$}
		\end{subfigure}
		\vspace{-0.1in}
		\captionsetup{justification=centering, font=small}
		\caption{FTPM-Runtime Comparison on SC (real-world)}
		\label{fig:runtimebaselineSC}
	\end{minipage}%
	\vspace{-0.02in}  
\end{figure*}  
\begin{figure*}[!t]
	\vspace{-0.1in}
	\begin{minipage}[t]{1\columnwidth} 
		\centering
		\begin{subfigure}{0.45\columnwidth}
			\centering
			\resizebox{\linewidth}{!}{
				\begin{tikzpicture}[scale=0.2]
					\begin{axis}[
						compat=newest,
						xlabel={$\sigma_{\min}$ (\%)},
						ylabel={Memory Usage (MB)},
						label style={font=\huge},
						ticklabel style = {font=\huge},
						xticklabel style = {xshift=2mm}, 
						xmin=20, xmax=100,
						ymin=10, ymax=100000,
						xtick={20,40,60,80,100},
						legend columns=-1,
						legend entries = {A-FTPM, E-FTPM, A-HTPGM, E-HTPGM, Z-Miner, TPMiner, IEMiner, H-DFS},
						legend style={nodes={scale=0.5,  transform shape}, font=\Large},
						legend to name={legendcomparison},
						ymode=log,
						log basis y={10},
						ymajorgrids=true,
						grid style=dashed,
						line width=1.75pt
						]
						\addplot[
						color=blue,
						mark=square,
						mark size=4pt,
						]
						coordinates {
							(20,752)(40,297)(60,240)(80,118)(100,30)
						};
						\addplot[
						color=teal,
						mark=asterisk,
						mark size=4pt,
						]
						coordinates {
							(20,1521)(40,662)(60,452)(80,252)(100,54)
						};
						\addplot[
						color=red,
						dashed,
						mark=asterisk,
						mark size=4pt,
						]
						coordinates {
							(20,1075)(40,404)(60,360)(80,157)(100,45)
						};
						\addplot[
						color=brown,
						dashed,
						mark=oplus*,
						mark size=3pt,
						]
						coordinates {
							(20,1748)(40,1005)(60,612)(80,314)(100,86)
						};
						\addplot[
						color=cyan,
						mark=pentagon,
						mark size=4pt,
						]
						coordinates {
							(20,91875)(40,17026)(60,12016)(80,1934)(100,754)
						};
						\addplot[
						color=red,
						mark=triangle,
						mark size=4pt,
						]
						coordinates {
							(20,5658)(40,1662)(60,828)(80,455)(100,176)
						};
						\addplot[
						color=black,
						mark=*,
						mark size=4pt,
						]
						coordinates {
							(20,7241)(40,2468)(60,1206)(80,657)(100,281)
						};
						\addplot[
						color=violet,
						mark=diamond,
						mark size=5pt,
						]
						coordinates {
							(20,11976)(40,6294)(60,2108)(80,877)(100,385)
						};		
					\end{axis}
				\end{tikzpicture}
			}
			\captionsetup{justification=centering, font=scriptsize}
			\caption{\scriptsize Varying $\sigma_{\min}$}
		\end{subfigure}
		\begin{subfigure}{0.45\columnwidth}
		\centering
		\resizebox{\linewidth}{!}{
			\begin{tikzpicture}[scale=0.2]
				\begin{axis}[
					compat=newest,
					xlabel={$\delta$ (\%)},
					ylabel={Memory Usage (MB)},
					label style={font=\huge},
					ticklabel style = {font=\huge},
					xticklabel style = {xshift=2mm}, 
					xmin=20, xmax=100,
					ymin=10, ymax=100000,
					xtick={20,40,60,80,100},
					legend columns=-1,
					legend entries = {A-FTPM, E-FTPM, A-HTPGM, E-HTPGM, Z-Miner, TPMiner, IEMiner, H-DFS},
					legend style={nodes={scale=0.5,  transform shape}, font=\Large},
					legend to name={legendcomparison},
					ymode=log,
					log basis y={10},
					ymajorgrids=true,
					grid style=dashed,
					line width=1.75pt
					]
					\addplot[
					color=blue,
					mark=square,
					mark size=4pt,
					]
					coordinates {
						(20,752)(40,364)(60,301)(80,130)(100,36)
					};
					\addplot[
					color=teal,
					mark=asterisk,
					mark size=4pt,
					]
					coordinates {
						(20,1521)(40,982)(60,578)(80,302)(100,79)
					};
					\addplot[
					color=red,
					dashed,
					mark=asterisk,
					mark size=4pt,
					]
					coordinates {
						(20,1075)(40,543)(60,420)(80,262)(100,62)
					};
					\addplot[
					color=brown,
					dashed,
					mark=oplus*,
					mark size=3pt,
					]
					coordinates {
						(20,1748)(40,1358)(60,747)(80,421)(100,120)
					};
					\addplot[
					color=cyan,
					mark=pentagon,
					mark size=4pt,
					]
					coordinates {
						(20,91875)(40,19026)(60,16082)(80,5241)(100,886)
					};
					\addplot[
					color=red,
					mark=triangle,
					mark size=4pt,
					]
					coordinates {
						(20,5658)(40,3121)(60,1134)(80,700)(100,204)
					};
					\addplot[
					color=black,
					mark=*,
					mark size=4pt,
					]
					coordinates {
						(20,7241)(40,4261)(60,1426)(80,905)(100,302)
					};
					\addplot[
					color=violet,
					mark=diamond,
					mark size=5pt,
					]
					coordinates {
						(20,11976)(40,7882)(60,3681)(80,1343)(100,426)
					};		
				\end{axis}
			\end{tikzpicture}
		}
		\captionsetup{justification=centering, font=scriptsize}
		\caption{\scriptsize Varying $\delta$}
	\end{subfigure}
		\vspace{-0.1in}
		\captionsetup{justification=centering, font=small}
		\caption{FTPM-Memory Usage Comparison on {\footnotesize NIST (real-world)}}
		\label{fig:memorybaselineRE}
	\end{minipage}    
	\hspace{0.2in}  
	\begin{minipage}[t]{1\columnwidth} 
		\centering
		\begin{subfigure}{0.45\columnwidth}
			\centering
			\resizebox{\linewidth}{!}{
				\begin{tikzpicture}[scale=0.2]
					\begin{axis}[
						compat=newest,
						xlabel={$\sigma_{\min}$ (\%)},
						ylabel={Memory Usage (MB)},
						label style={font=\huge},
						ticklabel style = {font=\huge},
						xticklabel style = {xshift=2mm}, 
						xmin=20, xmax=100,
						ymin=10, ymax=10000,
						xtick={20,40,60,80,100},
						legend columns=-1,
						legend entries = {A-FTPM, E-FTPM, A-HTPGM, E-HTPGM, Z-Miner, TPMiner, IEMiner, H-DFS},
						legend style={nodes={scale=0.5,  transform shape}, font=\Large},
						legend to name={legendcomparison},
						ymode=log,
						log basis y={10},
						ymajorgrids=true,
						grid style=dashed,
						line width=1.75pt
						]
						\addplot[
						color=blue,
						mark=square,
						mark size=4pt,
						]
						coordinates {
							(20,142)(40,85)(60,70)(80,28)(100,20)
						};
						\addplot[
						color=teal,
						mark=asterisk,
						mark size=4pt,
						]
						coordinates {
							(20,490)(40,158)(60,106)(80,35)(100,32)
						};
						\addplot[
						color=red,
						dashed,
						mark=asterisk,
						mark size=4pt,
						]
						coordinates {
							(20,361)(40,110)(60,87)(80,30)(100,27)
						};
						\addplot[
						color=brown,
						dashed,
						mark=oplus*,
						mark size=3pt,
						]
						coordinates {
							(20,710)(40,201)(60,162)(80,50)(100,36)
						};
						\addplot[
						color=cyan,
						mark=pentagon,
						mark size=4pt,
						]
						coordinates {
							(20,2490)(40,1162)(60,648)(80,209)(100,134)
						};
						\addplot[
						color=red,
						mark=triangle,
						mark size=4pt,
						]
						coordinates {
							(20,1002)(40,352)(60,210)(80,61)(100,47)
						};
						\addplot[
						color=black,
						mark=*,
						mark size=4pt,
						]
						coordinates {
							(20,1397)(40,554)(60,301)(80,95)(100,60)
						};
						\addplot[
						color=violet,
						mark=diamond,
						mark size=5pt,
						]
						coordinates {
							(20,1893)(40,725)(60,405)(80,137)(100,98)
						};		
					\end{axis}
				\end{tikzpicture}
			}
			\captionsetup{justification=centering, font=scriptsize}
			\caption{\scriptsize Varying $\sigma_{\min}$}
		\end{subfigure}
		\begin{subfigure}{0.45\columnwidth}
			\centering
			\resizebox{\linewidth}{!}{
				\begin{tikzpicture}[scale=0.2]
					\begin{axis}[
						compat=newest,
						xlabel={$\delta$ (\%)},
						ylabel={Memory Usage (MB)},
						label style={font=\huge},
						ticklabel style = {font=\huge},
						xticklabel style = {xshift=2mm}, 
						xmin=20, xmax=100,
						ymin=10, ymax=10000,
						xtick={20,40,60,80,100},
						legend columns=-1,
						legend entries = {A-FTPM, E-FTPM, A-HTPGM, E-HTPGM, Z-Miner, TPMiner, IEMiner, H-DFS},
						legend style={nodes={scale=0.5,  transform shape}, font=\Large},
						legend to name={legendcomparison},
						ymode=log,
						log basis y={10},
						ymajorgrids=true,
						grid style=dashed,
						line width=1.75pt
						]
						\addplot[
						color=blue,
						mark=square,
						mark size=4pt,
						]
						coordinates {
							(20,142)(40,101)(60,75)(80,30)(100,25)
						};
						\addplot[
						color=teal,
						mark=asterisk,
						mark size=4pt,
						]
						coordinates {
							(20,490)(40,204)(60,120)(80,48)(100,42)
						};
						\addplot[
						color=red,
						dashed,
						mark=asterisk,
						mark size=4pt,
						]
						coordinates {
							(20,361)(40,141)(60,102)(80,45)(100,37)
						};
						\addplot[
						color=brown,
						dashed,
						mark=oplus*,
						mark size=3pt,
						]
						coordinates {
							(20,710)(40,337)(60,186)(80,60)(100,58)
						};
						\addplot[
						color=cyan,
						mark=pentagon,
						mark size=4pt,
						]
						coordinates {
							(20,2490)(40,1467)(60,1069)(80,363)(100,250)
						};
						\addplot[
						color=red,
						mark=triangle,
						mark size=4pt,
						]
						coordinates {
							(20,1002)(40,578)(60,402)(80,119)(100,110)
						};
						\addplot[
						color=black,
						mark=*,
						mark size=4pt,
						]
						coordinates {
							(20,1397)(40,824)(60,581)(80,160)(100,126)
						};
						\addplot[
						color=violet,
						mark=diamond,
						mark size=5pt,
						]
						coordinates {
							(20,1893)(40,1126)(60,821)(80,249)(100,215)
						};		
					\end{axis}
				\end{tikzpicture}
			}
			\captionsetup{justification=centering, font=scriptsize}
			\caption{\scriptsize Varying $\delta$}
		\end{subfigure}
		\vspace{-0.1in}
		\captionsetup{justification=centering, font=small}
		\caption{FTPM-Memory Usage Comparison on SC (real-world)}
		\label{fig:memorybaselineSC}
	\end{minipage}    
	\vspace{-0.04in}  
\end{figure*}   
\subsubsection{A-RTPM: Evaluation of accuracy}\label{sec:rare_accuracy}
To evaluate A-RTPM accuracy, we compare the patterns extracted by A-RTPM and E-RTPM. Table \ref{tbl:rareAccuracyReal} shows the accuracies of A-RTPM for different $\sigma_{\min}$, $\delta$, and $\sigma_{\max}$ on the real world datasets. It is seen that A-RTPM obtains high accuracy ($\ge 83\%$) with lowest $\sigma_{\min}$ and $\delta$, and highest $\sigma_{\max}$, e.g., $\sigma_{\min}=1\%$,$\delta=60\%,\sigma_{\max}=20\%$, and very high accuracy ($\ge 93\%$) with higher  $\sigma_{\min}$ and $\delta$, and lower $\sigma_{\max}$, e.g., $\sigma_{\min}=3\%$, $\delta=70\%$, $\sigma_{\max}=10\%$. 
\subsection{Quantitative Evaluation of FTPM}
\begin{table}[!t]
	\centering
	\begin{minipage}{\linewidth}
		\caption{The Accuracy of A-FTPM (\%)}
		\vspace{-0.1in}
		\resizebox{\linewidth}{1.3cm}{
			\small
			\begin{tabular}{|c|c|c|c|c|c|c|c|c|}
				\hline 
				\multirow{3}{*}{\bfseries $\sigma_{min}$ (\%)} & \multicolumn{8}{c|}{\bfseries $\delta$ (\%)} \\  
				\cline{2-9}   
				& \multicolumn{4}{c|}{\bfseries NIST} & \multicolumn{4}{c|}{\bfseries SC}
				\\  \cline{2-9}  
				& {\bfseries 10} & {\bfseries 20} & {\bfseries 50}  & {\bfseries 80} & {\bfseries 10} & {\bfseries 20} & {\bfseries 50}  & {\bfseries 80}  \\
				\hline
				10 & 87  & 89   & 91 & 94 & 78  & 83 & 98 & 100    \\  \hline			
				20 & 96  & 89   & 91  & 94  & 83  & 83 & 98  & 100  \\  \hline				
				50 & 100  & 100   & 96  & 94   & 99  & 99 & 98  & 100   \\  \hline					
				80 & 100  & 100   & 100  & 100 & 100  & 100 & 100  & 100  \\  \hline					
			\end{tabular}
		}			
		\label{tbl:ratiopatterns_Appro_Exact}
	\end{minipage}
\end{table}
\subsubsection{FTPM: Baselines comparison on real world datasets}\label{sec:baselines}
We compare E-FTPM and A-FTPM against the baselines in terms of runtime and memory usage. Further, we also compare E-FTPM and A-FTPM against E-HTPGM and A-HTPGM from the conference version \cite{ho2022efficientvldb} to assess the performance improvement obtained by using the new data structure.
Figs. \ref{fig:runtimebaselineRE}, \ref{fig:runtimebaselineSC}, \ref{fig:memorybaselineRE}, and \ref{fig:memorybaselineSC} show the experimental results on NIST and SC. 

We can see from Figs. \ref{fig:runtimebaselineRE} and \ref{fig:runtimebaselineSC} that A-FTPM achieves the fastest runtime among all methods, and E-FTPM has faster runtime than the baselines. On the tested datasets, the range and average speedups of A-FTPM compared to E-FTPM is $[1.5$-$6.1]$ and $2.7$, and compared to the baselines is $[4.2$-$356.1]$ and $45.8$. The range and average speedup of E-FTPM compared to the baselines is $[2.6$-$130.4]$ and $24.7$.

Note that the time to compute MI and $\mu_{\min}$ for NIST and SC datasets in Figs. \ref{fig:runtimebaselineRE} and \ref{fig:runtimebaselineSC} are $32.6$ and $26.4$ seconds, respectively, making it negligible in the total runtime. 
Moreover, by using the improved hierarchical hash table instead of the hierarchical pattern tree in \cite{ho2022efficientvldb}, both E-FTPM and A-FTPM are more efficient than E-HTPGM and A-HTPGM. The speedup of E-FTPM over E-HTPGM is in the range [$1.1$-$4.7$], and A-FTPM over A-HTPGM is in the range [$1.3$-$5.6$].

Finally, A-FTPM is most efficient, i.e., achieves highest speedup and memory saving, when the support threshold is low, e.g., $\sigma_{\min}=20\%$. This is because typical datasets often contain many patterns with very low support and confidence. Thus, using A-FTPM to prune uncorrelated series early helps save computational time and resources. However, the speedup comes at the cost of a small loss in accuracy (discussed in Section \ref{sec:accuracy}).

In terms of memory consumption, as shown in Figs. \ref{fig:memorybaselineRE} and \ref{fig:memorybaselineSC}, A-FTPM uses the least memory, while E-FTPM uses less memory than the baselines. A-FTPM consumes $[1.4$-$3.6]$ (on average $1.9$) times less memory than E-FTPM, and $[6.8$-$112.6]$ (on average $15.4$) times less than the baselines. E-FTPM uses $[4.1$-$58.2]$ (on average $5.8$) times less memory than the baselines. Compared to E-HTPGM and A-HTPGM \cite{ho2022efficientvldb},  E-FTPM and A-FTPM are both more memory efficient. E-FTPM consumes $[1.1$-$2.8]$ times less memory than E-HTPGM, while A-FTPM uses $[1.2$-$3.1]$ times less memory than A-HTPGM.

We also perform other experiments on FTPM, including scalability evaluation on synthetic datasets, and evaluation of different pruning techniques on real-world datasets as in RTPM. These experiments are reported in the electronic appendix.

\subsubsection{A-FTPM: Evaluation of the accuracy}\label{sec:accuracy}
We proceed to evaluate the accuracy of A-FTPM by comparing the patterns extracted by A-FTPM and E-FTPM. Table \ref{tbl:ratiopatterns_Appro_Exact} shows the accuracies of A-FTPM for different support and confidence thresholds on the real-world datasets. It is seen that A-FTPM obtains high accuracy ($\ge 78\%$) when $\sigma_{\min}$ and $\delta$ are low, e.g., $\sigma_{\min}=\delta=10\%$, and very high accuracy ($\ge 95\%$) when $\sigma_{\min}$ and $\delta$ are high, e.g., $\sigma_{\min}=\delta=50\%$.    

\textbf{Other experiments:} 
We analyze the effects of the tolerance buffer $\epsilon$, and the overlapping duration t$_{\text{ov}}$ to the quality of extracted patterns. The analysis can be found in the electronic appendix.

\section{Conclusion and Future Work}\label{sec:conclusion}
This paper presents our comprehensive Generalized Frequent Temporal Pattern Mining from Time Series (GTPMfTS) solution that offers: (1) an end-to-end GTPMfTS process to mine both rare and frequent temporal patterns from time series, (2) an efficient and exact Generalized Temporal Pattern Mining (GTPM) algorithm that employs efficient data structures and multiple pruning techniques to achieve fast mining, and (3) an approximate GTPM that uses mutual information to prune unpromising time series, allows GTPM to scale on big datasets. Extensive experiments conducted on real world and synthetic datasets for rare temporal pattern mining (RTPM) and frequent temporal pattern mining (FTPM) show that both exact and approximate algorithms for RTPM and FTPM outperform the baselines, consume less memory, and scale well on big datasets. Compared to the baselines, the approximate A-RTPM is up to an order of magnitude speedup and the approximate A-FTPM delivers two orders of magnitude speedup. In future work, we plan to extend GTPM to prune at the event level to further improve their performance.

\bibliographystyle{IEEEtran}
\bibliography{references}

\end{document}